\documentclass{emulateapj}

\usepackage{graphicx}
\usepackage{natbib}
\usepackage[usenames]{color}
\usepackage{subfigure}
\usepackage{multirow}
\usepackage{amsmath}
\usepackage{svn-multi}
\usepackage{xspace}
\usepackage{flafter}
\graphicspath{{./}}
\usepackage{setspace}

\shorttitle{The QUIET Instrument}
\shortauthors{The QUIET Collaboration}

\begin{document}

\author{
The QUIET Instrument 
\\
QUIET Collaboration---C.~Bischoff\altaffilmark{1,20},
A.~Brizius\altaffilmark{1,2},
I.~Buder\altaffilmark{1,20},
Y.~Chinone\altaffilmark{3,4},
K.~Cleary\altaffilmark{5},
R.~N.~Dumoulin\altaffilmark{6},
A.~Kusaka\altaffilmark{1,19},
R.~Monsalve\altaffilmark{7},
S.~K.~N\ae ss\altaffilmark{8},
L.~B.~Newburgh\altaffilmark{6,19},
G.~Nixon\altaffilmark{19},  
R.~Reeves\altaffilmark{5},
K.~M.~Smith\altaffilmark{1,19},
K.~Vanderlinde\altaffilmark{1,23},
I.~K.~Wehus\altaffilmark{9,10},
M.~Bogdan\altaffilmark{1},
R.~Bustos\altaffilmark{7,11,12},
S.~E.~Church\altaffilmark{13},
R.~Davis\altaffilmark{14},
C.~Dickinson\altaffilmark{14},
H.~K.~Eriksen\altaffilmark{8,15},
T.~Gaier\altaffilmark{16},
J.~O.~Gundersen\altaffilmark{7},
M.~Hasegawa\altaffilmark{3},
M.~Hazumi\altaffilmark{3},
C.~Holler\altaffilmark{10},
K.~M.~Huffenberger\altaffilmark{7},
W.~A.~Imbriale\altaffilmark{16},
K.~Ishidoshiro\altaffilmark{3},
M.~E.~Jones\altaffilmark{10},
P.~Kangaslahti\altaffilmark{16},
D.~J.~Kapner\altaffilmark{1,21},
C.~R.~Lawrence\altaffilmark{16},
E.~M.~Leitch\altaffilmark{16},
M.~Limon\altaffilmark{6},
J.~J.~McMahon\altaffilmark{17},
A.~D.~Miller\altaffilmark{6},
M.~Nagai\altaffilmark{3},
H.~Nguyen\altaffilmark{18},
T.~J.~Pearson\altaffilmark{5},
L.~Piccirillo\altaffilmark{14},
S.~J.~E.~Radford\altaffilmark{5},
A.~C.~S.~Readhead\altaffilmark{5},
J.~L.~Richards\altaffilmark{5},
D.~Samtleben\altaffilmark{2,22},
M.~Seiffert\altaffilmark{16},
M.~C.~Shepherd\altaffilmark{5},
S.~T.~Staggs\altaffilmark{19},
O.~Tajima\altaffilmark{1,3},
K.~L.~Thompson\altaffilmark{13},
R.~Williamson\altaffilmark{6,1},
 B.~Winstein\altaffilmark{1,$\dagger$},
E.~J.~Wollack\altaffilmark{24},
J.~T.~L.~Zwart\altaffilmark{6,25}
}

\vspace{+0.2in}

\altaffiltext{1}{Kavli Institute for Cosmological Physics, Department of Physics, Enrico Fermi Institute, The University of Chicago, Chicago, IL 60637, USA}
\altaffiltext{2}{Max-Planck-Institut f\"ur Radioastronomie, Auf dem H\"ugel 69, 53121 Bonn, Germany}
\altaffiltext{3}{High Energy Accelerator Research Organization (KEK), 1-1 Oho, Tsukuba, Ibaraki 305-0801, Japan}
\altaffiltext{4}{Astronomical Institute, Graduate School of Science, Tohoku University, Aramaki, Aoba, Sendai 980-8578, Japan}
\altaffiltext{5}{Cahill Center for Astronomy and Astrophysics, California Institute of Technology, 1200 E. California Blvd M/C 249-17, Pasadena, CA 91125, USA}
\altaffiltext{6}{Department of Physics and Columbia Astrophysics Laboratory, Columbia University, New York, NY 10027, USA}
\altaffiltext{7}{Department of Physics, University of Miami, 1320 Campo Sano Drive, Coral Gables, FL 33146, USA}
\altaffiltext{8}{Institute of Theoretical Astrophysics, University of Oslo, P.O. Box 1029 Blindern, N-0315 Oslo, Norway}
\altaffiltext{9}{Department of Physics, University of Oslo, P.O. Box 1048 Blindern, N-0316 Oslo, Norway}
\altaffiltext{10}{Department of Astrophysics, University of Oxford, Keble Road, Oxford OX1 3RH, UK}
\altaffiltext{11}{Departamento de Astronom\'ia, Universidad de Chile, Casilla 36-D, Santiago, Chile}
\altaffiltext{12}{Departamento de Astronom\'ia, Universidad de Concepci\'on, Casilla 160-C, Concepci\'on, Chile}
\altaffiltext{13}{Kavli Institute for Particle Astrophysics and Cosmology and Department of Physics, Stanford University, Varian Physics Building, 382 Via Pueblo Mall, Stanford, CA 94305, USA}
\altaffiltext{14}{Jodrell Bank Centre for Astrophysics, Alan Turing Building, School of Physics and Astronomy, The University of Manchester, Oxford Road, Manchester M13 9PL, UK}
\altaffiltext{15}{Centre of Mathematics for Applications, University of Oslo, P.O. Box 1053 Blindern, N-0316 Oslo, Norway}
\altaffiltext{16}{Jet Propulsion Laboratory, California Institute of Technology, 4800 Oak Grove Drive, Pasadena, CA, USA 91109}
\altaffiltext{17}{Department of Physics, University of Michigan, 450 Church Street, Ann Arbor, MI 48109, USA}
\altaffiltext{18}{Fermi National Accelerator Laboratory, Batavia, IL 60510, USA}
\altaffiltext{19}{Joseph Henry Laboratories of Physics, Jadwin Hall, Princeton University, Princeton, NJ 08544, USA; send correspondence to L.~Newburgh, \textbf{newburgh@princeton.edu}}
\altaffiltext{20}{Harvard-Smithsonian Center for Astrophysics, 60 Garden Street MS 43, Cambridge, MA 02138, USA}
\altaffiltext{21}{Micro Encoder Inc., Kirkland, WA 98034, USA}
\altaffiltext{22}{Nikhef, Science Park, Amsterdam, The Netherlands}
\altaffiltext{23}{Department of Physics, McGill University, 3600 Rue University, Montreal, Quebec H3A 2T8, Canada}
\altaffiltext{24}{Goddard Space Flight Center, Greenbelt, MD 20771, USA}
\altaffiltext{25}{Physics Department, University of the Western Cape, 
Private Bag X17, Bellville 7535, South Africa}
\altaffiltext{$\dagger$}{Deceased}

\slugcomment{
Submitted to ApJ---This paper should be cited as ``QUIET Collaboration (2012)''
}
\journalinfo{Submitted to ApJ---Draft version \today}

\begin{abstract}
\noindent
The Q/U Imaging ExperimenT (QUIET) is designed to measure polarization
in the Cosmic Microwave Background, 
targeting the imprint of inflationary gravitational waves at large
angular scales ( $\sim$ 1$^\circ$). Between
2008 October and 2010 December, two independent receiver arrays were
deployed sequentially on a 1.4\,m side-fed Dragonian telescope. 
The polarimeters which form the focal planes use a highly compact design based on High
Electron Mobility Transistors (HEMTs) that provides simultaneous measurements of 
the Stokes parameters Q, U, and I in a single module.  
The 17-element Q-band polarimeter array, with a central frequency of 43.1\,GHz, 
has the best sensitivity (69\,$\mu\mathrm{Ks}^{1/2}$) and the lowest instrumental systematic errors ever achieved in this band, 
contributing to the tensor-to-scalar ratio at $r < 0.1$.
The 84-element W-band polarimeter array has a sensitivity of
87\,$\mu\mathrm{Ks}^{1/2}$ at a central frequency of 94.5\,GHz. It has the lowest systematic errors to date, contributing at $r < 0.01$ \citep{quiet:2012}
The two arrays together cover 
multipoles in the range $\ell \approx 25-975$.
These are the largest HEMT-based arrays deployed to date.
This article describes the design, calibration, 
performance of, and sources of systematic 
error for the instrument.

\end{abstract}


\maketitle

\nocite{*}

\section{Introduction}
\label{sec:intro}

The Cosmic Microwave Background (CMB) is a powerful probe of 
early universe physics.  Measurements of the temperature anisotropy 
power spectrum are critical in establishing the concordance 
$\Lambda$CDM model \citep[e.g.][and references therein]{liddle:2000}, 
and measurements of CMB polarization 
currently provide the best prospects for confirming inflation or 
constraining the level of the primordial gravitational wave
background. The CMB is polarized via Thomson scattering off 
temperature anisotropies. The curl-free component of the polarization 
field (E-mode polarization) is generated by the same density inhomogeneities 
responsible for the measured temperature anisotropy.  However a measurement of
the E-mode polarization 
can break degeneracies 
in cosmological parameters inherent to measurements of the temperature 
anisotropy spectrum alone.  
A divergence-free component of the 
polarization field (B-mode polarization) is generated from three 
possible sources. One is from lensing of E-mode 
polarization into B-mode polarization by intervening large-scale structure along the line-of-sight.  
It can be used to probe structure formation in the early universe. The second could come from a large  
class of inflationary models that predict a spectrum of gravitational waves generated 
during inflation which could produce a measureable B-mode amplitude around  $\ell\sim$ 100 
\citep{Seljak:1997,Kamionkowski:1997,Dodelson:2009}.  
The detection of these B-modes, parameterized by the tensor-to-scalar ratio $r$, 
would provide a 
measurement of the energy scale of inflation.  A third contribution
to both E-mode and B-mode polarization spectra is expected from
polarized foreground emission.   Understanding the spectral 
dependence and spatial distribution of foregrounds is 
critical for pushing the limits of B-mode polarization detection or constraint.
\addtocounter{footnote}{-1}
The goal of detecting or placing competitive constraints on the 
inflationary B-mode CMB polarization signature led 
us to optimize QUIET\footnote{
Bruce Winstein, who died in 2011 February soon after observations were completed, 
was the principal investigator for QUIET. His intellectual and scientific guidance 
were crucial to the experiment's success.}
 for both sensitivity and control of systematic errors. 
We demodulate the signal at two phase-switching rates 
(``double demodulation'') to reduce both the $1/f$ noise and instrumental
systematic effects. 
In addition, our scan strategy, consisting of constant 
elevation scans performed 
between regular elevation steps, frequent boresight rotations, and natural 
sky rotation reduces systematic errors. 
Using arrays with two widely separated 
bandpasses centered between atmospheric absorption features 
allows us to separate a cosmological signal from Galactic foreground signals. 

\begin{figure}[t]
\centering
\subfigure[]{
\includegraphics[width=3.5in]{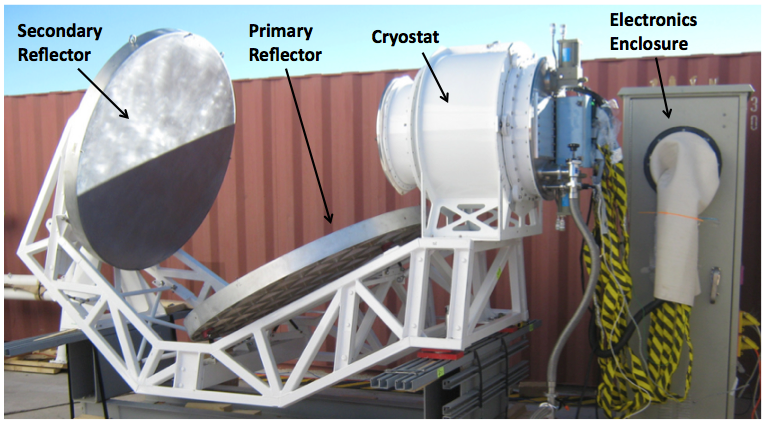}
\label{fig:overview-a}
}
\subfigure[]{
\includegraphics[width=3.5in]{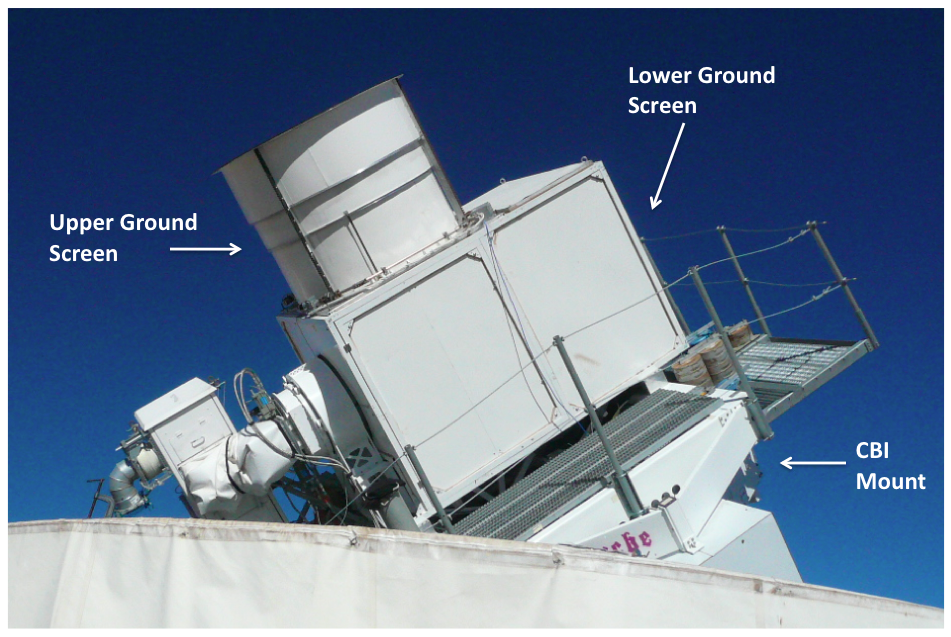}
\label{fig:overview-b}
}
\caption{{\em a:} The QUIET instrument before placement upon mount, showing the electronics enclosure, 
cryostat, and reflectors. {\em b:} The mounted instrument shown within an absorbing ground screen.}

\end{figure}

This paper describes the QUIET instrument, designed to measure the 
CMB polarization 
and the synchrotron foreground. 
Table~\ref{tab:exptsumm-a} 
lists the salient 
characteristics of the QUIET experiment. 
Figures~\ref{fig:overview-a} and \ref{fig:overview-b} show views of the receiver,
telescope, and electronics enclosure.
QUIET deployed two arrays of 19 and 90 HEMT-based coherent detector assemblies in the 
Chajnantor plateau in the Atacama Desert of Northern Chile. 
The extreme aridity of this region results in excellent
observing conditions for most of the year \citep{Radford/Holdaway:1998}. 
The arrays operate at central frequencies of 43\,GHz and 95\,GHz for the Q-band and W-band receivers, respectively, 
and are the largest HEMT-based arrays used to date.   In the focal plane,
each assembly contains 
passive waveguide components and a module, 
a small interchangeable HEMT-based electronics package.   
Within these two arrays,  17 (84) of the Q-band (W-band) 
assemblies are polarimeters, each measuring simultaneously the
Q, U, and I Stokes parameters.  The remaining 2 (6) assemblies measure the 
CMB temperature anisotropy (``differential-temperature assemblies'').  The Q-band and W-band assemblies are cooled 
to $\approx$ 20\,K and 27\,K, respectively, in a cryostat and placed at the focus of a 
1.4\,m side-fed Dragonian telescope enclosed in an absorbing ground 
screen.  The resulting full width at half maximum (FWHM)
angular resolution is 27.3$^{\prime}$ (11.7$^{\prime}$) for each Q-band (W-band) assembly.

\par
The following sections describe the observing site and strategy, optics,  
cryogenics and the optical window properties, polarimeter and differential-temperature assemblies, electronics, 
and calibration tools.  
Finally, we present a detailed description of the performance of both receivers.

\begin{table}
\centering
\caption{Instrument Overview}
{\scriptsize
\begin{tabular}{lll} 
Band                  & Q & W \\
\hline
Frequency (GHz) & 43.1 & 94.5 \\ 
Average Bandwidth (GHz) & 7.6 & 10.7 \\
\# of Polarization Assemblies & 17  & 84  \\ 
\# of Temperature Anisotropy Assemblies &  2 &      6 \\ 
FWHM Angular Resolution (arcmin) & 27.3 & 11.7 \\ 
Field of View ($^\circ$)  & 7.0 & 8.2 \\
$\ell$ range      & $\approx$ 25$-$475  & $\approx$ 25$-$975 \\
Instrument Sensitivity ($\mu\mathrm{K}{\rm s}^{1/2}$) & 69 & 87 \\
\hline
\end{tabular}
\label{tab:exptsumm-a}
}
\end{table}

\section{Observing Site and Strategy}

\begin{figure}
\centering
\includegraphics[width=3.5in]{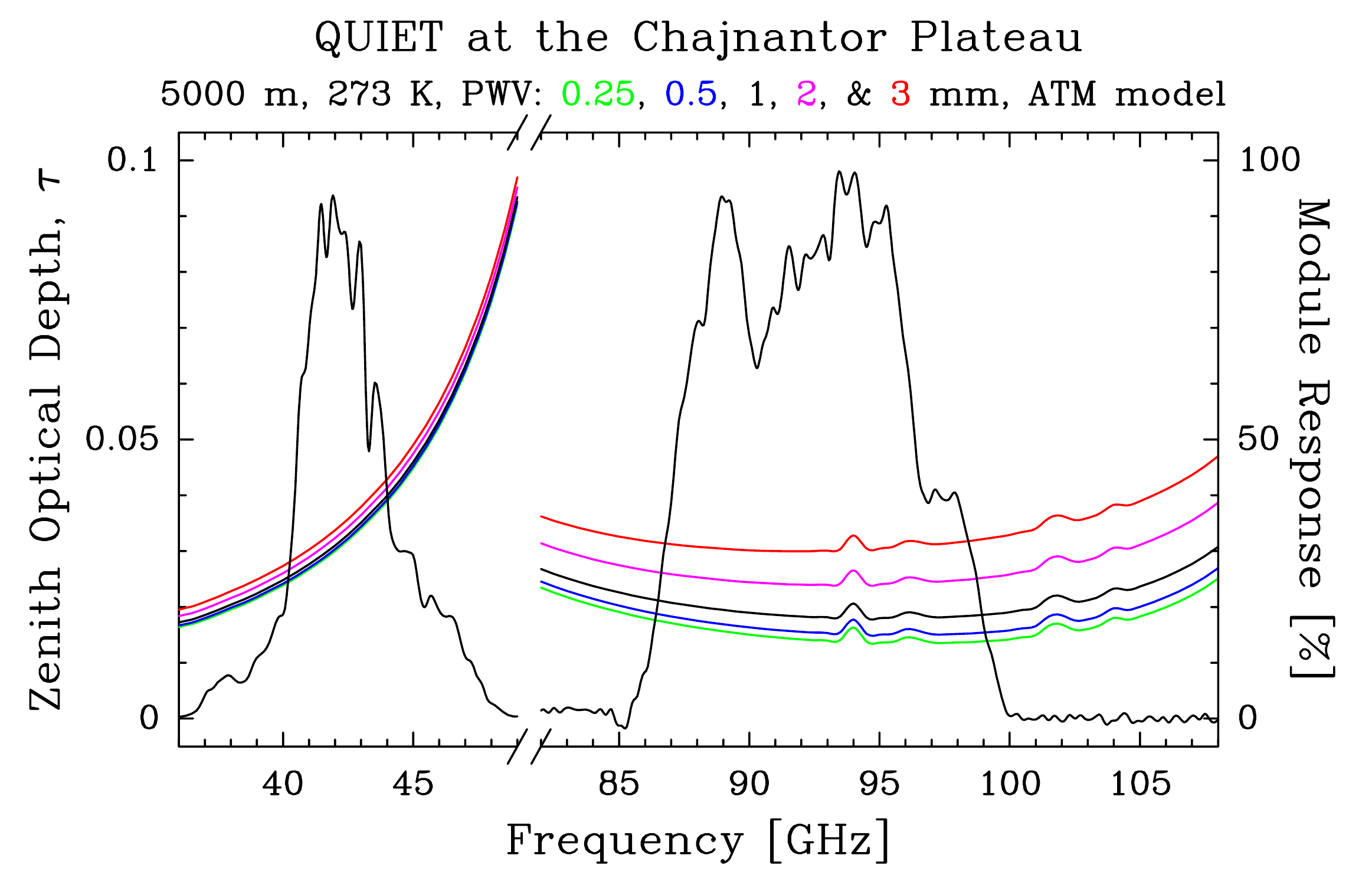}
\caption[Atmospheric opacity near the two QUIET frequency bands]{Zenith optical depth for typical atmospheric conditions at the  
Chajnantor plateau (left scale) and representative QUIET module  
bandpass responses (right scale). The atmospheric spectrum is calculated with the ATM model from ~\cite{Pardo2001}.
}
\label{fig:h2oabs}
\end{figure}

Observations (Table~\ref{tab:seasonoverview}) were performed at the Chajnantor plateau at 5080\,m
altitude in the Atacama Desert of Northern Chile ($67^{\circ}$45$^{\prime}$42$^{\prime}$$^{\prime}$W $23^{\circ}$1$^{\prime}$42$^{\prime}$$^{\prime}$S). 
Atmospheric conditions were monitored using data from a 183\,GHz line radiometer located at the APEX telescope \citep{Gusten:2006}, $\sim$ 1\,km away from the QUIET site. Typical atmospheric optical depths in our observing bands over all scanning elevations at Chajnantor are 0.02--0.1 (Figure~\ref{fig:h2oabs}). The median precipitable water vapor (PWV) was 1.2\,mm (0.9\,mm) during the Q-band (W-band) observing season. The data fraction surviving data selection for the Q-band (W-band) arrays are 82\% (75\%) of the data below the median PWV, and 59\% (54\%) of the data above the median PWV. Because the Q-band is affected primarily by the oxygen absorption line in the atmosphere, water vapor variations will typically have a greater effect on the W-band data quality than the Q-band data quality. 

\begin{table}[h]
\centering
\footnotesize{
\caption{Summary of Observations}
\begin{tabular}{l|ll} \hline
Band & Q & W \\ \hline
Season start & 2008 Oct 24 & 2009 Aug 12 \\
Season end & 2009 Jun 13 & 2010 Dec 22 \\
Total Observing Hours & 3458 & 7426 \\ 
CMB Observing (\%) & 77 & 72 \\ 
Galactic Observing (\%) & 12 & 14 \\
Calibration (\%) & 7 & 13 \\ 
Other (\%) & 4 & 1 \\ \hline
\end{tabular}
\label{tab:seasonoverview}
\begin{tabular}{llll}
CMB Fields & J2000 Center (RA, Dec) & Q (hours) & W (hours) \\ 
CMB-1 & 12$^h$04$^m$ $-39^{\circ}$ & 905 & 1855 \\
CMB-2 & 05$^h$12$^m$ $-39^{\circ}$ & 703 & 1444 \\
CMB-3 & 00$^h$48$^m$ $-48^{\circ}$ & 837 & 1389 \\
CMB-4 & 22$^h$44$^m$ $-36^{\circ}$ & 223 & 650 \\ \hline 
\end{tabular}
\tablecomments{The partition of the Q-band and W-band seasons by observation type (hours do not include data cuts obtained during data analysis for glitches, poor noise, etc.). `Other' includes data taken during engineering tests, aborted scans, etc. }
\label{tab:exptsumm-b}
}
\end{table}

We employed a fixed-elevation, azimuth-scanning technique: a $\sim15^{\circ}\times15^{\circ}$ field (the fields are given in Table~\ref{tab:exptsumm-b}) was scanned in azimuth as it drifted through the $\sim 7^{\circ}$ ($\sim 8^{\circ}$) field-of-view for the Q-band (W-band) array. These constant elevation scans (CES) typically lasted $\sim$ 40--90\,minutes. The telescope then re-tracked the field center and began another CES. By scanning at constant elevation for a given scan, we observed through a constant column density of atmosphere so that only weather variations within a scan contributed an atmospheric signal. Most calibration sources were observed at constant elevation, but occasionally we employed raster scans, changing elevation between azimuth slews to more rapidly observe a calibration source.

The infrastructure and three-axis driving mount previously used for the CBI experiment \citep{2002PASP..114...83P} was refurbished for QUIET, in part to enable rapid azimuth scanning. 
The mount control software is an augmented version
of the CBI control system. The principal modifications 
included the addition of support for rapid scanning of the
azimuth axis of the mount and for monitoring and archiving of data from
the QUIET receiver. This software consists of a central control and
data collection program, a graphical user interface program, a
real-time computer running the VxWorks\footnote{www.windriver.com} operating system to control the
telescope mount, and a real-time computer running Linux to control the
receiver. The mount was operated by a queue of non-interactive
observing scripts written in a custom control language. The
modifications 
supported high scanning accelerations without overwhelming the
counter-torque in the anti-backlash system of the azimuth
drive. Tracking accuracy is therefore sacrificed for high scanning speeds and
accelerations. However, accurate pointing information
can be reconstructed during the data analysis from frequent
readouts of the axis encoders and a dynamic model of the mechanical
response of the mount. To facilitate this, the CBI control system was
also modified to acquire encoder readouts at 100\,Hz. The modified
control system supports scans with coasting speeds of up to $6^{\circ}/\mathrm{s}$ and turnaround accelerations of up to $1.5^{\circ}/\mathrm{s^{2}}$. The accuracy of the encoder readout timestamps
is $\sim$0.5\,ms. The worst-case following error (the difference between the commanded trajector and the encoder-read trajectory) were $\sim 8^{\prime}$ at maximum acceleration during azimuth turn-arounds. Both the timing and the following errors resulted in negligible pointing errors during the observing seasons (pointing accuracy is discussed further in Section~\ref{sec:characterization:pointing}). 
We achieved a mean azimuthal scan speed of $\sim 5^\circ/\mathrm{s}$. The resulting scanning speed on the sky is elevation-dependent and corresponds to about $2^\circ / \mathrm{s}$, yielding azimuth scan frequencies of 45--100\,mHz. As each $15^{\circ}\times15^{\circ}$ observing field rises, its azimuthal extent with respect to the fixed telescope mount increases. As a result, the telescope azimuth slew size increases for higher elevation scans. Avoiding scanning through the azimuth limit leads to an upper elevation limit; the mount azimuth limit is $\sim440^{\circ}$ (80$^{\circ}$ past one full rotation), yielding an upper elevation limit of $75^{\circ}$ for CMB scans. The lower limit of the elevation range of the mount is $43^{\circ}$.

In addition to the azimuth and elevation axes, the mount provides a third rotation
axis through the boresight. We rotate this boresight angle (`deck angle') once per week in order to separate the polarization on the sky from that induced by systematic errors such as leakage from temperature to polarization.

\section{Optics}
\label{sec:optics}

The optical chain consists of a classical side-fed Dragonian antenna~\citep{Dragone} coupled to a platelet array of diffusion-bonded corrugated feed horns cooled to $\simeq$ 20 K ($\simeq$ 27 K) inside the Q-band (W-band) cryostat. The outputs of these optical elements are directed into the polarimeter and differential-temperature assemblies described in Sections~\ref{sec:polarimeter-assemblies} and~\ref{sec:TT}, respectively.  The main reflector (MR) and sub-reflector (SR) as well as the aperture of the cryostat are enclosed by an ambient temperature ($\simeq 270$ K), absorbing ground screen.  The design and characterization of the telescope, feed horns and ground screen are described in Sections~\ref{sec:telescope}, \ref{sec:instrument-feeds}, and \ref{sec:groundshield}, respectively.  The optical performance, as measured by the main beam, the sidelobes and the instrumental polarization, is described in Sections~\ref{sec:mainbeam}, \ref{sec:sidelobe}, and \ref{sec:leakagebeams}, respectively.

\begin{table}[t]
\begin{center}
\caption{Telescope Design Parameters}
\label{tab:telescope_parameters}
\begin{tabular}{c|c}
\hline
\hline
 
description, parameter  & design/actual value   \\ 

\hline
MR circular aperture diameter, D & 1470/1400 mm  \\
SR edge $\angle$, $\theta_e$ & $17^{\circ}$/$20^{\circ}$     \\
MR-SR separation, $\ell$   & 1270 mm       \\
MR offset $\angle$, $\theta_0$   & $-53^{\circ}$     \\
$\angle$ between MR and horn axes, $\theta_p$   & $-90^{\circ}$      \\
\hline
                                      & calculated value  \\
\hline
MR focal length, F  & 4904.1 mm \\
SR eccentricity, e  &  2.244 \\
$\angle$ between SR and MR axis, $\beta$  & $-63.37^{\circ}$  \\
SR interfocal distance, 2c & 6516.1 mm \\
MR offset distance, $d_0$ & 4890.2 mm \\
\hline

\end{tabular}
\tablecomments{The design values in the top half of the table were used to establish the calculated values in the lower half of the table.  For the first two parameters, the actual values listed supersede the design values for the purpose of fabrication.  Negative angles are measured clockwise with respect to the vertical axis shown in Figure~\ref{fig:tel-schem}.}
\end{center}
\end{table}

\subsection{Telescope}
\label{sec:telescope}

The telescope design requirements include: a wide field of view, excellent polarization characteristics, minimal beam distortion, minimal instrumental polarization, minimal spillover, and low sidelobes that could otherwise generate spurious polarization.  The latter requirements have often been met by CMB experiments by using either classical, dual offset Cassegrain antennas \citep[e.g.,][]{barkats:2005a}, Gregorian antennas \citep[e.g.,][]{meinhold:1993}, or shaped reflectors \citep[e.g.,][]{page/etal:2003a}.  
QUIET is the first CMB polarization experiment to take advantage of the wide field of view enabled by a classical Dragonian antenna~\citep{imbriale:2011}.  An additional advantage of the classical Dragonian antenna is that it satisfies the Mizuguchi condition~\citep{Mizuguchi} which, when combined with the very low cross-polar characteristics of the conical corrugated feed horns, yields very low antenna contribution to the instrumental polarization.  As pointed out by~\citet{chang:2004}, a classical Dragonian antenna affords two natural geometries, a front-fed design and a side-fed (or crossed) design.  QUIET uses the side-fed design because it allows for the use of a larger cryostat, and hence focal plane array, without obstructing the beam.

\begin{figure}[t]
\centering
\includegraphics[width=3in]{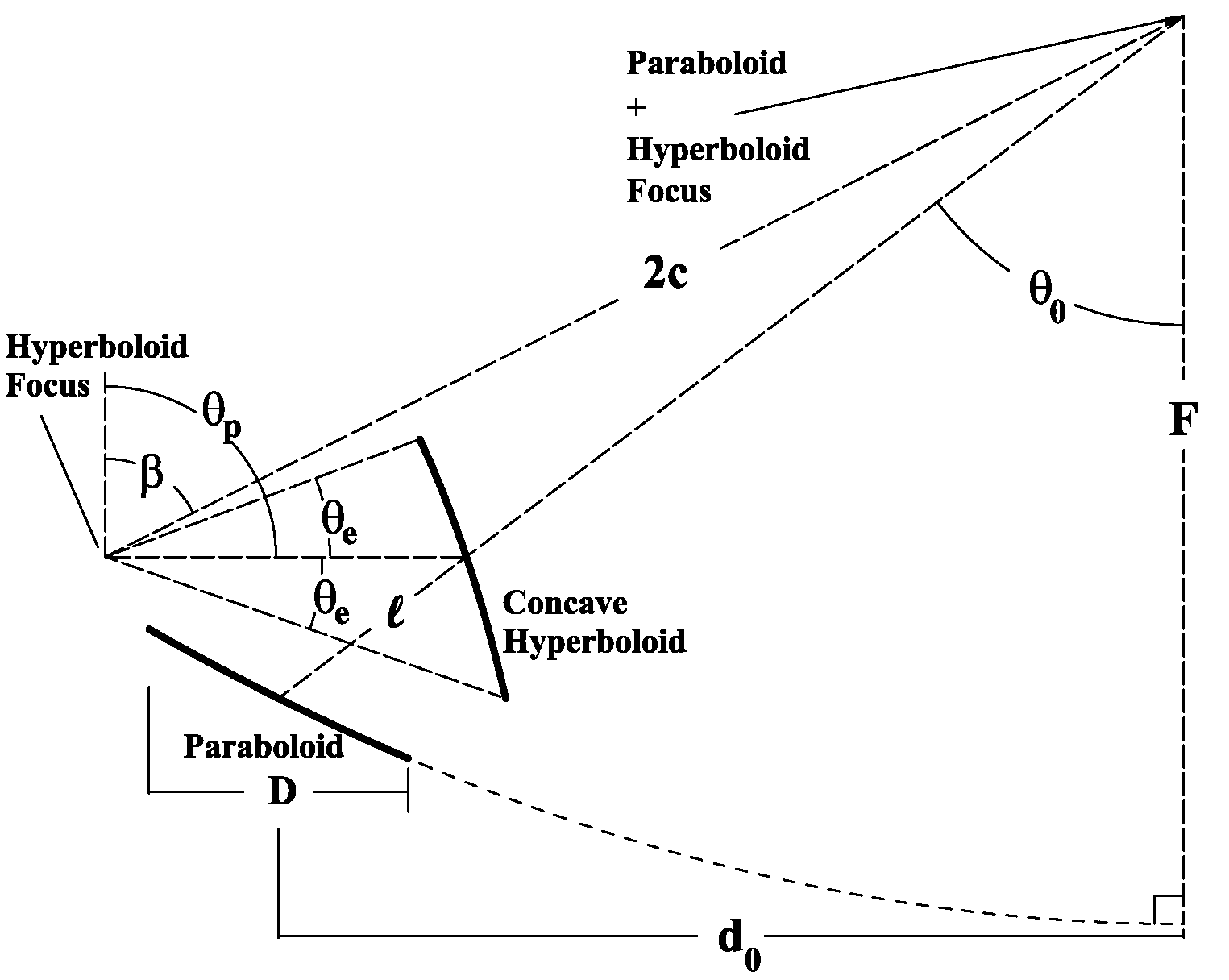}
\caption{This scaled schematic of the QUIET side-fed Dragonian antenna shows a number of the useful design parameters.  Table~\ref{tab:telescope_parameters} provides a description of each parameter and their values.}
\label{fig:tel-schem}
\end{figure}

\subsubsection{Telescope Design}

The design of the reflectors follows the procedure outlined by~\citet{chang:2004} and is augmented with a physical optics program~\citep{imbriale:1991} to predict beam patterns.  This procedure relies on the specification of the first five design parameters given in the top half of Table~\ref{tab:telescope_parameters} 
and shown in Figure~\ref{fig:tel-schem}.   Once these parameters are specified, a number of other useful parameters can be calculated including the MR focal length and the SR eccentricity, and these are listed in the lower half of Table~\ref{tab:telescope_parameters}.  The actual MR circular diameter was decreased slightly to 1400 mm, as noted by the actual value in Table~\ref{tab:telescope_parameters}.  Similarly, the actual SR circular diameter was increased slightly, also to 1400 mm, and this resulted in an increased value of the actual SR edge angle given by $20^{\circ}$ in Table~\ref{tab:telescope_parameters}.  The oversized SR reduces feed spillover for the horns on the edge of the array.  The design values (not the actual values) shown in the top half of Table~\ref{tab:telescope_parameters} were used to establish the calculated values shown in the lower half of Table~\ref{tab:telescope_parameters}.

\begin{table*}[t]
\begin{center}
\caption{Platelet Array Design Parameters}
\label{tab:plateletarray_parameters}
\begin{tabular}{c|c|c|c|c|c|c|c|c}
\hline
\hline
 
Frequency  & \# of & L$\times$W$\times$H & Mass & Aperture & Throat & \# of & Horn  & Semi-flare \\
 (band/GHz)	& Feeds &    (mm $\times$ mm $\times$ mm) & (kg) & Diameter & Diameter & Grooves & Separation & Angle  \\
       &    &    &    &  (mm) & (mm) &  & (mm) & (degrees) \\

\hline
Q/39--47 & 19 & 281.7$\times$427.3$\times$370.1 & 43.7 & 71.78 & 6.69 & 104 & 76.20 &  7.6$^{\circ}$\\
W/89--100 & 91 & 129.1$\times$427.8$\times$370.5 & 20.6 & 31.62 & 2.97 & 103 & 35.56  &  7.6$^{\circ}$\\

\hline
\end{tabular}
\end{center}
\end{table*}

\subsubsection{Telescope Fabrication and Alignment}

The telescope consists of two reflectors, the receiver cryostat (Figure~\ref{fig:overview-a}) and the structure that supports them (the `sled'). The reflectors are machined from solid pieces of aluminum 6061-T6, light-weighted on the reverse side leaving narrow ribs on a triangular grid, and attached with adjustable hexapod struts to the sled. The sled in turn is mounted on a deck structure (Figure~\ref{fig:overview-b}), which also supports the ground screen, the receiver electronics enclosure, the telescope drive crates, the uninterruptable power supply, and the expanded steel walkways. The deck is attached directly to the deck bearing. 

After the fabrication of the reflectors and sled, the telescope was assembled and pre-aligned
using a MetricVision MV200 laser radar.  This system was used to measure both the reflector surfaces as well as 
the absolute positions of tooling balls on the perimeter of each reflector once the reflectors were aligned to the focal plane.  
The rms deviations from the MR and SR design surfaces are 38 $\mu$m and 28 $\mu$m, respectively, once a small fraction $(<1\%)$ of the outlier measurements from the perimeter of each reflector are removed. 

In order to align the reflectors after assembling them at the site, an
animated 3-D model of the telescope was constructed which also accounted
for additional tooling balls on the cryostat face.  These tooling balls have well measured
displacements from the platelet array. Using the model, a transformation
matrix was established that mapped turnbuckle adjustments to tooling
ball displacements for each reflector~\citep{monsalve:2012}.
After assembly at the site, the distances between the tooling balls
were measured with a custom-built vernier caliper with a range of 2.4 m.
The transformation matrix was inverted and applied to the tooling ball
displacements in order to establish the proper turnbuckle adjustments. The
turnbuckles were then adjusted to bring the system into alignment. This
method enabled convergence to an aligned state after just three
iterations.  The 17 measurements used to establish the position of the SR with respect to the cryostat (for both the Q- and W-band systems) yielded an rms error of $<400$ $\mu$m when compared to the ideal positioning.  Similarly the 14 measurements used to establish the position of the MR with respect to the SR yielded an rms error of $<500$ $\mu$m when compared to the ideal positioning as established using the laser radar.  Tolerance studies allowing for comparable displacements show that this level of alignment error has minimal impact on the optical performance.

\subsection{Feed Horns}
\label{sec:instrument-feeds}

The requirements for the feed horns include high beam symmetry, efficiency, gain and bandwidth, as well as low sidelobes and cross-polarization. These requirements are satisfied by conical, corrugated feed horns~\citep{kay:1962,clarricoats:1984}. 
Standard production techniques for corrugated feed horns (e.g. computer-numerically-controlled lathe machining and electroforming) are prohibitively costly for the large number of feeds for the W-band array. A lower-cost option is described in the next subsection.

\begin{table*}
\begin{center}
\caption{Measured Platelet Array Performance}
\label{tab:plateletarray_measurements}
\begin{tabular}{c|c|c|c|c|c}
\hline
\hline
 
 Frequency & FWHM & Gain & Crosspol & Reflection & Insertion \\
(band/GHz)  & (deg) & (dB) & E/H & Strength & Loss\\
      &   &  & (dB) & (dB) & (dB)\\

\hline
 Q/39--47 & 8.3--6.9 & 27.2--28.5 & $<-34/-29$ & $<-25$ & $<-0.1$  \\
W/89--100 & 8.3--7.4 & 27.1--28.0 & $<-31/-29$ & $<-24$ & $<-0.1$   \\

\hline
\end{tabular}
\end{center}
\end{table*}

\subsubsection{Platelet Array Design}
A 91-element W-band and a 19-element Q-band platelet array of hexagonally-packed, conical, corrugated feed horns were designed for QUIET~\citep{gundersen:2009,imbriale:2011}.  Each array is machined from aluminum 6061-T6 and consists of a number of thin platelets each with a single corrugation, a number of thick plates each with multiple corrugations, and a base plate.  The assembly of platelets and plates is then diffusion bonded together. 
Table~\ref{tab:plateletarray_parameters} provides the parameters of each array.

Due to the side-fed geometry of the telescope, the feed horns must have relatively high gain $(\simeq 27$ to $28$ dB) in order to provide a low edge taper of 
$\leq -30$ dB for both the Q and W-band systems.  This dictates the aperture size of the feed horns and hence the horn-to-horn spacing.  For the W-band horns, this spacing is commensurate with the size of the modules.  Most of the dimensions of the Q-band horns are scaled by the ratio of the frequencies $(\sim 90/40=2.25)$ which results in a Q-band horn spacing that is larger than the Q-band modules.  These horn spacings give rise to angular separations of $1.75^{\circ}$ $(0.82^{\circ})$ between adjacent beams in the Q (W) systems and result in fields of view of $7.0^{\circ}$ and $8.2^{\circ}$ for the Q and W systems, respectively. 

The number of corrugations is fixed at three per wavelength for each horn and a semi-flare angle of $7.6^{\circ}$ is chosen using a design procedure that ensures both acceptable cross-polar levels and return loss ~\citep{hoppe:1987,hoppe:1988}.  This optimization procedure also adjusts the depth of the first six corrugations of each horn in order to reduce the predicted reflection coefficient to better than -32 dB over the full anticipated band of operation.

\subsubsection{Platelet Array Testing}
\label{sec:feedtesting}
\begin{figure}[t]
\centering
\includegraphics[width=3in]{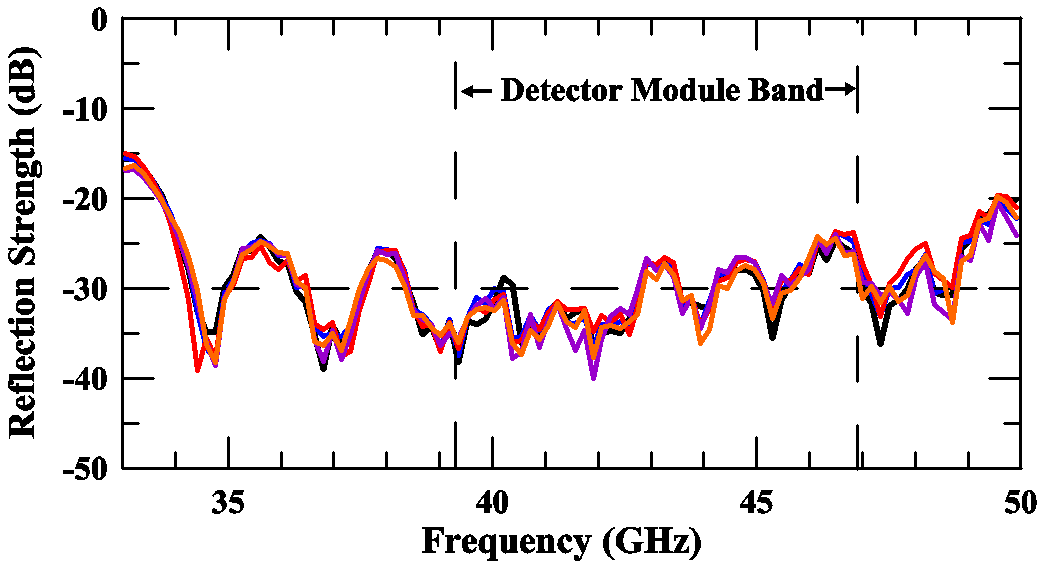}
\caption{Return loss measurements for five of the 19 Q-band horns.}
\label{fig:returnloss}
\end{figure}

A vector network analyzer (VNA) was used to measure the return loss of each horn in each array.  Each measurement consisted of attaching one horn in a platelet array to one port of the VNA using a commercially available circular-to-rectangular transition. A sheet of microwave absorber was placed at $45^{\circ}$ in front of the horns at a distance of $\simeq 1$ m.  The return losses for five of the 19 Q-band horns are shown in Figure~\ref{fig:returnloss} and are similar for the W-band feed horns.   Maximum reflection strengths (negative return loss) are listed in Table~\ref{tab:plateletarray_measurements}.  For comparison, individual electroformed horns that are identical in design to the Q and W-band horns were fabricated.  The array values in Table~\ref{tab:plateletarray_measurements} are comparable to but not quite as good as the electroformed horns or the theoretical predictions both of which were $<-30$ dB across the band.  
 
Beam patterns were measured for all 91 horns in the W-band array and 13 out of 19 horns in the Q-band array.  
A synthesizer combined with $\times 3$ and $\times 6$ multipliers generated the source signals at 40 and 90 GHz respectively.  A standard gain horn was used as a source antenna. 
The platelet arrays were mounted on an azimuth-elevation mount 
so that the source was in the far-field of the platelet array horns.  The source signals were modulated at 1 kHz and a lock-in amplifier connected to a detector diode on the platelet array detected the signal.  
A coaligned, alignment laser ensured that the source horn and platelet array horn were parallel and axially aligned to each other.
A digital protractor with an accuracy of $\pm0.05^{\circ}$ ensured that the source and receiver horn's polarization axes were coincident for the copolar patterns or perpendicular to each other for the crosspolar patterns.  Several measurements were made on each horn including E- and H-plane copolar patterns as well as their corresponding crosspolar patterns.  The patterns were taken by keeping the source horn static and rotating the platelet array horn in azimuth about a vertical axis that intersected the horn's phase center.  A detailed description of this procedure is given in~\citet{clarricoats:1984}.

\begin{figure}[t]
\centering
\includegraphics[width=3.2in]{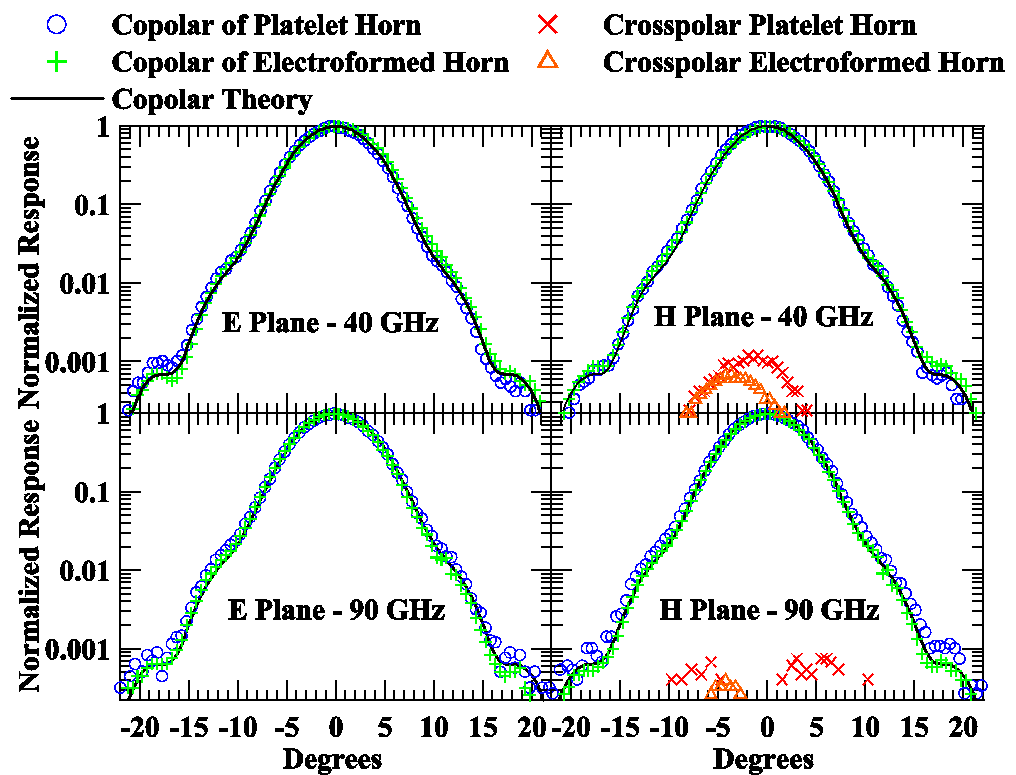}
\caption{Beam pattern measurements of a typical Q-band (W-band) horn in each platelet array along with an electroformed equivalent horn are shown in the top (bottom) two figures.  The left-hand subfigures show the E-plane results and the right-hand subfigures show the H-plane results.  The solid line in each case shows the theoretical prediction of the copolar responses.  The theoretical predictions of the crosspolar responses are all below -40 dB and are not shown.  Upper limits of -34 (-31) dB are placed on the E-plane crosspolar responses for Q-band (W-band).  The  H-plane crosspolar responses are measured at the -30 to -33 dB level for both Q and W-band platelet array horns as well as for their electroformed equivalents.}
\label{fig:hornpatterns}
\end{figure}

The beam patterns of typical Q-band and W-band horns are shown in Figure~\ref{fig:hornpatterns}.  This figure shows both E- and H-plane copolar patterns as well as crosspolar patterns for the platelet feeds and for an electroformed feed with identical design parameters.  The figure also shows the theoretical model responses.  In all cases the E- and H-plane copolar patterns are consistent with both the model and the electroformed feed measurements out to the $-30$ dB level.  Upper limits of $-34$ (-31) dB are set on the E-plane crosspolar levels for Q-band (W-band).  The H-plane crosspolar patterns are not in as good agreement with the model, which predicts both E- and H-plane crosspolar levels at the $< -40$ dB level.  The largest discrepancies are similar in shape to the Q-band H-plane crosspolar measurements shown in Figure~\ref{fig:hornpatterns} and have a non-null crosspolar boresight response.  This type of response is typical of angular misalignment between the source and receiver probes, and the level of the response is consistent with the precision of the digital level.  The W-band H-plane crosspolar response does have a null on boresight and is likely the true crosspolar response.  The fact that the platelet arrays' crosspolar responses are consistently higher than the corresponding electroformed horns' responses suggests that either the machining or the diffusion bonding process leads to somewhat compromised performance.  However, none of the measured feeds has crosspolar levels $>-29$ dB.  Table ~\ref{tab:plateletarray_measurements} summarizes the results of the beam pattern measurements.

Upper limits on the insertion loss were obtained during the return loss measurements of both the W-band and Q-band platelet arrays by placing a flat aluminum plate in front of the horn and generating an effective short.  In both cases the measured reflection strength allows a lower bound to be set on the feeds' room temperature transmission efficiency of $>99\%$.  Assuming solely ohmic losses, this transmission efficiency is expected to increase to $>99.5\%$ upon cooling to 25 K as the electrical resistivity of the horns decreases with temperature~\citep{clark:1970}. 

\subsection{Ground Screen}
\label{sec:groundshield}
The side-fed Dragone design minimizes but does not eliminate sidelobe power. Simulations show that a number of sidelobes are expected.  The performance of the ground screens is described in detail in Section~\ref{sec:sidelobe}.  In order to minimize the radiation from the ground and from celestial sources entering the receiver through sidelobes, an absorbing, comoving ground screen is employed. 
This shields the instrument from varying ground and Sun pick-up and provides a stable, essentially unpolarized emission source that does not vary during a telescope scan. The ground screen structure (Figure~\ref{fig:overview-b}) 
consists of two parts: the lower ground screen is an aluminum box that encloses both reflectors and the front half of the cryostat; the upper ground screen (UGS) is a cylindrical tube that attaches to the lower ground screen directly above the MR. The external surface of the ground screen is coated in white paint in order to reduce diurnal temperature variations and to minimize radiative loading. Following the approach used by the BICEP experiment~\citep{Takahashi:2007}, the interior of the ground screen is coated with a broadband absorber\footnote{The absorber is Emerson Cumming HR-10 (www.eccosorb.com) and is covered with Volara made by Sekisui Voltek (www.sekisuivoltek.com).  The Volara is transparent at QUIET observing frequencies and acts as weatherproofing.} that absorbs radiation and re-emits it at a constant temperature, allowing the ground screen to function as an approximately constant Rayleigh-Jeans source in both Q and W bands.  The UGS was not in place for the Q-band measurement.  It was installed in January 2010, approximately a third of the way through the W-band season.   

\begin{figure}[t]
\centering
\includegraphics[width=3.2in]{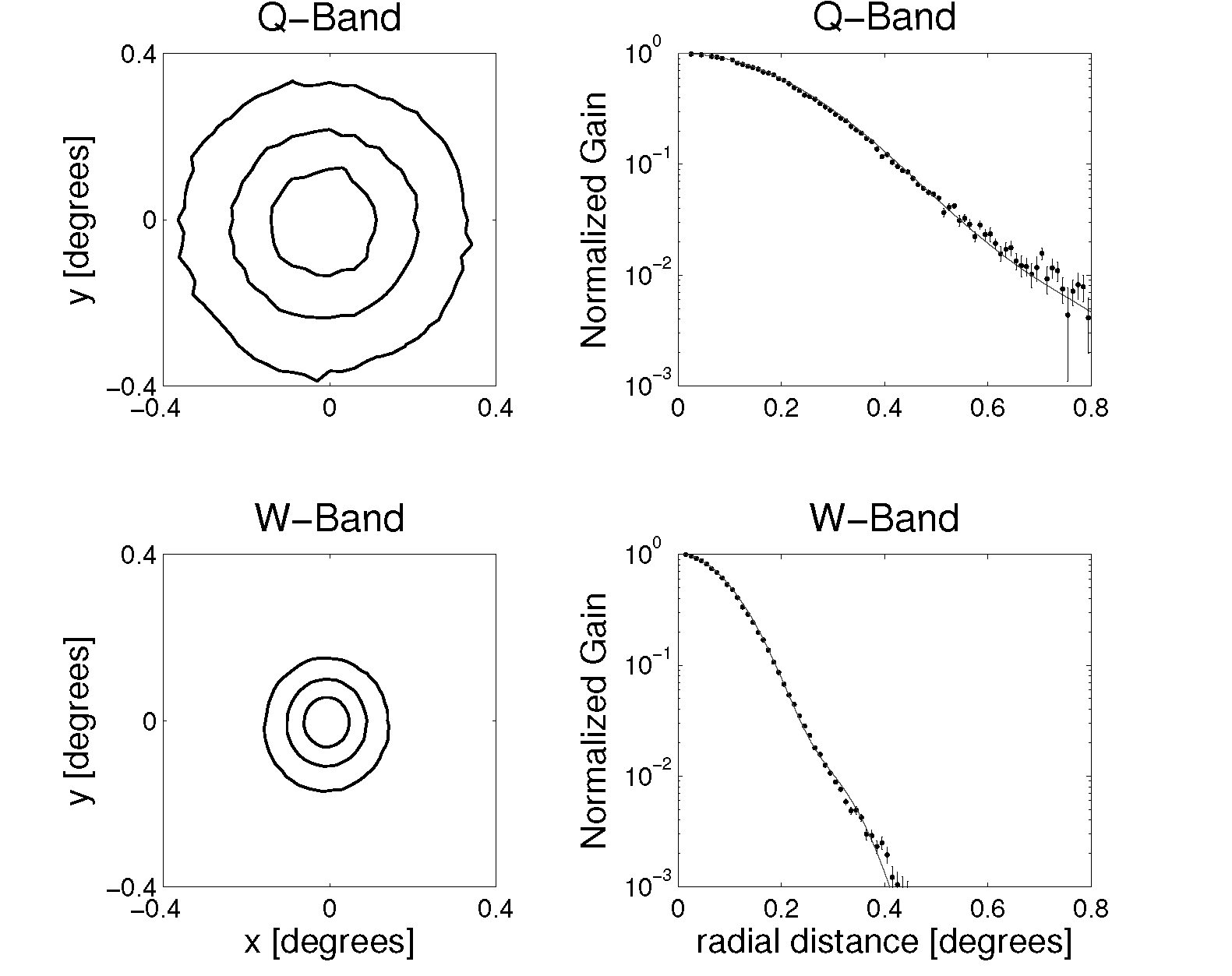}
\caption{Normalized beam maps of Jupiter are shown on the left for representative differential-temperature assemblies for the Q- and W-band systems with contours at 20\%, 50\%, and 80\% of the peak power.  The corresponding azimuthally-averaged beam profiles for each map are shown on the right in comparison with the theoretical prediction (solid line).  Similar maps and profiles of Tau~A were measured using the polarimeter assemblies but at a reduced signal-to-noise.}
\label{fig:beammaps}
\end{figure}

\subsection{Main Beam Performance}
\label{sec:mainbeam}
The main beam profiles are primarily determined from observations of Jupiter.  Additional observations of Taurus A (hereafter Tau~A) are performed to check the main lobe response, to measure the polarized responsivity, to determine the polarization angles and to characterize instrumental polarization.
Tau A and Jupiter are used for main beam characterization since they are, respectively, the brightest polarized and unpolarized, compact sources in the sky.  Figure~\ref{fig:beammaps} shows beam patterns of Jupiter for a differential-temperature assembly in each of the Q and W-band arrays.  
These measurements are consistent with the lower signal-to-noise main beam profiles measured using Tau~A once the slightly different instrumental bandpasses, source spectra, and positions in the focal plane are taken into account.  The main beam is used to compute the main beam solid angle $\Omega_B$, the main beam forward gain, $G_{\mathrm{m}}=4\pi / \Omega_{\mathrm{B}}$, and the telescope sensitivity,

\begin{equation}
\label{eqn:gamma}
 \Gamma=\frac{10^{-20}c^2}{2k_{\mathrm{B}}\nu_{\mathrm{e}}^2\Omega_{\mathrm{B}}}\; \mu\mathrm{K\,Jy}^{-1}
\end{equation}

\noindent in terms of the effective frequency,

\begin{equation}
\label{eqn:nue}
\nu_{\mathrm{e}}=\frac{\int\nu f(\nu)\sigma(\nu)d\nu}{\int f(\nu)\sigma(\nu)d\nu}
\end{equation} 

\noindent for a given instrumental bandpass $f(\nu)$ and source spectrum $\sigma(\nu)$.  Equations~\ref{eqn:gamma} and~\ref{eqn:nue} explicitly assume $G_m\propto\nu^2$.  The source spectra of Tau~A and Jupiter are based on the WMAP measurements~\citep{weiland2010}.  A Tau~A source spectrum with $\sigma\propto\nu^{-0.302}$ is assumed for the calculation of the effective frequency for the Tau~A measurements. An empirical fit to WMAP's measurements of Jupiter's brightness temperatures yields a source spectrum of the form

\begin{equation}
\label{eqn:jupspec}
\begin{split}
\sigma(\nu_{\mathrm{GHz}})=&\frac{2k_{\mathrm{B}}\nu_{\mathrm{GHz}}^2}{c^2}(96.98+2.175\nu_{\mathrm{GHz}}\\ 
&-2.219\times10^{-2}\nu_{\mathrm{GHz}}^2+8.217\times 10^{-5}\nu_{\mathrm{GHz}}^3).
\end{split}
\end{equation}

\noindent Similarly a source spectrum of the form 

\begin{equation}
\label{eqn:cmbspec}
\sigma(\nu)\propto \nu^4e^x(e^x-1)^{-2}
\end{equation}

\noindent is used to compute the effective frequency for unresolved CMB fluctuations, where $x=h\nu/k_{\mathrm{B}}T_{\mathrm{CMB}}$.  

Table~\ref{tab:taua_parameters} provides a summary of the mean values of these quantities for the Q-band and W-band polarization and total-power modules for a source spectrum of the form given in Equation~\ref{eqn:cmbspec}.  
The Q-band total power values are for the lone Q-band differential-temperature assembly, while the Q-band polarization values are for the central pixel which is typical for the array.  Both the W-band total power and polarization values shown in Table~\ref{tab:taua_parameters} are averaged over the respective differential-temperature and polarization array elements using an inverse-variance weighting.

\begin{table}[t]
\begin{center}
\caption{Main Beam Performance Parameters}
\label{tab:taua_parameters}
\begin{tabular}{c|c|c|c|c|c}
\hline
\hline
 & $\nu_{\mathrm{e}}$ & FWHM & $\Omega_{\mathrm{B}}$  & $G_{\mathrm{m}}$  & $\Gamma$ \\
& (GHz)  & (deg) & ($\mu$sr) & (dBi) & ($\mu$K$\,$Jy$^{-1}$) \\ 

\hline
Q$_{\mathrm{P}}$ & 43.0 & 0.455 & 74.3 & 52.3  & 237 \\
Q$_{\mathrm{T}}$  & 43.4 & 0.456 & 78.0 & 52.1  & 222 \\
W$_{\mathrm{P}}$ & 94.4 & 0.195 & 13.6 & 59.6  & 269 \\
W$_{\mathrm{T}}$  & 95.7   & 0.204    & 15.6   & 59.1     & 228     \\
\hline

\end{tabular}
\tablecomments{Mean effective frequencies, FWHM beam sizes, main beam solid angles, main beam forward gains and telescope sensitivities for both the polarization (subscript P) and differential-temperature (subscript T) assemblies assuming a CMB-like, broadband source with a spectrum given by Equation~\ref{eqn:cmbspec}.}
\end{center}
\end{table}

The shape of the main beam and its uncertainties are used to compute the instrumental 
window function and its associated uncertainties~\citep{monsalve:2010}.  Initially, an 
arbitrarily oriented, 2-D, elliptical gaussian beam is fit to the data shown in 
Figure~\ref{fig:beammaps}.  If $\sigma_a$ and $\sigma_b$ represent the beam widths of the semi-major and semi-minor axes of the elliptical gaussian (with $\sigma_a\ge\sigma_b$), then the elongation is defined by $\epsilon=(\sigma_a-\sigma_b)/(\sigma_a+\sigma_b)$.  Typical elongations were found to be $<0.02$ and averaged about $0.01$.  This low elongation, and the fact that the CMB scans use a combination of natural sky rotation and deck angle rotation, imply that the beams are well described by an axially-symmetric beam.  
The symmetrized beam is is expressed as a Hermite expansion~\citep{monsalve:2010}, and this expansion is used to compute the transfer function and covariance matrix~\citep{page/etal:2003b}.

\subsection{Sidelobe Characterization}
\label{sec:sidelobe}

Two different methods are used to measure sidelobes.  These included pre-deployment antenna range measurements and in-situ measurements of a bright, near-field source.  In addition, unintentional measurements of the sun in the sidelobes also enabled their characterization.  These three measurements and their results are discussed in more detail here.

\begin{figure}[t]
\centering
\includegraphics[width=3.2in]{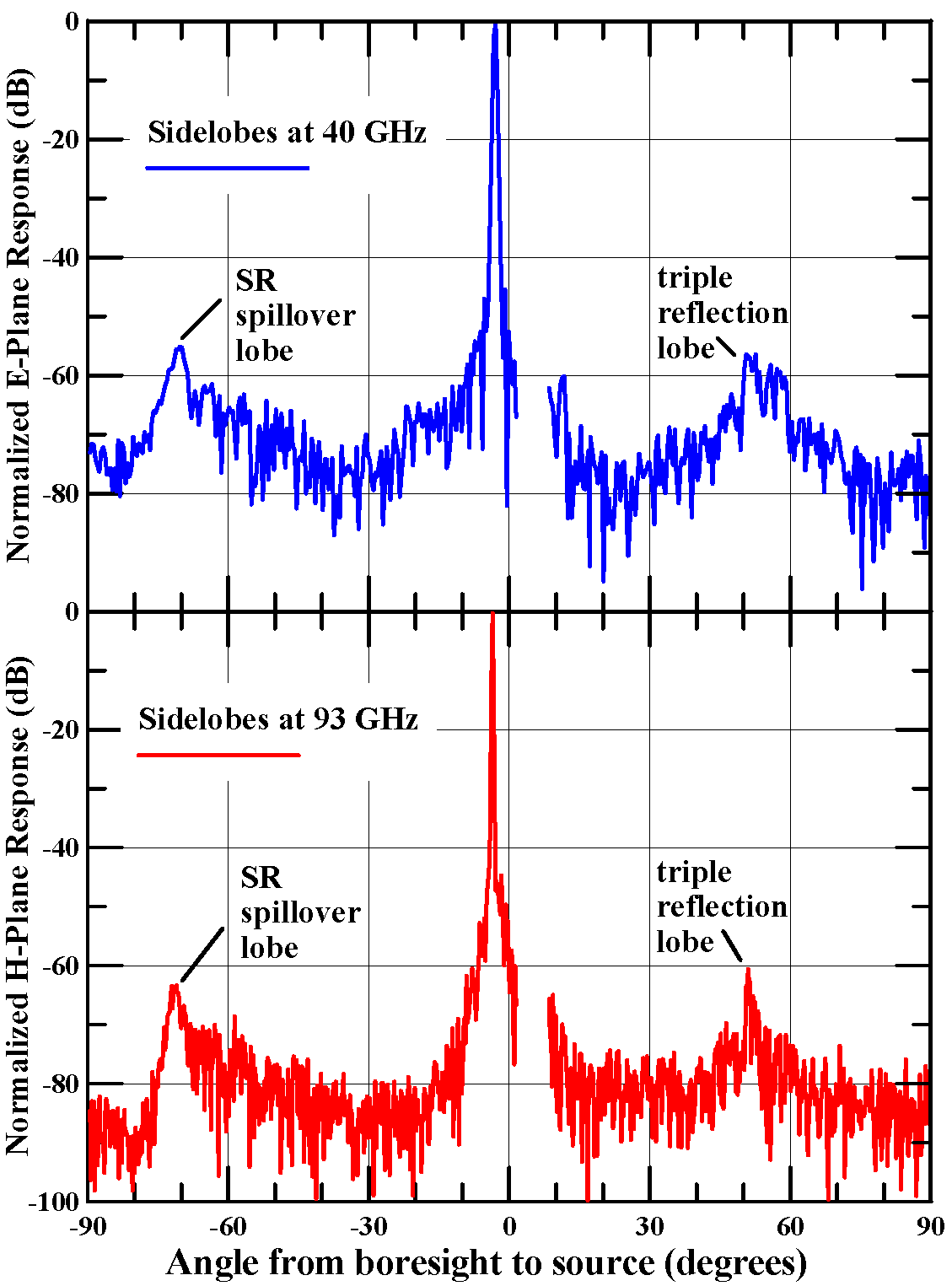}
\caption{Results from the antenna range measurements with no ground screens in place.  The top measurements are the 40 GHz E-plane results for a horn located in the top row, 20.46 mm above the central horn.  The bottom measurements are the 90 GHz H-plane results for a horn located in the top row, 23.87 mm above the central horn.  
The gap in the measurements from boresight angles of $+1.5^{\circ}$ to $+8.5^{\circ}$ is due to mount-related elevation angle limitations.  The two most prominent far sidelobes are the triple reflection sidelobe and the SR spillover lobe as indicated in each figure.  The optical paths associated with these lobes are shown in Figure~\ref{fig:sunsidelobe}.  Top row horns, such as these, are most susceptible to each of these lobes due to their location in the focal plane. 
 }
\label{fig:rangemeasurements}
\end{figure}

\begin{figure*}[t]
\centering
\includegraphics[width=7.0in]{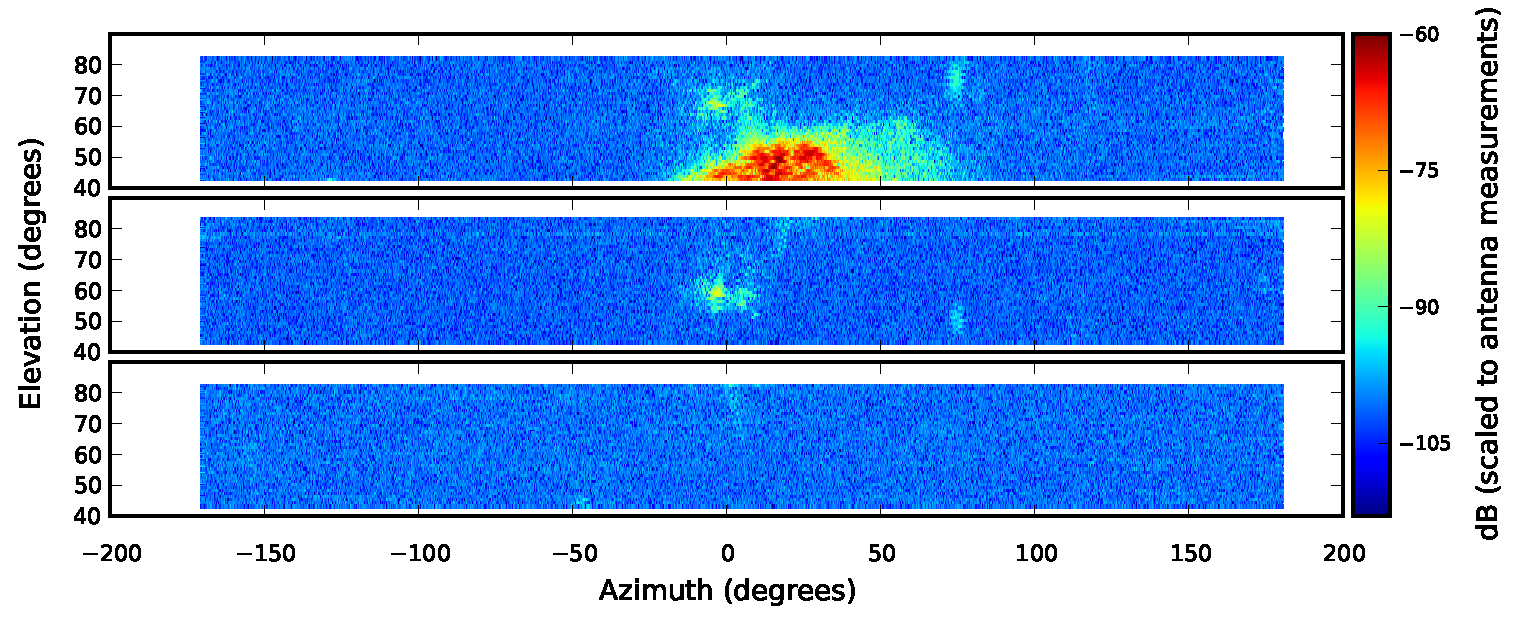}
\caption[Sidelobe measurements at the observing site]{Sidelobe measurements for W-band module 40, located on the edge of the array, with the deck angle set at $-180^{\circ}$ and the near-field source located at an azimuth of $\simeq$ 20$^{\circ}$ and an elevation of $\simeq-5^{\circ}$. {\textit top:} Measurements with only the lower ground screen.  The lobe seen at the bottom of the map is from spillover past the SR.  This lobe is removed after the installation of the UGS. \textit{middle:} Measurements with the lower ground screen and UGS installed. The lobe at the top is due to holes in the absorber from the ground screen structure, and is present before the UGS was added as well, but its position has shifted slightly because the source was moved between measurements. \textit{bottom: } Results with the complete ground screen installed and with additional absorber placed over holes in the floor of the lower ground screen.  The color scale is the same between all three measurements and has been normalized to match the antenna range measurements.  The UGS reduces the far sidelobes by at least an additional 20 dB below the levels shown in Figure~\ref{fig:rangemeasurements}.}
\label{fig:sourcemeasurements}
\end{figure*}

\subsubsection{Antenna Range Measurements of Sidelobes}

The telescope was installed on the Jet Propulsion Laboratory's Mesa Antenna Measurement Facility for measurements of both the main lobe and far sidelobes at both 40 and 90 GHz.  The telescope was mounted on an elevation-over-azimuth positioner with $4''$ pointing accuracy.  
Individual electroformed versions of the Q and W-band horns, described in Section~\ref{sec:feedtesting}, were used for the range measurements.  
The range measurements were conducted before the ground screens were fabricated, so the sidelobe results are only appropriate for the telescope in its bare configuration.  The measurements made use of the facility's Scientific Atlanta model 1797 heterodyne receiver system which enabled repeatable measurements down to $-90$ dB of the peak power level.  A combination of a source synthesizer, multiplier and amplifier was used to generate $\simeq 100-200$ mW of power at each frequency.  The sources were separately connected to corrugated feeds at the focus of a small cassegrain antenna at a distance of 914 m from the telescope.  Due to limitations of the mount, only a simple principal plane cut within $\pm 90^{\circ}$ of the telescope boresight (in the plane shown in Figure~\ref{fig:tel-schem}) was performed for a number of arrangements of the source/receiver antennas.  These arrangements included moving the receiver horn to a few positions in the focal plane and rotating the source and receiver horns for both E- and H-plane cuts. 

\begin{figure*}[t]
\centering
\includegraphics[width=7.0in]{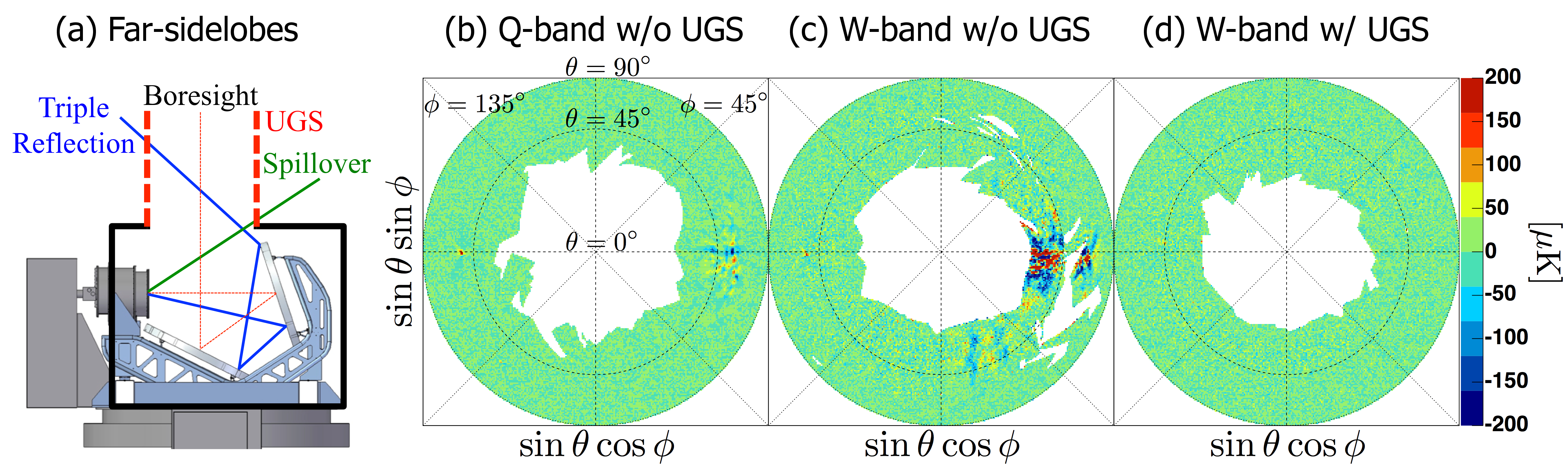}
\caption[sunsidelobe]{Sidelobe characterization using the sun. 
(a) The optical paths that give rise to the triple reflection and spillover sidelobes are shown before the installation of the UGS.  (b) The telescope boresight-centered map of the sun (see text) is shown before the installation of the UGS for a Q-band feed horn in the top row, nearest to the vertical centerline. 
The sharp spike induced by the triple reflection is seen at ($\theta$, $\phi$)~$\simeq$~($50^{\circ}$, $180^{\circ}$), while
the large area of sidelobe contamination just under the $\phi= 0^{\circ}$ line is induced by the SR spillover.
(c) The telescope boresight-centered map of the sun is shown for a horn in a similar position in the W-band array before the UGS installation.
(d) The same map is shown for the same W-band horn after the UGS installation and after the holes in the lower ground screen floor were filled with absorber.
}
\label{fig:sunsidelobe}
\end{figure*}

The results for one feed horn position for each of the Q- and W-band arrays are shown in Figure~\ref{fig:rangemeasurements}.  In each case the feed horn position that was tested corresponds to the top row of the respective platelet array, furthest from the MR and directly above the central feed horn.   
Cross polar measurements were not made on the antenna range since they are made during routine calibrations.  The main lobe beamsizes compare well with initial theoretical predictions~\citep{imbriale:2011}; however, the near-in (i.e., within $\pm 5^{\circ}$ of the main lobe) sidelobe levels do not.  As described by~\citet{imbriale:2011}, this is due to the reflector surface imperfections, which were not included in the initial theoretical predictions.  As shown in Figure 18 from~\citet{imbriale:2011}, once the measured reflector surface is incorporated in the theoretical pattern predictions, the predicted envelope of near-in sidelobes matches well with the observations.  The surface imperfections caused the near-in sidelobe levels to increase by as much as 15 dB in some regions.  The two dominant far sidelobes are the SR spillover lobe and the `triple reflection' lobe.  The SR spillover lobe is broad and arises from direct coupling into the feed horn.  It is located $\sim 70^{\circ}$ from boresight as predicted by~\citet{imbriale:2011}.  The triple reflection lobe is due to an additional reflection off the SR (as indicated in Figure~\ref{fig:sunsidelobe}) and it is located $\simeq 50^{\circ}$ from boresight in the opposite direction from the SR spillover lobe.  This position also matches the prediction shown in Figure 10 of ~\citet{imbriale:2011}.  The amplitude of each lobe for the W-band case is $-60$ to $-62$ dB, while they are $-58$ to $-59$ dB for the Q-band measurement. These amplitudes are both 5--7 dB above the uncorrected predictions of ~\citet{imbriale:2011}.  As with the increased near-in sidelobe levels, this increase in the far sidelobes can be attributed to reflector surface imperfections. 

\subsubsection{Source Measurements of Sidelobes}

The performance of the UGS was assessed using the W-band array in 2010 January.   For these measurements, a polarized, modulated 92 GHz oscillator was placed in the near field of the telescope at a distance of approximately 15 m.  The telescope was scanned over its entire azimuth and elevation range at four different deck angles (0$^{\circ}$, 90$^{\circ}$, $-$90$^{\circ}$, $-$180$^{\circ}$).    The top and middle panels in Figure~\ref{fig:sourcemeasurements} show measurements before and after the installation of the UGS, respectively.  The main sidelobe feature at the bottom of the top map corresponds to the line-of-sight over the SR.  This feature is clearly removed by the UGS.   The remaining sidelobes were caused by holes in the floor of the lower ground screen below the SR.  A third measurement taken after placing absorber over these holes (bottom panel in Figure~\ref{fig:sourcemeasurements}) verifies this and displays the sidelobe performance in the final ground screen configuration. The UGS was not in place during any of the Q-band observing season nor during the first third of the W-band observing season.

\subsubsection{Sun Measurements of Sidelobes}
\label{sec:sunmeasurements}
Before the installation of the UGS, the sun was occasionally detected in the sidelobes.  This is particularly apparent once the data are binned into maps in `telescope boresight-centered' coordinates~\citep{chinone:2011}. The cartesian basis of this coordinate system has $\hat{i}$ oriented along the feed horn boresight, $\hat{k}$ oriented along the telescope boresight, and $\hat{j}=\hat{k}\times\hat{i}$.  If $\hat{s}$ is directed toward the sun, the corresponding spherical coordinates of the sun are defined to be $\theta=\cos^{-1}(\hat{s}\cdot\hat{k})$, and $\phi =\tan^{-1}(\hat{s}\cdot\hat{j}/\hat{s}\cdot\hat{i})$.  
Figure~\ref{fig:sunsidelobe}a shows the optical path of these sidelobes before the installation of the UGS.  Figure~\ref{fig:sunsidelobe}b shows the telescope boresight-centered map for a feed horn on the top row of the Q-band array that is closest to the vertical centerline of the platelet array. The direction of the triple reflection far sidelobe is similar among feed horns. However, the direction of the spillover far sidelobe is different among feed horns because it couples directly to the feed horns and not through the reflectors.  
Therefore the far sidelobe response is characterized for each feed horn separately.
The far sidelobes for W-band were also measured before and after the UGS installation (Figures~\ref{fig:sunsidelobe}c and \ref{fig:sunsidelobe}d).  Figure~\ref{fig:sunsidelobe}d confirms that both far sidelobes are eliminated by the UGS.  The $\phi=0^{\circ}-180^{\circ}$ horizontal line in Figure~\ref{fig:sunsidelobe} corresponds to the principal plane measurement shown in Figure~\ref{fig:rangemeasurements}, and both show the SR spillover lobe and triple reflection lobe before the installation of the UGS.  The amplitudes of the two far sidelobes measured with the sun are consistent with the $\sim -60$ dB levels obtained with the range measurements shown in Figure~\ref{fig:rangemeasurements}.  Data with the moon or sun in the sidelobes were excised in the Q-band analysis~\citep{quiet:2011} as well as during the first third of the W-band season (in preparation).  The addition of the UGS for the W-band data, in combination with azimuth filtering and data rejection used for the Q-band data, makes the spurious polarization signal due to sidelobes a negligible effect on the B-mode measurements.

\subsection{Leakage Beams}
\label{sec:leakagebeams}

\begin{figure*}[t]
\centering
\includegraphics[width=7.0in]{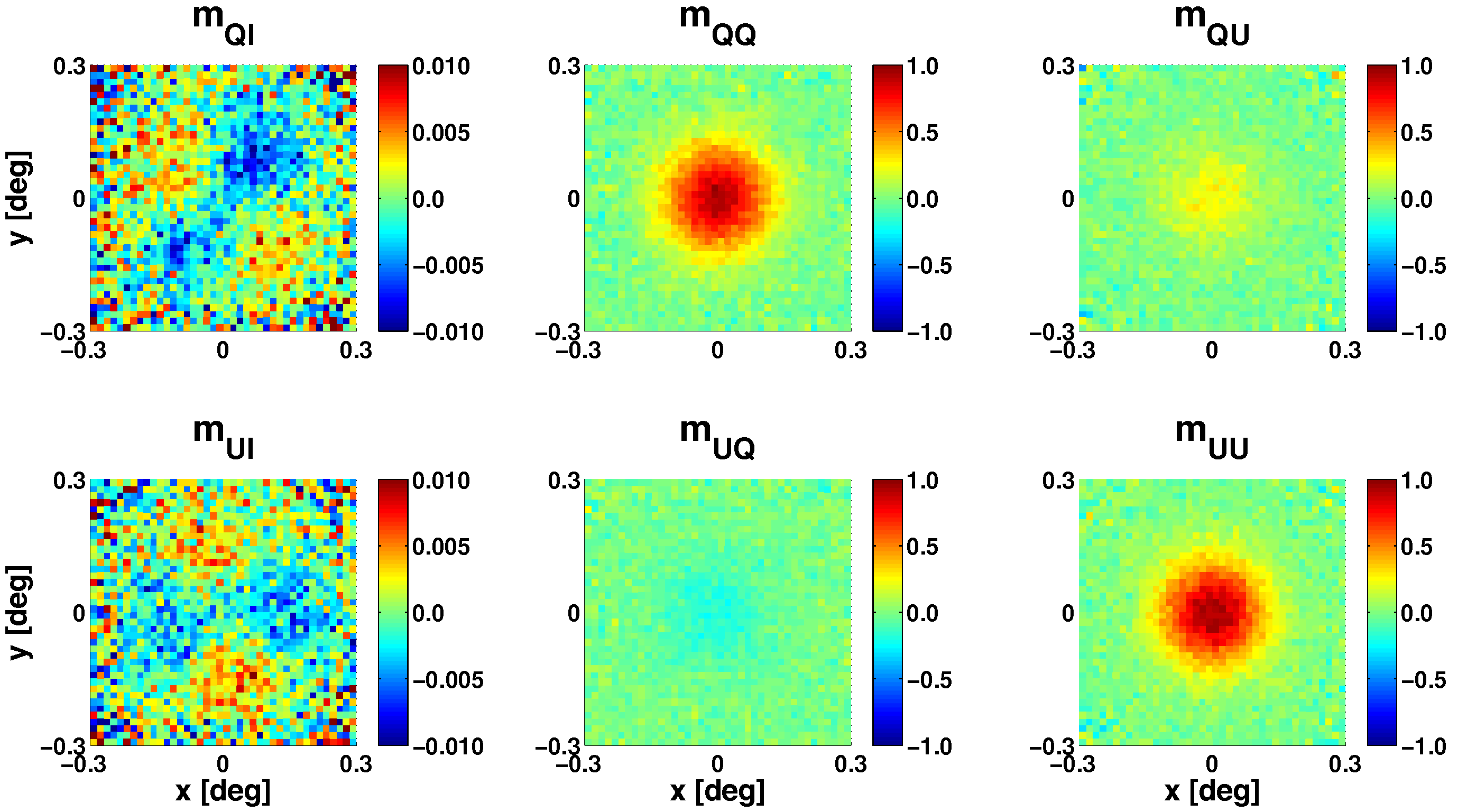}
\caption{The extracted Mueller fields are shown for a Q and U diode pair behind the central horn of the W-band array.  For the purpose of this figure, the $m_{\mathrm{QQ}}$ and $m_{\mathrm{UU}}$ fields have been normalized to one and the normalizations have been applied to the off-diagonal fields.  A $\simeq 0.4\%$ quadrupole term is evident in the $m_{\mathrm{QI}}$ and $m_{\mathrm{UI}}$ leakage beams, while no higher order structure is evident in the $m_{\mathrm{QU}}$ or $m_{\mathrm{UQ}}$ leakage beams at the $\simeq 0.1\%$ level.  As described in Section~\ref{sec:leakagebeams}, the monopole contribution to the $m_{\mathrm{QU}}$ and $m_{\mathrm{UQ}}$ leakage beams can be absorbed into the detector angle which is measured during the calibration procedure.  Similar results for the Q-band central pixel are given in~\citet{monsalve:2010}.} 
\label{fig:leakage}
\end{figure*}

\begin{figure*}[t]
\centering
\includegraphics[width=7.0in]{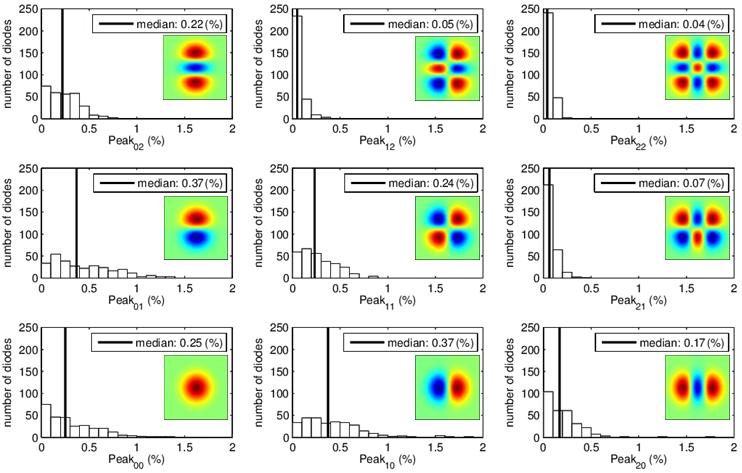}
\caption{These histograms show the number of W-band diodes that have a maximum absolute value of the product $|s_{ij}f_{ij}|$ (denoted Peak$_{ij}$ on the ordinate) in a given percentile range for both the $m_{QI}$ and $m_{UI}$ leakage beams.  The Hermite expansion term is also shown in each panel.  
A median value of all detector diodes is provided in each histogram and indicated with a vertical line.  Similar results for the central pixel of the Q-band system are given in~\citet{monsalve:2010}.}
\label{fig:leakagehist}
\end{figure*}

The leakage beams quantify both the Q and U detector diodes'\footnote{The detector diode nomenclature is described in Section~\ref{sec:modules}.} responses to an unpolarized source, as well as the leakage that can convert a sky Q into a measured U or a sky U into a measured Q.  In order to assess these various forms of leakage, daily observations of Jupiter and/or Tau~A were performed.  These produce beam maps that are subsequently decomposed into their respective beam Mueller fields following~\citet{odea:2007}.  The beam Mueller fields are related to the co- and cross-polar components of the dual, orthogonal polarizations supported by the feed system.  For a linearly polarized source with Stokes parameters $I_{\mathrm{src}},\, Q_{\mathrm{src}},\, U_{\mathrm{src}}$ (assuming $V_{\mathrm{src}}=0)$, degree of linear polarization $p=(Q_{\mathrm{src}}^2+U_{\mathrm{src}}^2)^{1/2}/I_{\mathrm{src}}$, and position angle $\gamma_{\mathrm{PA}}=(1/2)\tan^{-1}(-U_{\mathrm{src}}/Q_{\mathrm{src}})$, the output voltage $d_{\mathrm{Q}}$ of a Q diode as a function of instrumental flux density gain $g_Q$ and instrumental position angle $\psi$ is given by

\begin{equation}
\label{eqn:qout}
\begin{split}
 d_{\mathrm{Q}}=g_{\mathrm{Q}}\,e^{-\tau}\,I_{\mathrm{src}}\,\{m_{\mathrm{QI}}&+p\,m_{\mathrm{QQ}}\,\cos(2[\gamma_{\mathrm{PA}}-\psi])  \\
&+p\, m_{\mathrm{QU}}\,\sin(2[\gamma_{\mathrm{PA}}-\psi])\},
\end{split}
\end{equation}

\noindent where $m_{\mathrm{QI}}$ and $m_{\mathrm{QU}}$ are the Mueller fields representing the I-to-Q and U-to-Q leakage beams, $m_{\mathrm{QQ}}$ is the extracted Q polarization beam and $\tau$ is the opacity with typical values given in Figure~\ref{fig:h2oabs}.  Similarly, the output voltage of a U diode is given by 
\begin{equation}
\label{eqn:uout}
\begin{split}
d_{\mathrm{U}}=g_{\mathrm{U}}\, e^{-\tau}\, I_{\mathrm{src}}\, \{m_{\mathrm{UI}}&+p\, m_{\mathrm{UU}}\,\sin(2[\gamma_{\mathrm{PA}}-\psi])  \\
&+p\, m_{\mathrm{UQ}}\,\cos(2[\gamma_{\mathrm{PA}}-\psi])\}, 
\end{split}
\end{equation}

\noindent where $m_{\mathrm{UI}}$ and $m_{\mathrm{UQ}}$ are the corresponding leakage beams, and $m_{\mathrm{UU}}$ is the U polarization beam.  In each of these expressions, the factor $g$ is the product of the receiver responsivity $R$ (described in Section~\ref{sec:arraytesting-responsivity}) and the telescope sensitivity $\Gamma$ given by Equation~\ref{eqn:gamma}.  The instrumental position angle is given by $\psi=\eta+\phi_{\mathrm{d}}$ where $\eta$ is the parallactic angle of the beam center and $\phi_{\mathrm{d}}$ is the deck angle.\footnote{For reference, when $\phi_{\mathrm{d}}=0^{\circ}$ or $\phi_{\mathrm{d}}=180^{\circ}$, $\hat{j}$ (defined in Section~\ref{sec:sunmeasurements}) is parallel to the ground.  In the event that the parallactic angle of a given beam is similarly zero (so that the beam is observing the local meridian), then the $\hat{i}-\hat{j}$ plane is perpendicular to the local meridian, yielding an instrumental position angle $\psi=0^{\circ}$.  The $\hat{i}-\hat{j}$ plane is coincident with the plane of the septum polarizers described in Section~\ref{sec:polarimeter-assemblies}.}  For a number of sources, Tau~A in particular, the parallactic angle coverage is not very large, so beam maps at various deck angles are necessary in order to vary the outputs of the Q and U detector diodes.  Figure~\ref{fig:leakage} shows the results of this extraction of the leakage and polarized beams for a Q and U diode pair behind the central W-band horn.  A similar figure is shown in~\citet{monsalve:2010} for the Q-band system.  

The $m_{\mathrm{QI}}$ and $m_{\mathrm{UI}}$ Mueller fields are of particular importance since they characterize the instrumental polarization.  Instrumental polarization can be generated by any of the elements in the optical path including the reflectors, the curved cryostat window, the IR blocker, the feed horns, the septum polarizers and the modules themselves.  In the Appendix, specific expressions are derived for these leakage terms for the modules and the septum polarizers.  These two elements are the primary cause of the monopole leakage contribution to the $m_{\mathrm{QI}}$ and $m_{\mathrm{UI}}$ Mueller fields. The median W-band monopole leakage is $0.25\%$ and is lower than the median Q-band monopole leakage.  These Q and W-band leakages measured with Jupiter and Tau~A are consistent with those obtained from skydip measurements that are described in Section~\ref{sec:arraytesting-responsivity}.  As reported in~\citet{quiet:2011}, the Q-band monopole leakage is the largest systematic error in the B-mode measurement at $\ell \sim 100$ where it begins to dominate the constraint on $r$ at levels of $r<0.1$.  A na\:{\i}ve estimate of the impact of this leakage would cause it to dominate at a much higher level; however, a combination of sky rotation and frequent boresight rotation suppresses this systematic by some two orders of magnitude.  The origins of the Q-band monopole leakage are described in more detail in Section~\ref{sec:polarimeter-assemblies}. 

The monopole leakage refers to the $s_{00}$ term in the Gauss-Hermite expansion of these leakage beams given by $b_{\mathrm{leak}}(x,y)$~\citep{monsalve:2010}. Here and in Figure~\ref{fig:leakage} the coordinates $(x=\sin\theta\sin\phi,y=\sin\theta\cos\phi)$ are telescope boresight-centered coordinates defined in Section~\ref{sec:sunmeasurements}.  The leakage beams can be expressed as

\begin{equation}
b_{leak}(x,y)=\sum_{j=0}^2\sum_{i=0}^2 s_{ij} \, f_{ij}(x,y),
\label{eqn_leak_model}
\end{equation}

\noindent where $s_{ij}$ are the fit coefficients and the normalized basis functions $f_{ij}(x,y)$ are 

\begin{equation}
f_{ij}(x,y)=\left(\frac{1}{\sqrt{2^{i+j}i!j!\pi\sigma^2}}\right){\mathrm{e}}^{-\frac{1}{2\sigma^2}\left[x^2+y^2 \right]}H_{i}\left(\frac{x}{\sigma}\right)H_{j}\left(\frac{y}{\sigma}\right),
\label{leak_basis}
\end{equation}

\noindent where $\sigma$ is the gaussian width of the symmetrized beam described in Section~\ref{sec:mainbeam} and the $H_i$ and $H_j$ are Hermite polynomials.  

Higher-order leakage terms, including dipole $(s_{01}$ or $s_{10})$ and quadrupole leakages $(s_{11}$ or $(s_{20}-s_{02})/2$), can also arise due to the off-axis nature of the telescope and the imperfectly matched E- and H-plane feed horn patterns.  The full array drift scans of Jupiter are particularly useful in measuring these quantities for every diode in the W-band array.  Histograms of the peak amplitudes complete to $i=j=2$ are shown in Figure~\ref{fig:leakagehist} for the W-band array.  (Similar results are provided for the central pixel of the Q-band array in~\citet{monsalve:2010}.)  Additional terms in the expansion are also included, but they are consistently less than $0.1\%$.   Leakages above $1\%$ are quite rare and typical values are in the $0.2-0.4\%$ range.  The W-band dipole and quadrupole leakages are typically slightly higher than those in Q-band.  The systematic effects that these leakage beams generate for power spectrum estimation are provided for the Q-band results~\citep{quiet:2011} and in the W-band analysis \citep{quiet:2012}.

The $m_{\mathrm{UQ}}$ and $m_{\mathrm{QU}}$ Mueller fields measure the leakage of the incident Q Stokes parameter into the measured U Stokes parameter or the incident U Stokes parameter into the measured Q Stokes parameter.  Curved reflector surfaces, imperfections in the septum polarizer, and imperfections in the phase switch are potential sources of this leakage.  These primarily give rise to monopole leakage and effectively rotate the instrumental position angle.  In the case that the ratios $m_{\mathrm{QU}}/m_{\mathrm{QQ}}$ and $m_{\mathrm{UQ}}/m_{\mathrm{UU}}$ are constant over the extent of the beam, the $m_{\mathrm{UQ}}$ and $m_{\mathrm{QU}}$ Mueller fields can be absorbed into the expressions for the two diode outputs with the definition of detector angles $\psi_{\mathrm{Q}}$ and $\psi_{\mathrm{U}}$.  The detector angles are defined by replacing the last two terms in each of equations~\ref{eqn:qout} and \ref{eqn:uout} with a single term as follows:

\begin{equation}
\label{eqn:psi_q}
\begin{split}
&p\, m_{\mathrm{QQ}}\,\cos(2[\gamma_{\mathrm{PA}}-\psi-\psi_{\mathrm{Q}}])\equiv \\
&p\, m_{\mathrm{QQ}}\,\cos(2[\gamma_{\mathrm{PA}}-\psi])+p\, m_{\mathrm{QU}}\,\sin(2[\gamma_{\mathrm{PA}}-\psi])
\end{split}
\end{equation}

\noindent and

\begin{equation}
\label{eqn:psi_u}
\begin{split}
&p\, m_{\mathrm{UU}}\,\sin(2[\gamma_{\mathrm{PA}}-\psi-\psi_{\mathrm{U}}])\equiv \\
&p\, m_{\mathrm{UU}}\,\sin(2[\gamma_{\mathrm{PA}}-\psi])+p\, m_{\mathrm{UQ}}\,\cos(2[\gamma_{\mathrm{PA}}-\psi]), 
\end{split}
\end{equation}

\noindent respectively.  
A Hermite decomposition of the $m_{\mathrm{QU}}$ and $m_{\mathrm{UQ}}$ Mueller fields shown in Figure~\ref{fig:leakage} shows that they are simply related by a multiplicative factor to the $m_{\mathrm{QQ}}$ and $m_{\mathrm{UU}}$ fields.  Thus they can be represented in terms of single-valued detector angles, $\psi_{\mathrm{Q}}$ and $\psi_{\mathrm{U}}$ and
are not a source of systematic error. 
In order to achieve the maximum benefit of simultaneous Q/U detection, it is an important feature that the detector angles are separated by nearly integer multiples of $45^{\circ}$ for each of the four diodes in a given module.  This is shown to be the case in Section~\ref{sec:detangles}.

\section{Cryostats}
\label{sec:cryogenics}

\subsection{Cryostat Design}
\begin{figure}
\centering
\includegraphics[width=3in]{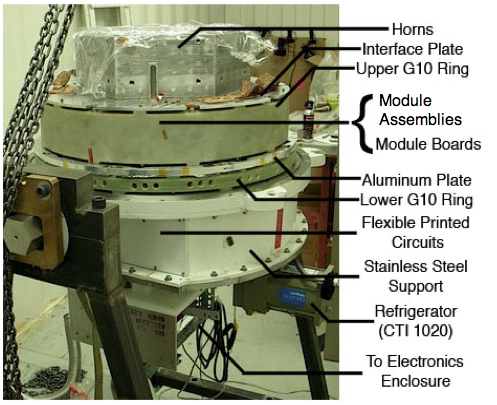}
\caption{The W-band cryostat with the vacuum shell and radiation shields removed.}
\label{fig:innercryo}
\end{figure}

The Q-band and W-band receiver arrays each has a dedicated cryostat (Figure~\ref{fig:innercryo}). In each cryostat, cryogenic temperatures are achieved with two Gifford-McMahon dual-stage refrigerators. The first stage of the refrigerators provide cooling power to a radiation shield, maintained at $\sim$ 50\,K ($\sim$ 80\,K) for the Q-band (W-band) cryostat.  The difference in shield temperature between the W-band and Q-band instruments was not anticipated from the cryostat design, but ultimately did not greatly impact the module temperatures. Infrared radiation is reduced with 10\,cm thick, 3\,lb density polystyrene foam (Table~\ref{tab:finalwindowvals}) attached to the top of the radiation shield. The first stages of the refrigerators also provide a thermal break for the electrical cables. The second stages of the refrigerators provide cooling power for the feed horn array and the modules. The two stages are thermally isolated by G-10 rings. 

\subsection{Cryostat Performance}
The cryogenic performance of the Q-band array is consistent with the design goals of (i) 20\,K module temperatures and (ii) that the module temperatures remain constant during a scan to within $\pm$0.1\,K. A temperature sensor located on an edge module in the Q-band cryostat had a mean temperature of 20.0\,K with a standard deviation of 0.3\,K throughout the season and a deviation of 0.02\,K within a scan. 

For the W-band array, additional heat loads from the active components and conduction through cabling from a factor of five more modules contribute to slightly higher module temperatures compared with the Q-band array. Taking this into consideration, the W-band modules were still warmer than expected by $\sim$ 3\,K, likely as a result of both higher shield temperatures and a minor vacuum leak. A temperature sensor placed directly on the central polarimeter of the W-band array had a mean temperature of 27.4\,K with a standard deviation of 1.0\,K throughout the season, and a mean variation within a scan of 0.12\,K. For each receiver array, both the variation of the module temperatures within a scan and throughout the season had a negligible impact on the responsivity \citep{quiet:2011}.

\subsection{The Cryostat Window}
\label{sec:cryostat-window}
The vacuum windows for the Q-band and W-band cryostats are each $\sim 56$\,cm in diameter, the largest vacuum window to date for any CMB experiment. The vacuum windows must be strong enough to withstand atmospheric pressure while maximizing transmission of signal and minimizing instrumental polarization. 

Ultra-high molecular weight polyethylene (UHMW-PE) was chosen as the window material after stress-testing a variety of window materials and thicknesses. The index of refraction was expected to be $1.52$ \citep{lamb:1996}. To make a well-matched anti-reflection coating for the UHMW-PE in the QUIET frequency bands, the window was coated with expanded teflon, which has an index of refraction of $1.2$ \citep{Benford/Giadis/Kooi:03}. The teflon was adhered to the UHMW-PE window by placing an intermediate layer of low-density polyethylene (LD-PE) between the teflon and the UHMW-PE. The plastics were heated above the melting point of LD-PE while applying pressure with a clamping apparatus in a vacuum chamber to avoid trapping air bubbles between the material layers (the window material properties are summarized in Table~\ref{tab:finalwindowvals}). The band-averaged transmission was expected to improve from 90\% to 99\% for the Q-band array and from 91\% to 98\% for the W-band array by adding this anti-reflection coating to the windows.  

An anti-reflection coated sample for the W-band window was measured using a VNA. The envelope of the transmission and reflection response were fit to obtain values for the optical properties and material thicknesses. The expected contributions to the system noise from loss were computed using published loss tangent values ~\citep{lamb:1996}: $\sim$ 3\,K ($\sim$ 4\,K) for the Q-band (W-band) windows. These values were confirmed within $\sim$ 1\,K by placing a second window over the main receiver window and measuring the change in instrument noise.

\begin{table}[h!]
\caption{Cryostat Window Material Parameters} 
\centering
{\scriptsize
\begin{tabular}{lllll}
\hline
Material & Index of & \multicolumn{2}{c}{Thickness (mm)} & Vendor\\
 & refraction & Q-band & W-band  & \\ \hline
UHMW-PE & 1.52 & 9.52 & 6.35 & McMaster-Carr\\
LD-PE & 1.52 & 0.127 & 0.127 & McMaster-Carr\\
Teflon & 1.2 & 1.59 & 0.54 & Inertech \\ 
Polystyrene foam & -- & 101.6 & 101.6 & Clark Foam \\ \hline
\end{tabular}
\tablecomments{Values for the index of refraction for teflon and UHMW-PE come from the best-fit values to VNA measurements at 90\,GHz.}
\label{tab:finalwindowvals}
}
\end{table}

The curvature of the window under vacuum pressure could introduce cross-polarization by presenting a variable material thickness to the incoming radiation. A physical optics analysis of the W-band window was performed with the General Reflector Antenna Analysis (GRASP)\footnote{http:\//\//www.ticra.com} package to investigate the effect of the curved surface on the transmission properties of the window.
For these simulations we use a window curvature determined from measurements of the deflection of the window under vacuum, $\sim$ 7.5\,cm. 
With a curved window, the central feed horn has negligible instrumental polarization. The edge pixel has 0.16\% additional cross-polarization, where this is defined as leakage from one linear polarization state into the other linear polarization state. This -28\,dB cross-polarization is of the same order as expected cross-polarization from the horns alone and would contribute indirectly to the cross polarization coefficients $m_{QU}$ and $m_{UQ}$ given in Section~\ref{sec:leakagebeams}.

\section{QUIET Polarimeter and Differential-Temperature Assemblies}
\label{sec:assemblies}

QUIET uses HEMT-based low-noise amplifiers (`LNAs') with phase sensitive techniques, 
following the tradition of recent polarization-sensitive 
experiments such as DASI~\citep{leitch/etal:2002}, CBI~\citep{2002PASP..114...83P}, WMAP~\citep{jarosik/etal:2003}, COMPASS~\citep{farese:2004}, and PIQUE and CAPMAP~\citep{barkats:2005a}, 
Unlike those other experiments, however, 
QUIET uses a miniaturized design ~\citep{2004SPIE.5498..220L} suitable 
for large arrays. 
\par
\begin{figure}[h!]
\centering
\includegraphics[width=3.2in]{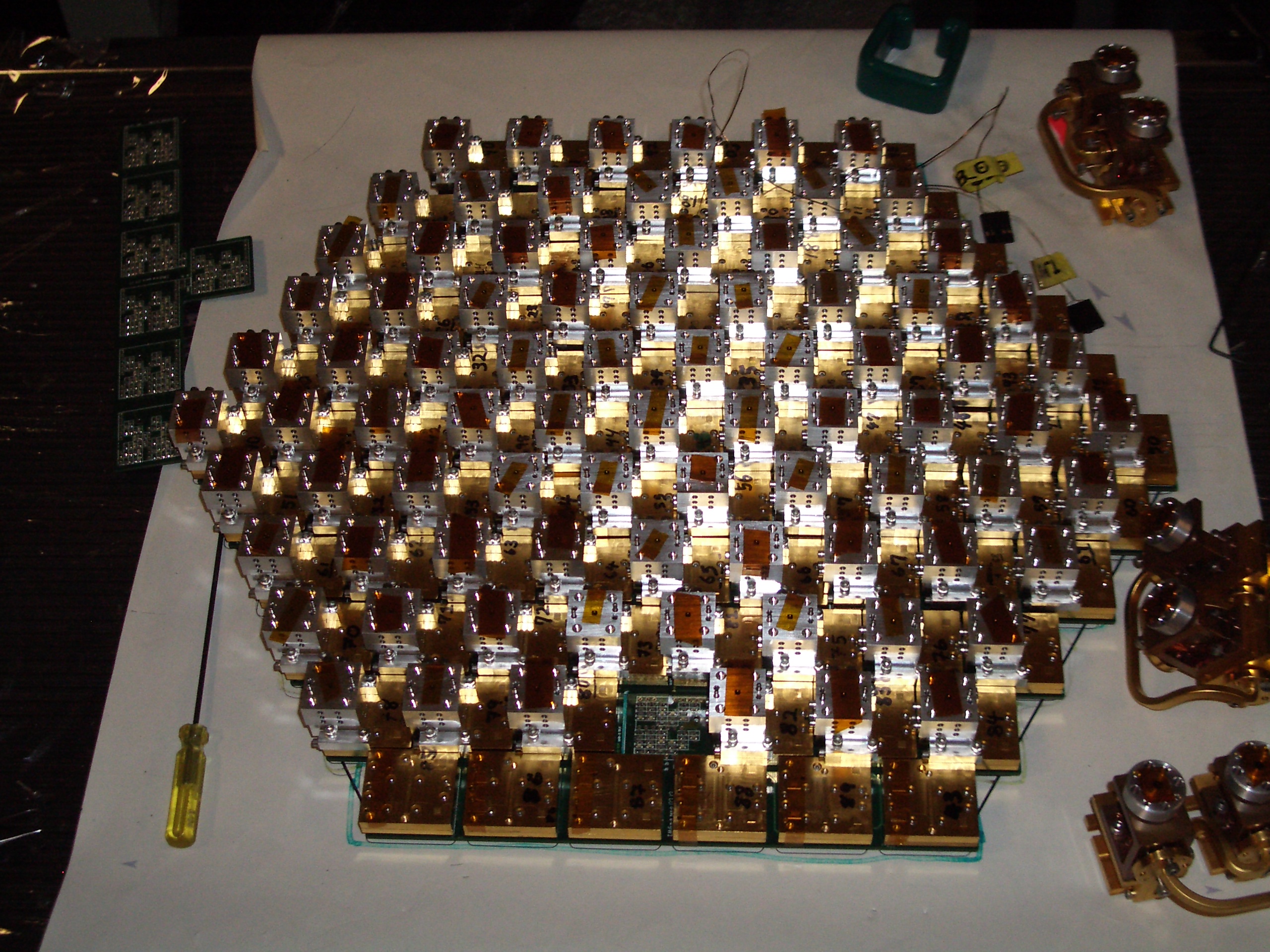}
\caption{The W-band array polarimeter and differential-temperature assemblies.  The 
latter are shown on the right hand side, yet to be installed.  This is 
the largest HEMT-based array ever assembled to date.} 
\label{fig:w-band-array}
\end{figure}

The QUIET Q-band (W-band) array contains 19 (90) assemblies, where 17 (84) are
polarization-measuring assemblies.  The remaining 2 (6) measure the CMB temperature
anisostropy (`differential-temperature assemblies') and are described in Section \ref{sec:TT}.
At the heart of these assemblies are the modules (see 
Section \ref{sec:modules}), 
a highly integrated package that replace many waveguide-block components 
with strip-line-coupled monolithic microwave integrated circuit (MMIC) devices containing
HEMTs. 
The modules have a footprint of 
3.18\,cm$\times$2.90\,cm (W-band) and 5.08\,cm$\times$5.08\,cm (Q-band). 
Figure \ref{fig:w-band-array} shows the W-band array assemblies.  

\subsection{Polarimeter Assemblies}
\label{sec:polarimeter-assemblies}

\begin{figure}[h!]
\centering
\includegraphics[width=3.2in]{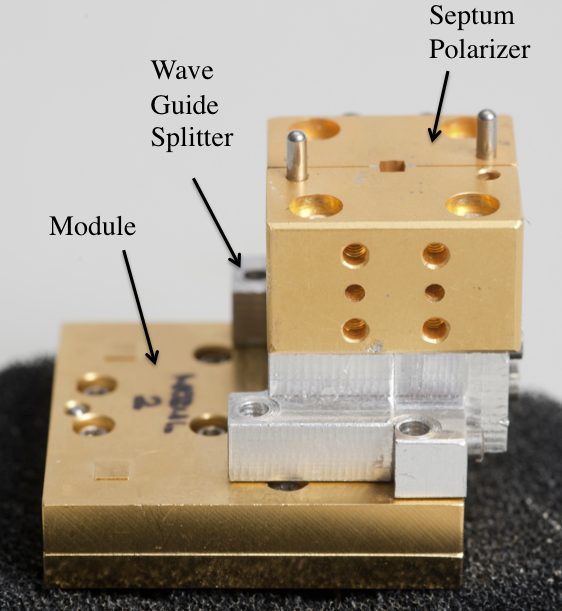}
\caption{A W-band polarimeter assembly.  The module is more compact than previous generation correlators
by an order of magnitude. }
\label{fig:polarimeter-assembly}
\end{figure}

Each QUIET polarimeter assembly consists of (i) a septum polarizer, 
(ii) a waveguide splitter, and (iii) a module containing the highly integrated 
package of 
HEMT-based MMIC devices (see Figure \ref{fig:polarimeter-assembly}).  
The septum polarizer consists of a short circular-to-square
transition into a square waveguide containing 
a septum (a thin aluminum piece with a stepped profile) in the center, which adds 
a phase lag to one of the propagating modes \citep{Bornemann}. 
Given an incident electric field with linear orthogonal components 
$E_{\rm x}$ and $E_{\rm y}$, where the $\it x$ and $\it y$ axis orientations are defined
by the septum, the septum polarizer assembly sends a 
left-circularly polarized component $L=(E_{\rm x} + i E_{\rm y})/\sqrt 2$ to
one output port, and a right-circularly polarized component 
$R=(E_{\rm x} - i E_{\rm y})/\sqrt 2$ to the other output port.   Thus the septum's spatial
orientation is used to define the instrumental position angle.  The output ports 
of the septum polarizer are attached to a waveguide splitter which transitions from the 
narrow waveguide spacing of the septum-polarizer component to the wider 
waveguide separation of the module waveguide inputs.  A more thorough mathematical description
of the septum polarizer is given in Appendix~\ref{sec:omtleakage}.
\par
The scattering matrices, gains, and the temperature-to-polarization (monopole)
leakage terms of both the Q-band and W-band septum polarizers 
are derived from VNA measurements. 
Spectrum analyzer measurements of the Q-band modules in the laboratory show a
degradation in the return loss 
near the low frequency end of the module's bandpass.
When this return loss power is reflected off the septum polarizer and back into the module, it is amplified in the LNAs in the 
module legs and sent back out
of the module to reflect again. This sets up an oscillation which renders the module incapable of measuring input signals.  
Therefore, a bandpass mismatch between the septum polarizer and module 
is deliberately introduced to send this return loss to the sky and
prevent oscillations in the module output.   The bandpass mismatch leads to an enhancement 
in the differential loss between the $E_x$ and $E_y$ transmissions at 47\,GHz, causing
a temperature-to-Stokes Q leakage of $\sim$ 1\%, averaged over the module's bandpass. 
This estimate is consistent with 
leakage values derived from Tau A measurements (Section~\ref{sec:leakagebeams}). 
W-band VNA measurements show no return loss degradation, and therefore no 
bandpass adjustments are needed.  The VNA measurements predict a 
smaller leakage of $\sim$ 0.3\%, so that it is subdominant to leakage due to 
optics.   These measurements are consistent with monopole leakage values obtained from 
on-sky calibrators (see Section \ref{sec:leakagebeams} and 
Figure \ref{fig:leakagehist}).  Note that since the optics leakage has a 
random direction relative to the polarimeter assembly leakage, the combined leakage averages to 
a smaller value and is randomly distributed both in sign and amplitude among modules.

\subsection{Modules}
\label{sec:modules}
The QUIET modules are used in the polarimeter and differential-temperature assemblies 
(see Sections \ref{sec:polarimeter-assemblies} and \ref{sec:TT}), functioning as pseudo-correlation
receivers so that the output is a product (rather than sum or
difference) of gain terms.  While the modules employ a high speed switching technique
to reduce $1/f$ noise, they are an improvement on classical Dicke-switched 
radiometers \citep{Dicke}:  
they do not have 
an active switch at the amplifier input, and there is an additional
improvement of $\sqrt{2}$ in sensivitity since the modules continually measure the sky 
signal \citep{Mennella}.  
\par
In a polarimeter assembly, the module receives as inputs the left ($L$) and 
right ($R$) circularly polarized components of the incident radiation, and measures 
the Stokes parameters $Q$, $U$ and $I$, defined as:

\begin{eqnarray}
I &=& |L|^2 + |R|^2, \nonumber \\
Q &=& 2 \, {\rm Re}(L^{*}R), \nonumber \\
U &=& -2 \, {\rm Im}(L^{*}R), \nonumber \\
V &=& |L|^2 - |R|^2,  \nonumber \\
\end{eqnarray}
where the $^*$ denotes complex conjugation and we expect $V$ to be zero but do not measure it.
\begin{figure}[t]
\centering
\subfigure[]{
\includegraphics[width=3in]{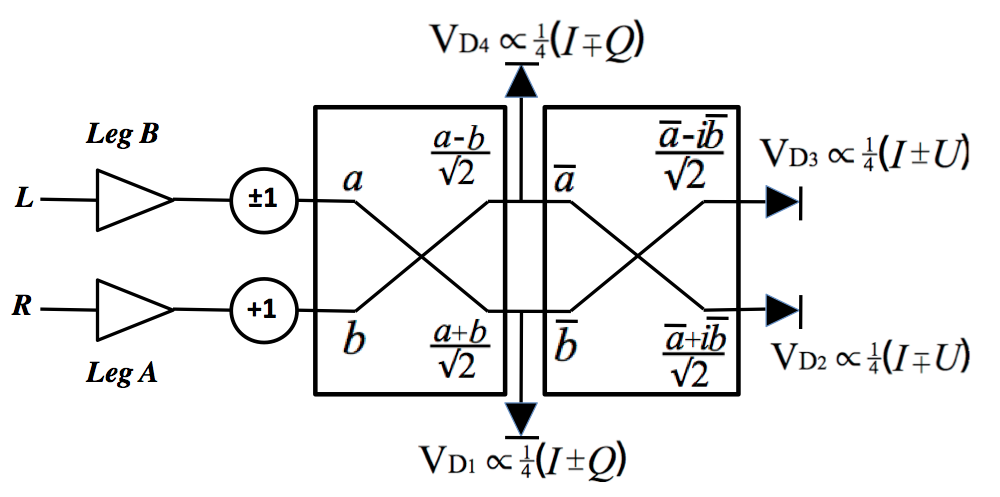}
\label{fig:mod-schem-ideal}
}
\subfigure[]{
\includegraphics[width=2.5in]{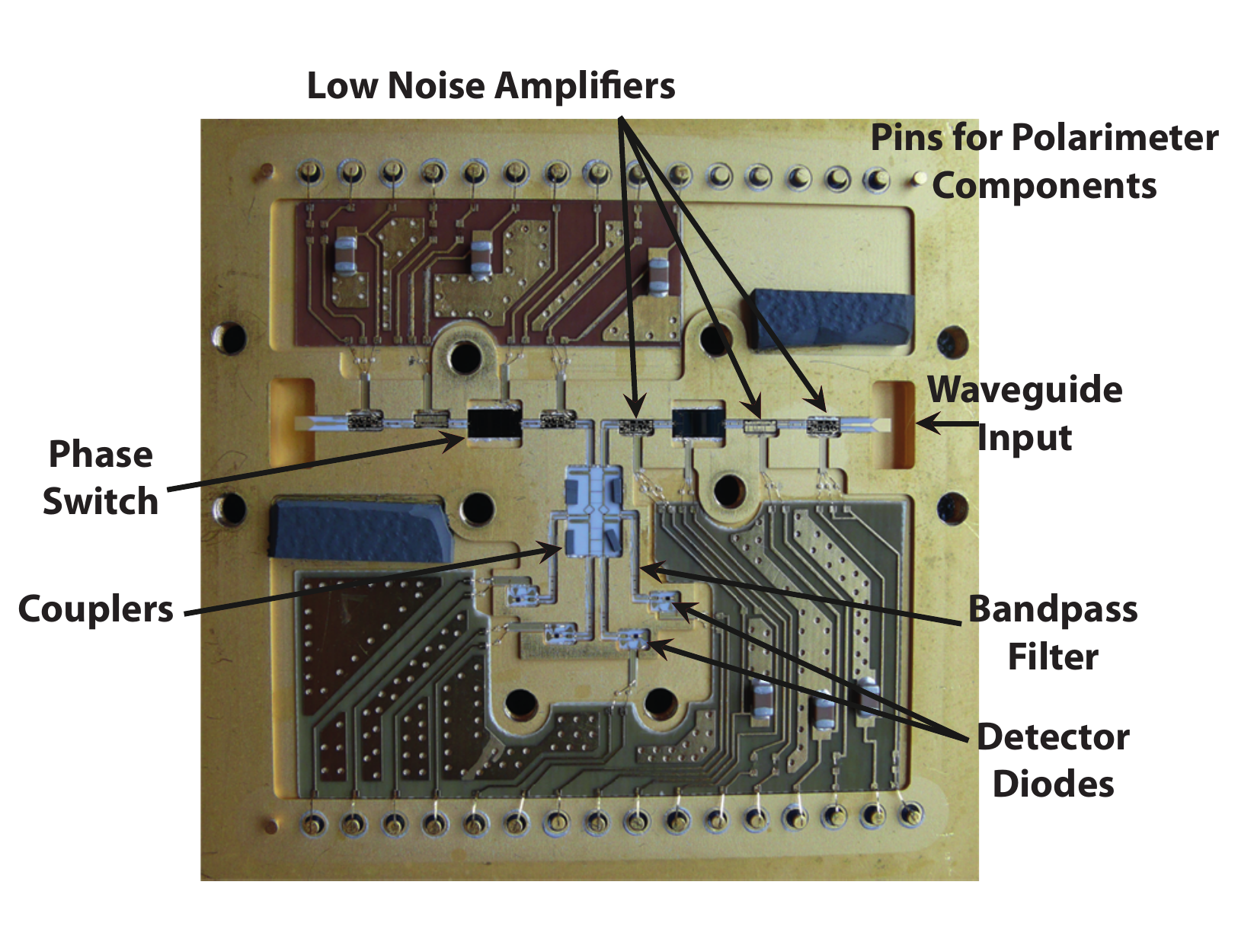}
\label{fig:Qmodulepic}
}
\caption[Signal processing components in a Q-band module]{{\em a:}
Signal processing schematic for an ideal module in a polarimeter
assembly. The diode raw signals are given for the two ($\pm 1$) 
leg B states, and for the leg A state fixed (+1).  For simplicity, details of 
the 3 LNAs and bandpass filters are not shown.
{\em b:} Internal components of a 5\,cm$\times$5\,cm Q-band module.}
\end{figure}

Figure~\ref{fig:mod-schem-ideal} shows a schematic of the QUIET module, 
in which $L$ and $R$ traverse separate amplification ``legs'' (called legs A and B). 
A phase switch in each leg allows the phase to be switched 
between 0$^\circ$($+1$) and 180$^\circ$($-1$)\footnote{The phase switch acts uniformly
across the bandwidth of the module.}. The outputs of the two amplification legs are 
combined in a 180$^\circ$ hybrid coupler which, for voltage inputs $a$ and $b$, 
produces $(a+b)/\sqrt{2}$ and $(a-b)/\sqrt{2}$ at its outputs. The hybrid coupler 
outputs are split, with half of each output power going to 
detector diodes $D_{1}$ and $D_{4}$, respectively. The other halves of the output powers are sent to  
a 90$^\circ$ coupler which, for voltage inputs $\bar{a}$ and $\bar{b}$, 
produces $(\bar{a}+i\bar{b})/\sqrt{2}$ 
and $(\bar{a}-i\bar{b})/\sqrt{2}$ at its outputs. The outputs of this 90$^\circ$ coupler are each
detected in 
diodes $D_{2}$ and $D_{3}$, respectively.  The detector diodes are operated in the 
square-law regime, and so their output voltages are 
proportional to the squared input magnitudes of 
the electric fields.

Table~\ref{ideal_diode_outputs.tbl} shows the idealized detector 
diode outputs for the 
two states of leg B with the leg A state held fixed. 
The diode outputs are averaged and demodulated by additional warm electronics 
(see Section \ref{sec:electronics}).  Given a diode output of $I \pm Q(U)$, the 
averaging and demodulation operations return $I$ and $Q(U)$ 
respectively.\footnote{When referring to diodes $D_1$, $D_2$, 
$D_3$, and $D_4$, the naming convention 
$Q_1$, $U_1$, $U_2$, and $Q_2$ diodes respectively is used.} 
The Stokes parameters can be self-consistently expressed in  
units of temperature as follows \citep{Staggs}.  Let $T_x$ ($T_y$) be the brightness
temperature of a source that emits the observed value of $<E^2_x>$ ($<E^2_y>$).
The Stokes parameters in temperature units become
\begin{eqnarray}
\label{stokes-temperature}
I_T = \frac{1}{2} \cdot (T_x + T_y), \nonumber \\
Q_T = \frac{1}{2} \cdot (T_x - T_y).
\end{eqnarray}
For completeness, the voltage $V_{Q1}$ appearing at the $Q_1$ diode would measure
\begin{equation}
V_{Q1} = g \cdot \left( \frac{1}{2}(T_x + T_y) \pm \frac{1}{2}(T_x - T_y) \right),
\end{equation}
where $\pm$ indicates the states of leg B, and
$g$ is the responsivity constant extracted using 
calibration tools and procedures described in Sections 
\ref{sec:calibrators} and \ref{sec:characterization}.
\par
In practice, the phase of leg B is switched at 4\,kHz, reducing the $1/f$ knee frequency from the LNAs once the signal is demodulated in the $Q$ and $U$ outputs. 
However the phase switches do not reverse 
the sign of $I$; therefore the $I$ output suffers from significant 
$1/f$ noise and so is not used to measure the temperature anisotropy.
The choice of circularly-polarized inputs thus allows for the simultaneous
measurement of both Stokes 
$Q$ and $U$, giving an advantage in detector sensitivity over incoherent detectors.
\par
The amplifier gains and transmission coefficients are represented by the 
proportionality symbols in Table~\ref{ideal_diode_outputs.tbl}.  In practice, 
the transmission through leg B is not exactly identical between the two leg B 
states, leading to additional free parameters needed to characterize the module.
If the leg B transmission differences are not accounted for, they lead to instrumental
(i.e.~false) polarization. This is resolved by modulating the phase of 
leg A at 50\,Hz during data taking, and
performing a double demodulation procedure on the offline data.
Imperfections in the optics and the septum polarizer introduce 
additional offsets and terms proportional to $I$.  These effects are
discussed in Appendices~\ref{sec:appendix:dd} and ~\ref{sec:omtleakage}.

\begin{table}
\caption{Idealized Detector Diode Outputs for a Polarimeter Assembly} 
\label{ideal_diode_outputs.tbl}
\begin{center}       
\begin{tabular}{|c|c|c|c|c|} 
\hline
\rule[-1ex]{0pt}{3.5ex} Diode & Raw Output & Average & Demodulated  \\
\hline
\hline
\rule[-1ex]{0pt}{3.5ex} $D_1$  & $\propto \frac{1}{4}(I \pm Q)$ & $\propto$ $\frac{1}{4}I$ & $\propto$ $\frac{1}{2}Q$ \\
\rule[-1ex]{0pt}{3.5ex} $D_2$  & $\propto \frac{1}{4}(I \mp U)$ & $\propto$ $\frac{1}{4}I$ & $\propto$ $-\frac{1}{2}U$ \\
\rule[-1ex]{0pt}{3.5ex} $D_3$  & $\propto \frac{1}{4}(I \pm U)$ & $\propto$ $\frac{1}{4}I$ & $\propto$ $\frac{1}{2}U$ \\
\rule[-1ex]{0pt}{3.5ex} $D_4$  & $\propto \frac{1}{4}(I \mp Q)$ & $\propto$ $\frac{1}{4}I$ & $\propto$ $-\frac{1}{2}Q$ \\
\hline
\end{tabular}
\tablecomments{Results are shown for
the two states of leg B, with the leg A state held fixed. }
\end{center}
\end{table} 
\par
In practice, the signal pseudo-correlation is implemented in a single 
small package as 
shown in Figure \ref{fig:Qmodulepic} \citep{kangaslahti2006,cleary:77412H}. 
The LNAs, phase switches and hybrid couplers are all produced using the same 
Indium-Phosphide (InP) fabrication process.  Three LNAs, each with gain 
$\sim $ 25\,dB, are used in each of the two legs.
When the input amplifiers are packaged in individual amplifier blocks and cryogenically
cooled to $\sim$ 20\,K, they exhibit noise temperatures of about 18\,K (50--80\,K) for the
Q-band (W-band).  The phase switches operate by 
sending the signal down one of two paths within the phase switch 
circuit, one of which has an added length of $\frac{\lambda}{2}$ 
(i.e., 180$^\circ$ shift).  Two InP PiN (p-doped, 
intrinsic-semiconductor, n-doped) diodes control which path the signal takes. 
The signals go through band-defining passive filters made from alumina substrates, and are then
detected by commercially-available 
Schottky detector diodes downstream of the hybrid couplers.
The amplifers and phase switches are specific to each band, and hence unique to 
each array.  The detector diodes are capable of functioning at both 40\,GHz 
and 90\,GHz, and so are identical between the two arrays.  
\par
The module components are packaged into clamshell-style brass housings, precision-machined for accurate
component placement and signal routing.  To provide bias for active components 
and readout of diodes, the
housing has feedthrough pins connecting to the module components via microstrip lines on
alumina substrates and wire bonds.  Miniature 
absorbers and an epoxy gasket between the two 
halves of the clamshell are used to suppress 
cross talk between the RF and DC components.  All Q-band modules and 
roughly 40\% of W-band modules were assembled by hand.
For the remaining W-band modules, the components and substrates were 
automatically placed 
in the housings by a commercial contractor using a pick-and-place machine; the wire 
bonding, absorber and epoxy gasket were then finished by hand.

\subsection{Differential-Temperature Assemblies}
\label{sec:TT}

\begin{figure}[h!]
\centering
\subfigure[]{
\includegraphics[width=3.0in]{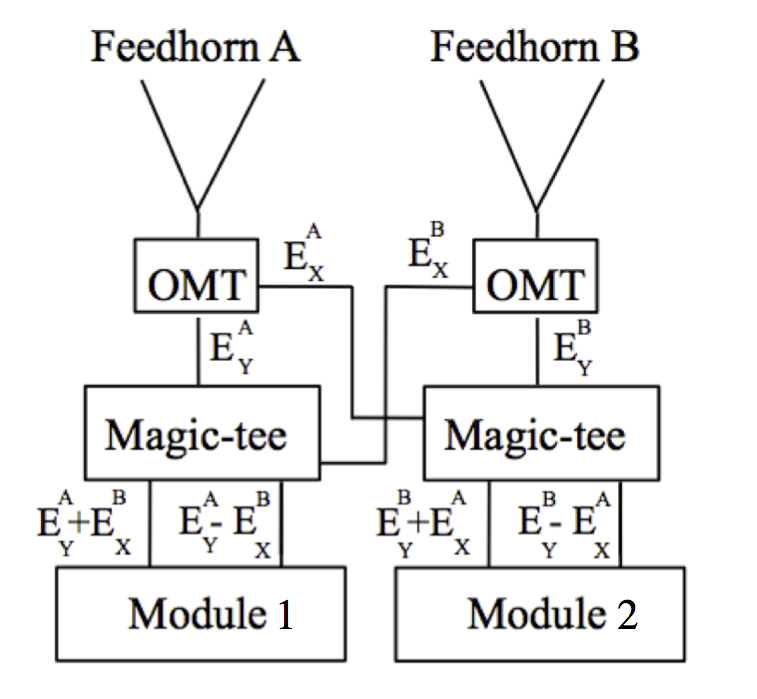}
\label{fig:TTassembly}}

\subfigure[]{
\includegraphics[width=3.0in]{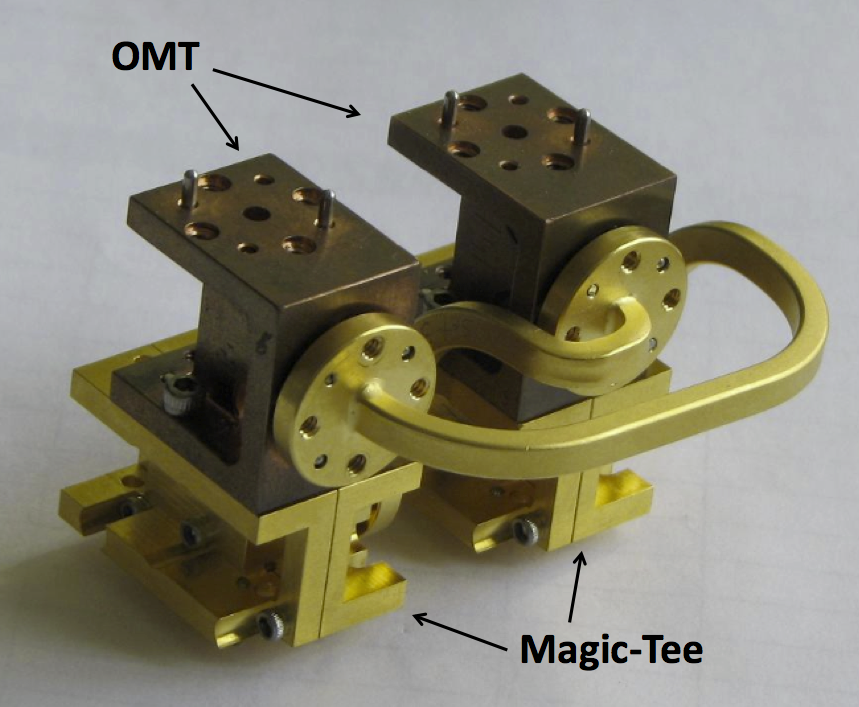}
\label{fig:TT-picture}}
\caption{{\em a:} Schematic of the waveguide coupling for the differential-temperature
assembly.
An Orthomode Transducer (OMT) located after feedhorn A outputs the linear 
polarizations, $E_{\rm Ax}$ and $E_{\rm Ay}$.  One of these polarizations, $E_{\rm Ay}$, 
enters a magic-tee 180$^{\circ}$ hybrid coupler and is combined with 
the orthogonal polarization from an adjacent feedhorn, $E_{\rm Bx}$.  
The factors of $1/\sqrt{2}$ for the magic-tee output labels have been omitted for simplicity.
{\em b:} Implementation of a W-band differential-temperature assembly (modules and 
feedhorns not shown). }

\end{figure}

\begin{table}
\caption{ Idealized Detector Diode Outputs for a Differential-Temperature Assembly
} 
\label{tt_diode_outputs.tbl}
{\scriptsize
\begin{center}       
\begin{tabular}{|c|c|c|} 
\hline
\rule[-1ex]{0pt}{3.5ex} & Mod 1 & Mod 2  \\
\hline
\hline
\rule[-1ex]{0pt}{3.5ex} ${D_{1}}$  & $\propto E^{2}_{{\rm A}y} (E^{2}_{{\rm B}x})$ & $\propto E^{2}_{{\rm B}y} (E^{2}_{{\rm A}x})$ \\
\rule[-1ex]{0pt}{3.5ex} ${D_{4}}$  & $\propto E^{2}_{{\rm B}x} (E^{2}_{{\rm A}y})$ & $\propto E^{2}_{{\rm A}x} (E^{2}_{{\rm B}y})$ \\
\hline
\rule[-1ex]{0pt}{3.5ex} $demod(D_{1},_{\rm Mod 1}) - $ & \multicolumn{2}{|c|}{} \\
\rule[-1ex]{0pt}{3.5ex} $demod(D_{1},_{\rm Mod 2})$  & \multicolumn{2}{|c|}{$(E^{2}_{{\rm A}x} + E^{2}_{{\rm A}y}) - (E^{2}_{{\rm B}x} + E^{2}_{{\rm B}y})$} \\
\hline
\end{tabular}
\tablecomments{Outputs of  
$D_{1}$ and $D_{4}$ corresponding to a leg B state of $+1(-1)$, with leg A fixed at $+1$. 
Also shown is the difference of the demodulated $D_{1}$ signals from two modules.
The outputs of $D_2$ and $D_3$ are zero for an ideal assembly (see text).}
\end{center}}
\end{table} 
The differential-temperature assemblies are grouped into pairs of 
assemblies, with waveguide components that mix two neighboring horn signals 
into two neighboring modules.  Figures \ref{fig:TTassembly} and \ref{fig:TT-picture} 
show the schematic and implementation of these assemblies.  
An orthomode transducer (OMT) located after 
feedhorn A outputs the linear polarizations $E_{\rm Ax}$ and $E_{\rm Ay}$. 
One of these polarizations, $E_{\rm Ay}$, enters a waveguide 180$^{\circ}$ coupler 
(a ``magic-tee'') and is combined with $E_{\rm Bx}$ from the adjacent feedhorn.
The magic-tee outputs are coupled to a module's inputs.  The 
OMTs were reused from CAPMAP \citep{barkats:2005a} while the 
waveguide routing and magic-tees were made by Custom Microwave.
Note that the differential-temperature assembly design resembles that of WMAP 
\citep{Jarosik-tt/etal:2003}, with the significant differences being in the feed horn
separaration and the implementation
of the LNAs.  While WMAP used a conservative design of discrete HEMT LNAs 
and waveguide components, advances in MMIC HEMT LNAs and planar circuitry enabled 
QUIET's cryogenically cooled integrated array design.
\par
For an ideal differential-temperature assembly, the demodulated Q diodes ($D_1$ and $D_4$) measure 
$E^{2}_{\rm Ax}-E^{2}_{\rm By}$, while their counterparts in the 
adjacent differential-temperature assembly measure $E^{2}_{\rm Ay}-E^{2}_{\rm Bx}$.  
The difference of demodulated Q diode outputs from adjacent differential-temperature assemblies measure
the beam-differenced total power $(E^{2}_{{\rm A}x} + E^{2}_{{\rm A}y}) - 
(E^{2}_{{\rm B}x} + E^{2}_{{\rm B}y}) = I_{A} - I_{B}$ (see Table~\ref{tt_diode_outputs.tbl}).  
The demodulated U diodes ($D_2$ and $D_3$) would measure zero for an 
ideal assembly. However, unequal path lengths ($\phi$) in the two legs of a
module mix some of the temperature difference signal from the Q diodes to
the U diodes.  The Q(U) diode signals vary as $\cos(\phi)$(sin$(\phi)$).  For
the differential-temperature assemblies, $\phi$ is $\sim10^\circ-20^\circ$ degrees which
transfers $\sim$ 15-30\% of the signal to the U diodes.  
\par
Finally, we note that the sum of demodulated Q diode outputs 
from adjacent modules is $Q_A+Q_B$, where $Q$
is the Stokes Q parameter seen by the respective horns. Thus one can 
in principle extract polarization information from the differential-temperature assemblies.  
However, as these assemblies form a small fraction of the array, the sensitivity gain
is marginal and so this was not explored further in the analyses.

\section{Electronics}
\label{sec:electronics}

Downstream of the modules are electronics for detector biasing, timing, preamplification,
digitization, and data collection.  These functions are accomplished by four systems: (i) Passive Interfaces, (ii) 
Bias, (iii) Readout, and (iv) Data Management.
The Passive Interfaces system (Section \ref{sec:electronics:interfaces}) 
forms the interface between the modules, the Bias system, and the Readout system.
The Bias system (Section \ref{sec:electronics:bias}) provides the 
necessary bias to each module's active components.
The Readout system (Section \ref{sec:electronics:readout}) amplifies and 
digitizes the module outputs. The Data Management system 
(Section \ref{sec:electronics:DAQ}) commands the other systems and 
records the data.  The Bias and Readout systems are housed in a weather-proof
temperature-controlled enclosure to protect them from the harsh conditions of the Atacama Desert.
The enclosure also serves as a Faraday cage to minimize radio-frequency 
interference.  Further description of these electronics can 
be found in \cite{Bogdan2007}.

\begin{figure}[h!]
\centering
\includegraphics[width=2.5in]{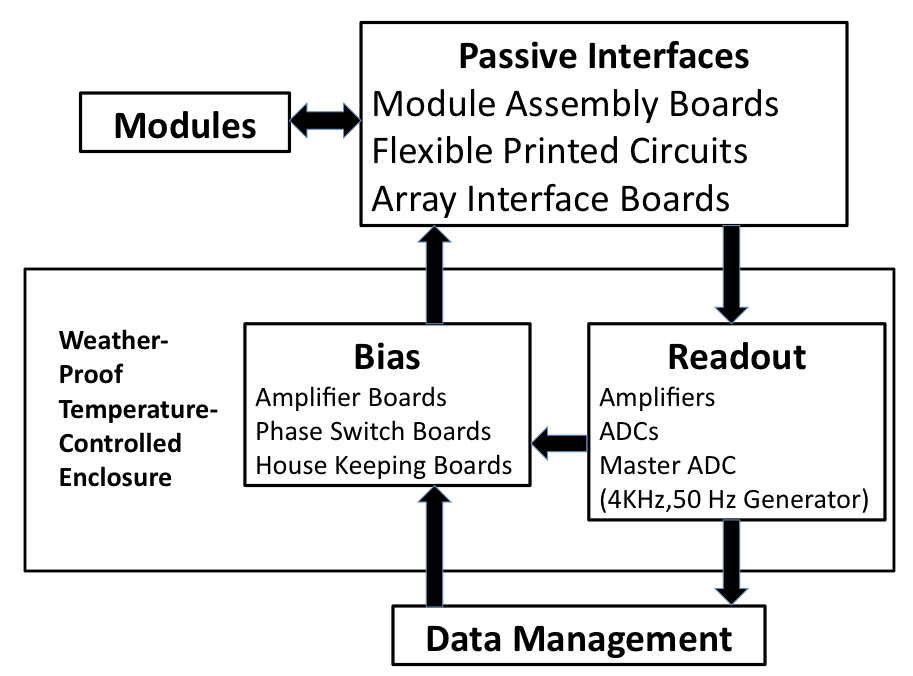}
\caption[QUIET electronics system]{Major components of the electronics.
Boxes outline the four main systems.  Arrows indicate the flow of bias 
commands and data signals.
}
\label{fig:elec-cartoon}
\end{figure}

\subsection{Passive Interfaces}
\label{sec:electronics:interfaces}

The electrical connection to, and protection of, the modules is provided by Module 
Assembly Boards (MABs).  Each MAB is a printed circuit board with pin 
sockets for seven modules.  
Voltage clamps and RC low-pass filters protect the 
sensitive components inside the module from damage.  The Q-band (W-band) modules require
28 (23) pins for grounding, biasing active components, and measuring the 
detector diode signals.  All of these electrical connections are 
routed to the outside the cryostat.
After the MAB protection circuitry, these signals travel on  high density flexible
printed circuits (FPC), which bring them out of the cryostat 
through Stycast-epoxy--filled hermetic seals.
An additional layer of electronic protection circuitry is provided by the 
array interface boards, which also adapt the FPC signals to 
board-edge connectors and route to the Bias and Readout systems.

\subsection{Bias System}
\label{sec:electronics:bias}
All biasing is accomplished by custom circuit boards.
The amplifier bias boards provide voltage and current to power the 
amplifiers in the modules.  Each of these bias signals is controlled by a 
10-bit digital-to-analog converter (DAC), which allows the biases to be tuned for optimal 
performance of each amplifier.
Phase switch boards provide control currents to the PiN diodes in the 
phase switches.  The control current is switched by the board at 4\,kHz for one 
phase switch and 50\,Hz for the other phase switch, generating the  
modulation described in Section~\ref{sec:modules}.
The data taken during the switch transition time are discarded in the 
Readout system.  A housekeeping board monitors the bias signals 
at $\approx$ 1\,Hz for each item being monitored.
The housekeeping board multiplexes these items, switching only during 
the phase switch transitions when data will be discarded.

The Q-band amplifier bias boards are designed to operate at $25\degr$C so 
the enclosure is thermally regulated at that 
temperature.
The W-Band amplifier bias boards use a different design 
that is much less temperature sensitive.  Therefore, the enclosure regulation 
temperature for the W-band is varied between 35$\degr$C and 40$\degr$C depending 
on the time of year to reduce the power needed for regulation.
For both the Q-band and W-band observing seasons, 
the enclosure temperature remained within the regulation setpoint 
for $\approx 90$\% of the time. For the Q-band system, the excursions
primarily affect the drain-current bias supplied by the amplifier 
bias boards, which changes the detector responsivity by $\approx 2\%/ \degr$C. 
This effect is taken into account with an enclosure-temperature 
dependent responsivity model \citep{quiet:2011}.

\subsection{Readout System}
\label{sec:electronics:readout}

The Readout system first amplifies each module's detector diode 
output by 
$\approx 130$ in order to match the voltage range of the digitizers.  
The noise of this warm preamplifier 
circuit does not contribute significantly to the total noise.  
This is determined {\it in~situ} at the 
site by selectively turning off the LNAs in the module and 
seeing that the total noise decreases by roughly two orders of magnitude.  
For the W-band array, the preamplifier noise contributes less than 2\% to the total noise in the quadrature sum.  
The amplifier chain also low-pass 
filters the signal at $\approx 160$\,kHz to prevent aliasing in digitization.
Each detector diode output is digitized by a separate 18-bit Analog Devices 
AD7674 (Analog-Digital Converter) ADC with 4V
dynamic range at a rate of 800\,kHz.
Each ADC Board has a field-programmable gate array (FPGA) that accumulates the 
samples from the 32 ADCs on that Board.  The FPGA on one ADC Board, designated 
the ``Master ADC Board'', generates the 4\,kHz and 50\,Hz signals used by the 
Bias system to modulate the phase switch control currents.  This signal is 
also distributed to all ADC Boards, and the FPGA on each ADC Board uses it to 
demodulate the detector diode data synchronously with the phase switch modulation.

\begin{figure}[h]
\centering
\includegraphics[width=\linewidth]{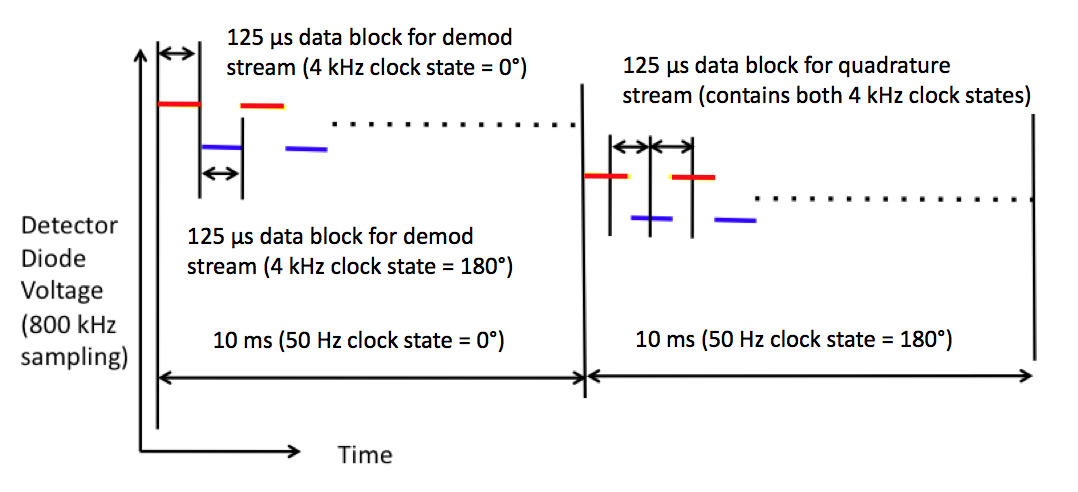}
\caption{Organization of 800\,kHz data to form the demodulated and 
quadrature 100\,Hz streams. The blue and red lines show detector diode data for the two 4\,kHz phase 
states.  Levels are exaggerated for clarity.}
\label{fig:100hz}
\end{figure}

Figure~\ref{fig:100hz} 
summarizes the organization of data performed by the FPGA. 
The FPGA organizes the 800\,kHz detector diode data into continuous 10\,ms blocks (i.e., 100\,Hz time streams), 
itself organized into continuous 125\,$\mu$s blocks.  
These 10\,ms blocks contain an equal sampling of both
4\,kHz clock states.  In the ``TP'' stream, the 800\,kHz data within a 10\,ms block are 
averaged, regardless of the 4\,kHz clock state. This stream is sensitive to Stokes 
$I$ and is used for calibration and monitoring
In the ``demodulated'' stream, data within a 125\,$\mu$s 
block have the same 4\,kHz phase state, and
are averaged.  Averaged data from sequential 125\,$\mu$s blocks are differenced, thus forming the 
polarization-sensitive data stream.   Offline, two adjacent 10 ms blocks in the demodulated stream are 
differenced to form the ``double-demodulated'' (50\,Hz) stream.  
The W-band ADC firmware was upgraded to include an additional specially demodulated 100\,Hz data stream, called the 
``quadrature stream.''  Unlike the usual demodulated stream, data within a 125\,$\mu$s block populate
equally both 4\,kHz phase states, and are averaged.  When these averaged data are differenced, the 
result has the same noise as demodulated data but has {\it no signal}.  The quadrature stream 
is used 
to monitor potential contamination and to understand the detector noise properties. 

As noted earlier, the data are  
masked at the phase switch transition. Masking 14\% 
of the samples around the transition is found to be 
adequate to remove contamination in the data stream.

\begin{figure}[h]
\centering
\includegraphics[width=2.5in]{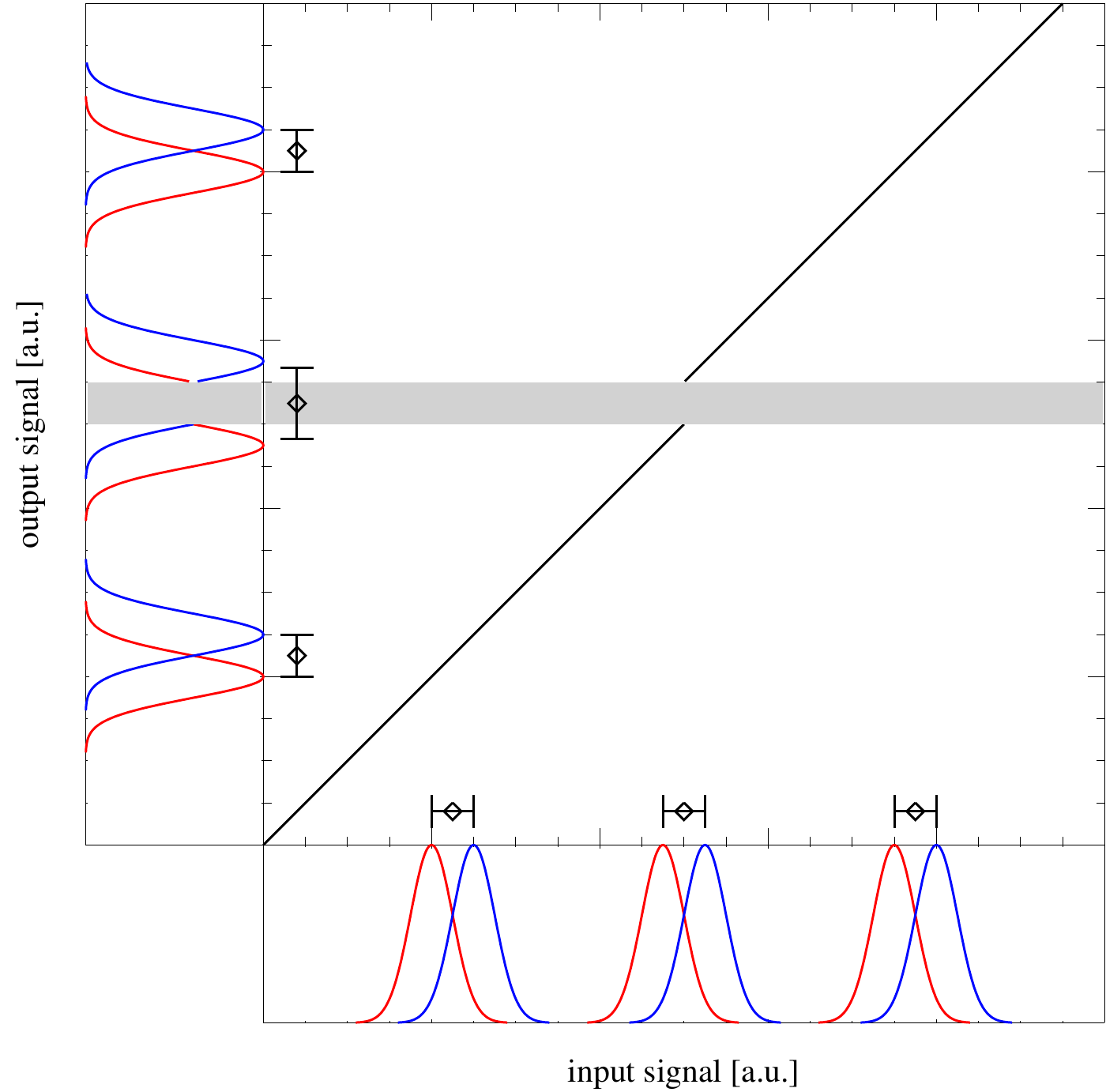}
\caption{Non-linear response of the ADC plotted in arbitrary units.
The horizontal axis shows the ADC input near the discontinuity.
The red and blue distributions show the fluctuations of 800\,kHz 
samples in two different 4\,kHz phase switch states.
When these distributions cross the ADC response discontinuity, the output 
distributions are split at the discontinuity.
When the red and blue states are differenced to create demodulated data, the split 
caused by the discontinuity is added to the result.
}
\label{fig:typeB_illustration}
\end{figure}

The ADC Boards have a small non-linearity in their response.
At intervals of 1024 counts, the ADC output has a jump discontinuity between
1 and 40 counts, affecting $\sim$ 14\% of the data. 
This jump is shown schematically in Figure \ref{fig:typeB_illustration}.
When the 800\,kHz 
data stream value falls at a discontinuity, the 
jump in the output signal will trickle into the 100\,Hz stream.  This 
non-linearity is corrected in the 100\,Hz stream.  The correction is statistical 
in nature, based on the width of the 800\,kHz noise and its proximity to the 
discontinuity \citep{Bischoff:2010}.  This nonlinearity, if uncorrected, causes a 
variation of responsivity during a CES and 
a systematic effect similar to the leakage of temperature to polarization.  
For the Q-band, the correction reduces the ADC nonlinearity to contribute at most
3\% to the leakage bias systematic error, 
and at most 50\% to the CES responsitivity systematic error.  
For the W-band, the residual ADC nonlinearity adds 40\% in 
quadrature to the leakage bias systematic error.
The effect on the CES responsivity is $ < 1\%$, negligible compared to other errors in the gain model. 
These affect $r$ at a level below 0.01 for the W-band.

The Readout system ensures that the housekeeping data and 100\,Hz data from the 
detectors are synchronized to each other and to the mount motion encoder readout.
Synchronization is achieved by distributing the same GPS-derived 
IRIG-B\footnote{Inter-range Instrumentation Group Mod B.} time 
code to both the receiver and mount electronics. In the Readout system, the time 
code is decoded by a Symmetricom TTM635VME-OCXO timing board.
One-Hz and 10-MHz clock signals, locked to the IRIG-B time code, synchronize the 
readout of all ADC Boards. The timing board provides the GPS-derived time to the Data 
Management system so that each datum is assigned a time stamp.
 
\subsection{Data Management}
\label{sec:electronics:DAQ}

The Data Management system sends commands to the Bias system to prepare for 
observation, acquires the data from the Readout  system, writes them to disk, 
and creates summary plots of the detector diode signals and housekeeping data 
for display in real time.  
The complete data are written to disk and DVDs 
in the control room at the observation site at a rate 
of $\simeq 8$\,GB\,day$^{-1}$ for the Q-band array.  W-band array data are written to 
blu-ray optical discs at a rate of $\simeq 35$\,GB\,day$^{-1}$.  
A subset of $\approx 10$\% of the data were transferred by internet every day to 
North Ameria for more rapid analysis and monitoring.  The DVDs or blu-ray discs 
were mailed weekly to North America.

\section{Artificial Calibrators}
\label{sec:calibrators}

\par

Both astronomical and artificial calibrators are used 
to characterize the instrument. Astronomical calibrators are 
described in Section~\ref{sec:mainbeam} and Section \ref{sec:characterization}.  
This section focuses on the artificial calibrators developed for QUIET 
for use in the laboratory and at the observation site.
\par

\subsection{The Optimizer}
\label{sec:optimizer}
\begin{figure}[h]
\centering
\includegraphics[width=2in]{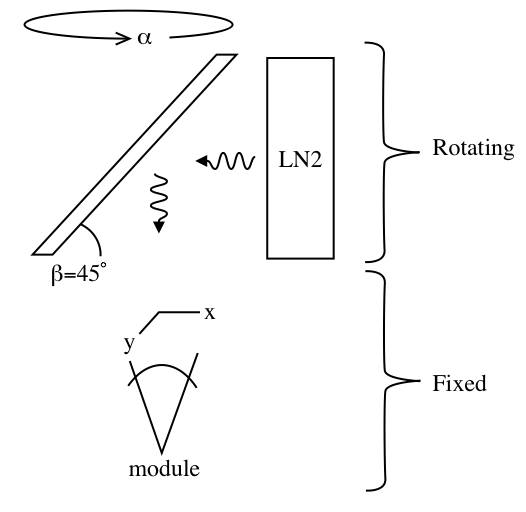}
\caption{The optimizer consists of a reflective metal plate and a cryogenic load, which co-rotate 
about the cryostat boresight axis.  The plate angle $\beta$ is 45$^{\circ}$. The 
reflected signal is polarized (given by Equation~\ref{eqn:optim-expected}) 
and the polarized component modulates at twice the angular frequency of the rotating apparatus. }
\label{fig:optim-rotate}
\end{figure}

The polarized response of the receiver in the laboratory is measured
with the `optimizer,' 
a reflective plate and cryogenic load that rotate around the boresight 
of the cryostat (Figure~\ref{fig:optim-rotate}). The optimizer was used to verify that the responsivities derived from 
unpolarized measurements with cryogenic loads were not substantially 
different from the polarized responsivities, and hence that the 
projections of instrument sensitivity (which were made from 
unpolarized measurements) are valid for the Q-band array. 
For the W-band array, the optimizer was used to select 
functioning modules for the final array configuration. 

The plate is oriented at angle $\beta$ from the plane of the feedhorns and reflects radiation
from the cryogenic load into the window of the cryostat with a Stokes Q in temperature
units given by \citep{barkats:2005a}:
\begin{eqnarray}
Q & = & \frac{1}{2} \cdot \frac{ 4 \pi \delta}{\lambda}(\cos\beta - 
\sec\beta)(T_{\rm plate} - T_{\rm load})\sin(2\alpha t),  \nonumber \\
\delta & = & \sqrt{\frac{\rho}{\mu_0 \pi \nu}},
\label{eqn:optim-expected}
\end{eqnarray}
where $\rho$ is the bulk resistivity of the metal plate, $\nu$ and $\lambda$ 
correspond to the center frequency and wavelength of the detector bandpass, 
$t$ is time, and 
$T_{\rm plate}$ and $T_{\rm load}$ are the 
temperatures of the plate and cryogenic load, respectively. 
Here, $\delta$ is the skin depth, and $\mu_0$ is the permeability of
free space.  
This apparatus rotates at an angular speed $\alpha$ around the boresight of the cryostat
so that the resulting polarized signal will rotate between the Stokes 
Q and U at an angular speed of 2$\alpha$. 
Polarization signals that do not rotate 
with the system (such as thermal emission from objects in the laboratory) 
will be detected at a rate of $\alpha$, and so can be removed. 

The predicted polarized emission from Equation~\ref{eqn:optim-expected} and 
the measured voltages on the detector diodes are used to 
calculate the polarized responsivities for 
polarimeters whose beams primarily sample the reflected cryogenic load. Various 
plate materials (aluminum, stainless steel, and galvanized steel) and two 
thermal loads (liquid nitrogen and liquid argon) are used to obtain 
multiple estimates of the polarized responsivity.  The loads are 
too small to fill the entire array beam, so only the measurements 
from the central polarimeter (Q-band) or inner two rings (W-band) are used. 

\subsection{The Wire Grid Polarizer}
\label{sec:wiregrid}
\begin{figure}[htb]
\begin{center}
\includegraphics[width=1.0\linewidth, keepaspectratio]{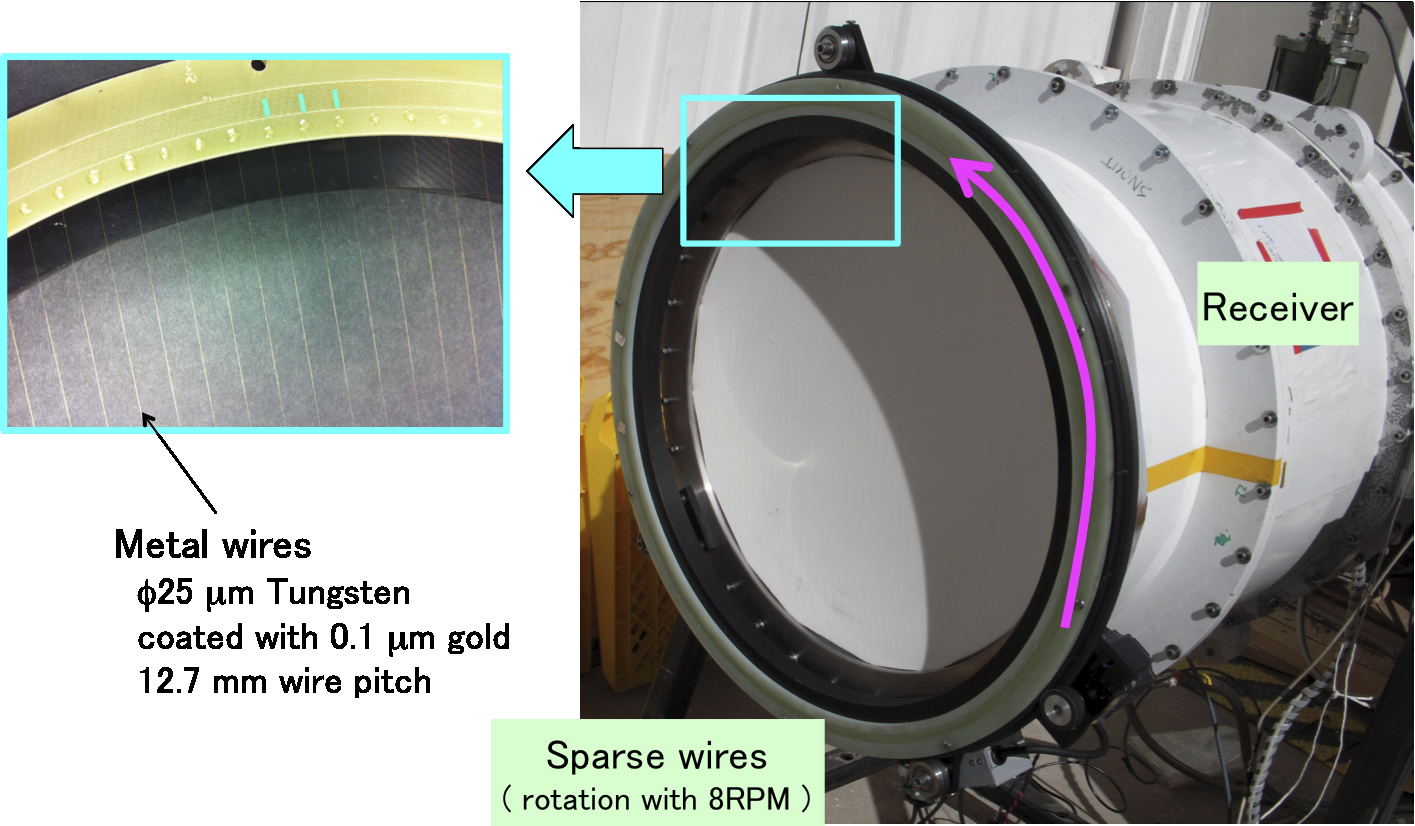}
\caption{
Sparse wire grid array mounted on the W-band cryostat (right), and the fine wire
detail (left). The grid rotates about the boresight axis of the cryostat.}
\label{fig:gridsetup}
\end{center}
\end{figure}

A `sparse wire grid' \citep{tajimaltd},  
a plane of parallel wires held in a large circular frame with the same 
diameter as the cryostat window,
is used to impose and modulate a polarization signal onto the array. 
For the polarization parallel to the grid wires, a fraction of the rays that would 
ordinarily pass through the telescope to the cold sky are instead scattered to 
large angles, mostly terminating on the warm ground shield.
The grid is placed as close to the cryostat aperture as possible to 
minimize interference with the telescope optics and to ensure that it
covers the field of view of each detector (Figure~\ref{fig:gridsetup}).
\par
With this geometry, the polarized signal directed parallel to the wires 
is empirically found to be $\sim$2\,K.  The circular frame rotates about 
the cryostat boresight axis via a small motor, 
allowing for modulation of the injected polarized signal at a constant frequency.
The wire grid was used for calibration measurements in the laboratory 
and three times during the observing season: 
at the end of the Q-band observing season and
at the beginning and end of the W-band season.  The grid 
was not mounted on the cryostat during sky observations.
\par
\begin{figure}[htb]
\begin{center}
\includegraphics[width=1.0\linewidth, keepaspectratio]{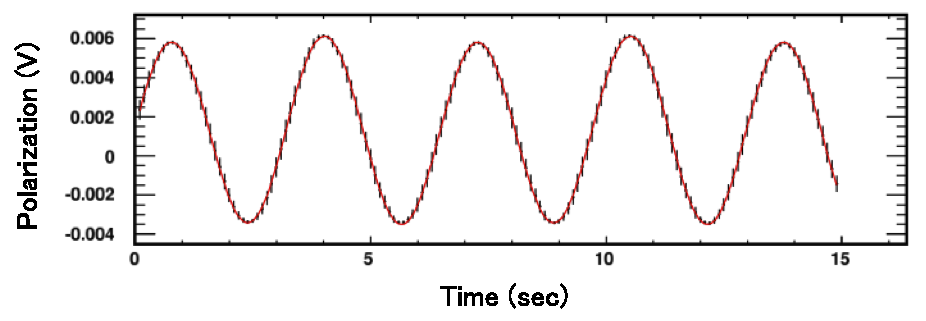}
\caption{
	Polarization response of a detector as a function of time, 
where the wiregrid was rotating at $\sim$8\,RPM.  A sinusoidal response 
is clearly observed at twice the rotation frequency.  The curve is the fit
to the data (dots) using Equation~\ref{Eqn:grid}.
}
\label{fig:labgrid_signal}
\end{center}
\end{figure}

An example of the data taken with the rotating grid is shown in 
Figure~\ref{fig:labgrid_signal}. 
In the ideal case in which the intensity of the reflected radiation is 
isotropically uniform over the array, the polarized signal $D(\theta)$ 
from each detector diode would exhibit a sinusoidal dependence at 
twice the frequency of $\theta$, the angle about the cryostat boresight axis between the wires and 
a fixed point on the cryostat.  The measured polarization signal 
has an additional dependence on $\theta$ due to rays terminating at 
different temperatures in the non-uniform ground screen. 
This variation appears in both the polarized data stream $D$ 
and the total power data stream $I$ as a function of 
$\theta$, and so this variation can be measured in the $I$
data stream and accounted for in the $D$ data stream. 
Each detector diode data stream is fitted to the form
\begin{eqnarray}
D(\theta) &=& D_0 + \Big( D_2 + \eta [I (\theta)-I_0] \Big) \cos[2(\theta - \gamma)],
\label{Eqn:grid}
\end{eqnarray}
where $D(\theta)$ and $I(\theta)$ are the double-demodulated polarization 
and total power signals, respectively.  Here, $I_0$ is the 
average of $I(\theta)$ over all angles $\theta$, and $D_0$ is an offset term
discussed in Appendix \ref{sec:omtleakage}.
The fit extracts $\gamma$, the angle $\theta$  that maximizes $D(\theta)$,  
$D_2$, the polarization amplitude (in mV), and $\eta$, a dimensionless constant relating the 
total power to polarization responsivity. 
Since the fixed point on the cryostat used to 
define $\theta$ can be arbitrarily chosen, only the 
relative $\gamma$s amongst the detector diodes are relevant; they 
are just the relative detector angles.
The values of $D_2$ indicate the spread of polarized responsivities.  
For the W-band, 
their relative ratios agree with ones derived from Tau A 
observations at the level of $20\%$.  The precision of this agreement
is limited by the statistical errors of Tau A observations for the off-center
detectors.
\par

\section{Receiver Characterization and Calibration}
\label{sec:characterization}

Each receiver diode (Table~\ref{tab:rxover}) is characterized by its bandpass, noise level, polarization angle, and total power and polarized responsivities. These quantities were measured for the receiver arrays in the laboratory prior to deployment, and at the site using astronomical calibrators, sky dip measurements, and polarizing grid measurements. 

This section describes methods of module bias optimization (Section~\ref{sec:biasoptim}) and module leakage remediation (Section~\ref{sec:doubled}) as well as characterizing module bandpasses (Section~\ref{sec:bandpasses}), responsivities (Section~\ref{sec:arraytesting-responsivity}), detector angles (Section~\ref{sec:detangles}), noise measurements (Sections~\ref{sec:arraytesting-nts} and \ref{sec:arraytesting-sn}), and sensitivity (Section~\ref{sec:sensitivity}).

\begin{table}[h]
\scriptsize
\centering
\caption{Detector yield for the Q-band and W-band arrays.}
\begin{tabular}{lcc} \hline
Band & Q & W \\ \hline
Number of  modules & 19 & 90 \\
Polarization modules & 17 & 84 \\
\hspace{0.5cm} Polarization diodes & 68 & 336 \\
\hspace{0.5cm} Working polarization diodes (Stokes Q) & 31 & 153 \\
\hspace{0.5cm} Working polarization diodes (Stokes U) & 31 & 155 \\
Total power modules & 2 & 6 \\
\hspace{0.5cm} TT diodes (Stokes Q only) & 4 & 12 \\ 
\hspace{0.5cm} Working TT diodes (Stokes Q only) & 4 & 12 \\ \hline
\end{tabular}
\label{tab:rxover}
\end{table}

\subsection{Detector Biasing and Optimization}
\label{sec:biasoptim}

For the Q-band array, the amplifiers were biased manually for gain balance between the module legs and for adequate signal level at the beginning of the observing season. The biasing was chosen for each module using a room temperature blackbody load in front of the cryostat. The phase switches were turned on separately, so that the signal only propagated through the module leg with the phase switch on. The amplifiers were then biased one leg at a time so that the first stage amplifier drain current was in the range 0-5\,mA, the second stage drain current was in the range 5-15\,mA, and the third stage amplifiers were in the range 15-30\,mA, and that the signal measured by the detector diodes was $\sim$ 5\,mV. This procedure was repeated, turning on only the phase switch for the other leg, and adjusting again to obtain a signal difference between the two legs of $<$ 0.6\,mV. This biasing scheme reduced the current through the first stage amplifier to $\sim$ 30\% of its operational value to keep its noise contribution low. The bias values for the phase switches were chosen to equalize the signal measured on the two separate legs of the module. These bias settings were chosen once at the beginning of the season, and kept fixed during the observing season. 

For the W-band array, biasing the modules by hand was not feasible due to the large number of modules compared to the Q-band array, and so an automatic method was developed. A sinusoidal polarized signal was injected during module biasing by continually rotating the sparse wire grid. Amplifier bias settings were found by maximizing the amplitude of the sinusoid relative to the time-stream noise. The bias settings were sampled via a computer-based downhill simplex algorithm and optimum values were found for all modules within a few hours. As with the Q-band array, the bias settings were kept fixed during the W-band observing season. Because the settings were chosen for signal-to-noise, balance between the legs was not explicitly prioritized (the condequences of this are discussed in the next section).   
 
\subsection{Temperature to Polarization Leakage Remediation}
\label{sec:doubled}
\begin{figure}[h]
\centering
\includegraphics[width=2.9in]{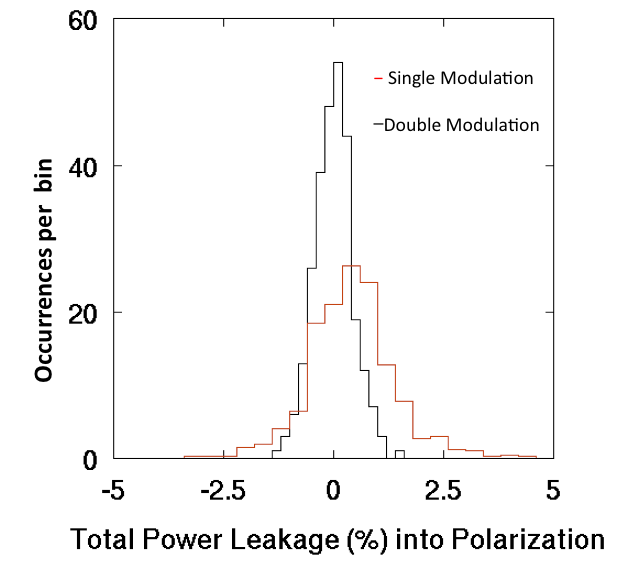}
\caption{A histogram of diode leakage values between total power and polarization channels during a large angle sky dip for the W-band array before and after double demodulation. Double demodulating reduces the total power leakage by a factor of $\sim$ 2 for the W-band array.}
\label{fig:baddoubdemod}
\end{figure}

One source of leakage from total power into polarization from the module stems from differential power transmission between the two phase switch states within a given leg (Appendix~\ref{sec:appendix}). We found that double demodulating (described in Section~\ref{sec:modules}) typically reduced the root-mean-square of leakage from 0.8\% to 0.4\% for the W-band modules (Figure~\ref{fig:baddoubdemod}). The improvement was smaller for the Q-band array, $<0.1\%$, likely because it was dominated by other sources of leakage (Section~\ref{sec:leakagebeams} and Section~\ref{sec:polarimeter-assemblies}) and because the phase switches had been balanced during bias optimization.

The module is not the only source of leakage between temperature and polarization. Instrumental polarization from optics, etc can be calibrated from large and small sky dips (elevation nods of $\pm 20^{\circ}$ and $\pm 3^{\circ}$ amplitude) with 0.3\% precision for each sky dip as the signal from the changing atmospheric temperature leaks into the polarized data stream. The median monopole leakage was 0.2\% for the W-band array, which is consistent with leakage measurements from Jupiter (Section~\ref{sec:leakagebeams}). The median monopole leakage was 1.0\%  and 0.2\% for the Q-diodes and U-diodes for the Q-band array, respectively, which are also consistent with measurements from other calibrators. The discrepancy in the monopole leakage between the two diodes for the Q-band array was anticipated from the measurements of the septum polarizers (Section~\ref{sec:polarimeter-assemblies}).  
  
\subsection{Bandpasses}
\label{sec:bandpasses}

Typical bandpasses for the Q-band and W-band arrays are shown with the spectrum of the atmosphere in Figure~\ref{fig:h2oabs}. Central frequencies and bandwidths are computed from discrete frequency steps as 
\begin{eqnarray}
\mathrm{Central~frequency} &  \equiv & \sum\limits_{i} I_{i}\nu_{i} \over \sum\limits_{i} I_{i} , \label{eqn:cf} \\
\mathrm {Bandwidth} & \equiv & [\sum\limits_{i} I_{i}]^{2} \Delta\nu \over \sum\limits_{i} I_{i}^{2} , \label{eqn:bw} 
\end{eqnarray}
\noindent where $I_{i}$ is the measured intensity from a detector diode for each frequency, $\nu_{i}$, and $\Delta\nu$ is the frequency step of the signal generator (100\,MHz). 

\begin{table*}[t]
\centering
\caption{Average bandwidths and central frequencies for the Q-band and W-band arrays.}
\begin{tabular}{l|ccc|ccc} \hline  
Band  &           \multicolumn{3}{c |}{Bandwidth (GHz)}  & \multicolumn{3}{c }{Central Frequency (GHz)} \\ 
 &     Value & Stat. error & System. error & Value & Stat. error & System. error \\ \hline
Q & 7.6 & 0.5 & 0.6 & 43.1 & 0.4 & 0.4 \\
W & 10.7 & - & 1.1 & 94.5 & - & 0.8 \\ \hline
\end{tabular}
\label{tab:rxbands}
\end{table*}

For the Q-band array, bandpasses were measured for each diode in the laboratory during the course of array testing and in end-of-season calibration measurements at the site. The laboratory measurement was performed by injecting a polarized carrier-wave signal from a signal generator with a standard-gain horn over a 35--50\,GHz range. The signal was injected into the receiver array through the cryostat window without additional imaging optics, with the horn approximately 3\,m away from the window. Sweeps were performed at least eight times. The average bandwidth and central frequency of the polarization modules are given in Table~\ref{tab:rxbands}. The statistical errors on this measurement are obtained by finding the standard deviation between the eight measurements for a given module, and then averaging that standard deviation for all modules.

Bandpasses were also measured at the site for the Q-band array by reflecting the swept signal from a small ($\sim$ 1\,cm$^{2}$) plate into the primary mirror. While measurements performed in the laboratory and at the site are consistent with each other, the variation in bandpass shape between the two days of data taking at the site showed that the systematic errors were larger in the experimental setup at the site, so laboratory measurements were used where available. Although the amplifier bias settings were different between the laboratory and the site measurements, a review of laboratory measurements revealed that changing the amplifer bias over the range of interest had no significant effect on the bandpasses. The systematic error in Table~\ref{tab:rxbands} is the average of the difference between the site and lab bandpasses. 

For the W-band array, bandpasses were measured at the site at the end of the observing season and the central frequency and bandwidth are also given in Table~\ref{tab:rxbands}. A standard-gain horn was mounted beside the secondary mirror, so it could illuminate the cryostat window from $\sim$ 1.5\,m away. The signal generator was swept over 72--120\,GHz, while the phase switches were held constant (no switching). In this configuration the signal can be sent down each module leg separately. The responses at each frequency bin for each module leg were combined to emulate the power combinations occurring in the module:

\begin{eqnarray}
P_{\rm pol} = P_{\rm A}P_{\rm B} \cos(2(\phi - \gamma)), 
\end{eqnarray}

\noindent where $P_{\rm A}$ and $P_{\rm B}$ are the measured bandpasses for the signal travelling through module legs A and B, respectively, $\phi$ is the detector angle (for example, a Q diode might have $\phi = 90^{\circ}$ and a U diode might have $\phi = 45^{\circ}$), and $\gamma$ is the angle of the polarized input from the signal generator. The measured signal is only dependent on the difference between the two angles. Systematic errors have two main sources: the accuracy with which the spike that was used to indicate the beginning of a sweep can be detected, and from reconstructing the bandpass for both module legs biased from data in which only one leg is biased. The first was computed by noting that the timing was accurate to 1.5\,ms, which corresponded to 0.7\,GHz during the sweep measurement. The second was computed by comparing measurements performed with both legs biased and the reconstruction from single-leg bandpasses from the total power stream. Because the total power stream does not have a dependence on detector angle $\phi$, the two should be identical and the difference represents the systematic error in the measurement. The systematic error was found to be 0.3\,GHz for the central frequency and 0.9\,GHz for the bandwidth.

\subsection{Responsivities}
\label{sec:arraytesting-responsivity}

The responsivities were characterized for the differential-temperature modules and the polarization modules separately with different calibration sources. Responsivities of the differential-temperature modules are computed from calibration observations of Jupiter, RCW38, and Venus, one of which was observed $\sim$ once per week for the Q-band receiver, and once a day for the W-band receiver. The average responsivity of the differential-temperature modules was 
2.2\,mV$\,\textrm{K}^{-1}$ for the Q-band array and 2.3\,mV$\,\textrm{K}^{-1}$ for the W-band array.

For the Q-band array, the absolute polarimeter responsivity for the central horn was determined from Tau A measurements performed every two days. Relative responsivity values among the polarization modules were measured from observations of the Moon (performed once per week). Sky dip measurements (elevation nods of $\sim 6^{\circ}$ for `normal' sky dips, and $\sim 40^{\circ}$ for `large' sky dips) are also used to obtain the relative total power responsivities of both the differential-temperature and polarized modules before each CES for the Q-band array (`flat fielding'). These frequent (once every $\sim$ 1.5 hours) responsivity measurements provide relative responsivity tracking for the differential-temperature and polarized modules on short time-scales. The relative responsivities were checked with an end-of-season wire-grid measurement and measurements of Tau A with off-center modules. 

For the W-band array, the Moon is too bright for relative responsivity calibration, so measurements from the wire-grid and Tau A from off-center modules were used. The average
responsivity for the polarized modules was 2.3\,mV$\,\rm{K}^{-1}$ for the Q-band
array \citep{quiet:2011} and 3.1\,mV$\,\rm{K}^{-1}$ for the W-band array \citep{quiet:2012}.
These responsivities are in terms of antenna temperature and include the gain factor of 130 from the preamplifier boards (Section~\ref{sec:electronics:readout}).

The responsivity depends on the amplifier bias settings.
The bias values for the Q-band array were found to be
dependent on the bias board temperature (typical values are 2\% of the average responsivity per\,$^{\circ}$C), which was
the motivation for thermally regulating the electronics enclosure within 1\,$^{\circ}$C of 25\,$^{\circ}$C. The final
responsivity model included a linear term for this temperature dependence, and it
was found to be a negligible systematic for scientific analysis \citep{quiet:2011}. The
bias circuit was upgraded in the W-band array, rendering the temperature
dependence of the boards negligible.

One potential concern when using amplifiers is signal compression: an input-dependent responsivity which is greatly reduced at high input powers. Compression is typically manifested as different responsivity values for different load temperatures, and has important consequences when using responsivities from calibration sources which are all usually much warmer than the CMB itself (for example, the Moon is $\sim$ 223\,K, \citet{moontemp}) and from extrapolating total power responsivities to polarization responsivities (for example, from sky dip measurements). For the Q-band array, responsivity measurements in the laboratory and at the site with different calibration sources were all consistent with each other, confirming that the modules were not operating in a compressed regime. Laboratory responsivity studies of the W-band modules using liquid nitrogen as a cold load showed some evidence for compression. In the field, the W-band modules exhibited compression during observations of the Moon. The emission from the $\sim 1^{\circ}$ Moon varies across its face \citep{moontemp}; polarized responsivities varied between the brightest and darkest portions of its face by 20\% (worst case 50\%). 

Compression affects the polarized signal and the
total power signal differently (Appendix~\ref{sec:appendix:compression}).  Since the sky dips measure total power responsivity only, this complicates the use of
sky dips to track relative polarized responsivity for the W-band array. As a result, daily Tau A measurements
of a single module were used to measure fast variations. Relative responsivities between the central module and the other modules are obtained from additional Tau A measurements and an end-of-season polarization grid measurement and these were used to extrapolate absolute responsivities to all modules.

\par
Additional laboratory studies performed after deployment explain why the W-band modules were operated in a compressed regime: passive components in the W-band modules had as much as twice the expected loss. To compensate for this loss, the amplifiers were biased higher than optimal. As a result, the bias power was large enough that it contributed a significant fraction of the power required to compress the amplifiers. Modules with new passive components having lower loss have been produced. These modules exhibit little compression and have noise temperatures closer to the $\sim$ 50\,K intrinsic W-band amplifier noise\citep{Reeves:2012}.

\subsection{Detector Angles}
\label{sec:detangles}
Absolute polarized detector angles were measured for the central module of each array through observations of Tau~A, whose position angle is known to
$0.2^{\circ}$ precision from IRAM measurements \citep{IRAM/TauA}. For the Q-band array, the absolute angle shifted by as much as 2$^{\circ}$ due to jumps in the pointing from a loose encoder during the first half of the Q-band season. The systematic uncertainties related to the encoder jumps are discussed in \citep{quiet:2011}.
The Q-band angle calibration relied on weekly Moon observations and an end-of-season sparse wire grid measurement to 
find the relative angles of the diodes. The relative angles between one of the diodes of the central module and every other diode from all $\sim 35$ Moon measurements deviated less than 0.2$^{\circ}$ from nominal, indicating that the relative angles remained nearly constant during the season. Relative detector angles are not affected by the encoder jumps.

The W-band array had a smaller, more efficient Tau A scan trajectory and was able to make measurements with all modules over the course of the season to obtain absolute angle calibration. The variance of detector angles for the central module from repeated measurements of Tau A is 0.3$^{\circ}$. The relative angles among the diodes were confirmed with end-of-season wire-grid measurements for both arrays to within 0.9$^{\circ}$. 

\begin{figure}[t]
\centering
\includegraphics[width=3in]{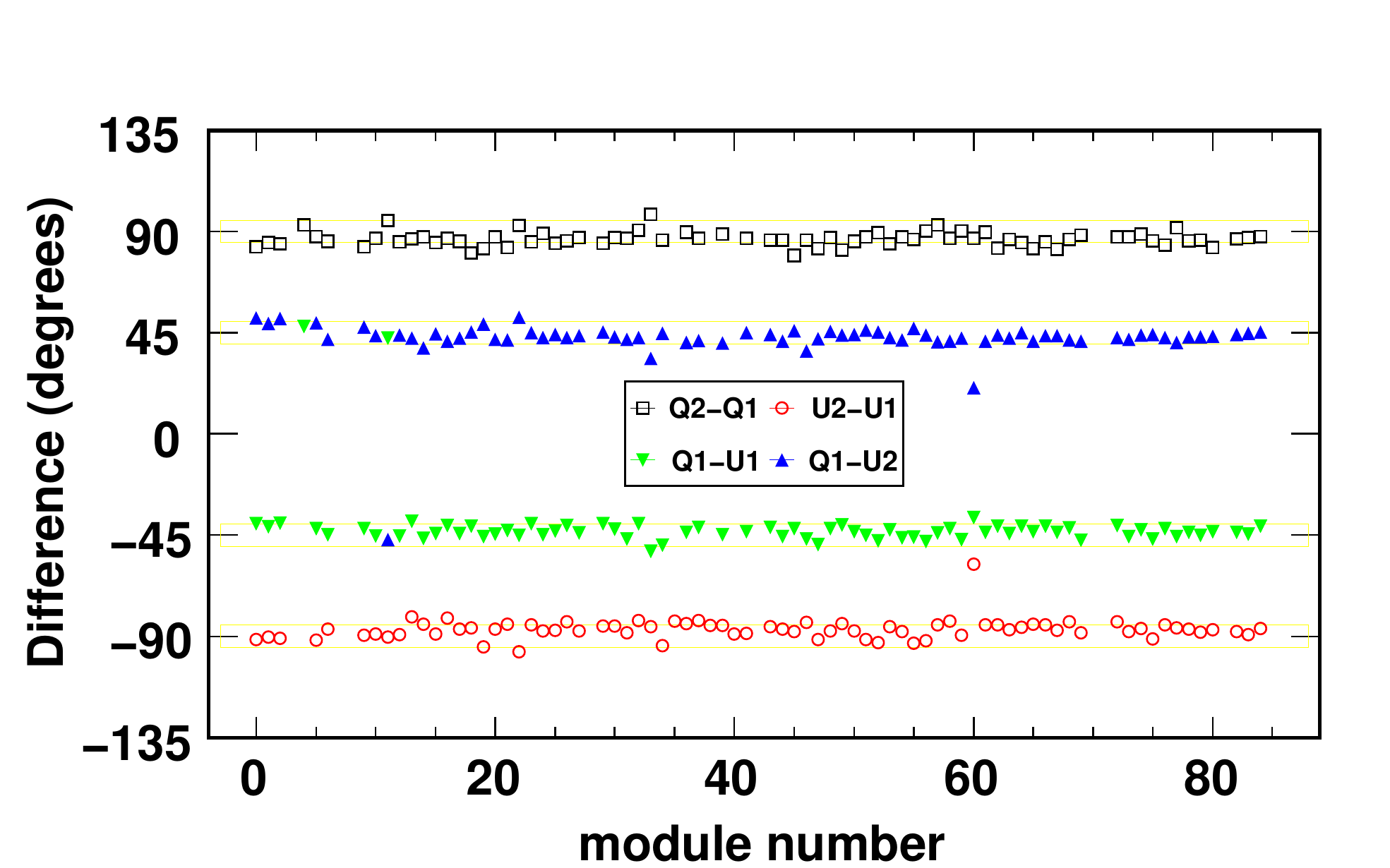}
\caption{Detector angle differences among diodes for each module in the W-band array.}
\label{fig:w91_angle_diff}
\end{figure}

Relative angles for all diodes in the W-band array are shown in Figure~\ref{fig:w91_angle_diff}. Systematic errors in the absolute angle are the largest source of systematic errors for the W-band array, which would limit the measurement of $r$ to 0.01 at $\ell\sim$ 100 \citep{quiet:2012}. 

\subsection{Pointing}
\label{sec:characterization:pointing}

The telescope pointing model is derived by fitting a physical model of the three-axis mount and telescope to astronomical observations \citep{naessphd}. 
The orientations of individual feed horns are determined by observations of the Moon and Jupiter. Then, holding the focal plane layout fixed, the parameters of the dynamical mount model are determined from observations of Jupiter, Venus, RCW38 (W-band only), and the Galactic plane\footnote{For preliminary Galactic maps from QUIET, see \citet{wehus:2012}.}. Optical observations are taken regularly with a co-aligned star camera and used to monitor the time evolution of the pointing model. Except for the mechanical problem with the deck-angle encoder during the first two months of Q-band observations \citep{quiet:2011}, no significant trends are found. 

The residual scatter after all pointing corrections is 3.5\,$^{\prime}$ rms in the Q-band observations \citep{quiet:2011} and 5.1\,$^{\prime}$ FWHM (2.2\,$^{\prime}$ rms) in the W-band observations \citep{quiet:2012}. This random scatter leads to an additional smearing factor in the final maps that may be modeled in terms of an effective window function, fully analogous to that of the instrument beam. Systematic errors are discussed for the Q-band \citep{quiet:2011} and W-band observations \citep{quiet:2012}. In order to validate the pointing model, a high-resolution W-band map of PNM J538-4405\citep{gold:2010} (a particularly bright point source in the QUIET observing field CMB-2) was produced and both its apparent position and angular size was found to be consistent with the assumed beam profile and estimated uncertainty.

\subsection{Noise Spectra}
\label{sec:arraytesting-nts}

\begin{figure}[t]
\centering
\includegraphics[width=2.9in]{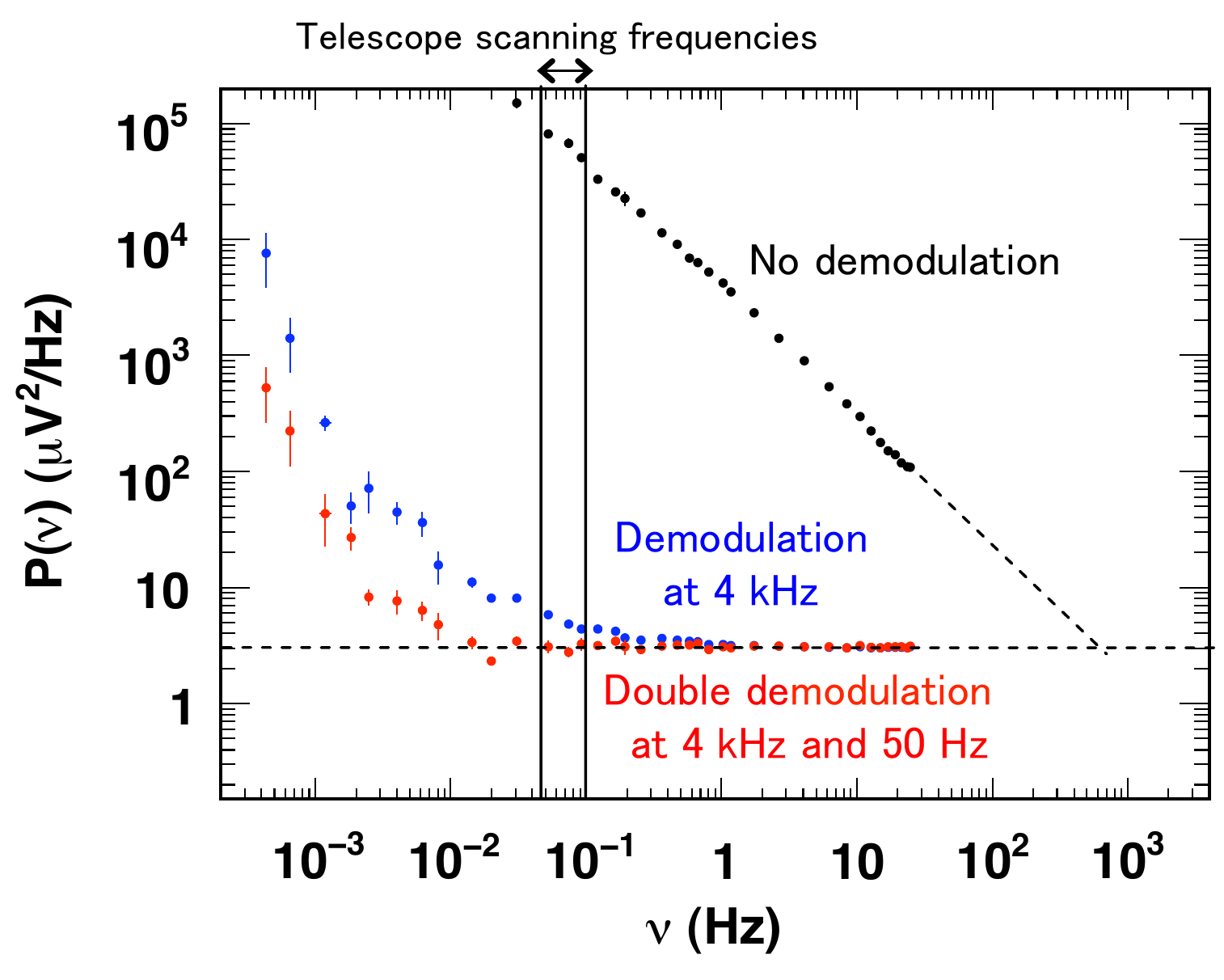}
\caption{Typical noise spectra of a W-band module with no demodulation, single demodulation, and double demodulation. Double demodulation reduces the knee frequencies below the telescope scan frequencies.}
\label{fig:w91_noise_spectrum}
\end{figure}

Noise measurements at the site were obtained from a noise spectrum fit to the Fourier-transform of the double-demodulated time stream for each CES. The measured noise floor should be proportional to the combination of module noise temperature, atmospheric temperature, contributions from optical elements, and CMB temperature. A power law with a flat noise floor was assumed for the functional form of the noise spectrum,
\begin{align}N(\nu) = \sigma_{0} \left[ 1+ \left( \frac{\nu}{\nu_{\mathrm{knee}}} \right)^{\alpha} \right] , \label{eqn:1fnoisemodel_lab}\end{align}\noindent 
where $N(\nu)$ and $\sigma_{0}$ have units V/$\sqrt{\mathrm{Hz}}$, $\nu$ is frequency, $\sigma_{0}$ is the white noise level, $\alpha$ is the slope of the low frequency end of the spectrum, and $\nu_{\mathrm{knee}}$ is the knee frequency. A typical noise power spectrum for a W-band module is given in Figure~\ref{fig:w91_noise_spectrum}, which also shows the effects on the noise of demodulating and double demodulating the time streams. After double demodulation, the median knee frequency is 5.5\,mHz (10\,mHz) for the Q-band (W-band) array; thus the noise is white at the scan frequencies of the telescope, 45--100\,mHz. 

The white noise is correlated among detector diodes within a given module. The correlation between Q and U diodes is expected \citep{Bischoff:2010}; the theoretical expectation and typical measured correlations are given in Table~\ref{tab:whitenoisecorr}. The measured correlation coefficients are larger than theoretically anticipated; the source is unknown but could come from unequal transmission in the coupling hybrid in the module, or from leakage of the atmosphere causing residual $1/f$ noise. However, the noise correlation among diodes is easily treated in the data analysis~\citep{quiet:2011}, and more importantly does not impact the measured polarized signal, which is a difference between diode signals: $(Q_{1}-Q_{2})$ and $(U_{1}-U_{2})$.
 
\begin{table}
\centering
\caption{Predicted and measured correlation coefficients among diodes.}
\begin{tabular}{c c c c}\hline
Diode $\times$ Diode & Design Value & \multicolumn{2}{c}{Typical Measured Value} \\ 
 & & Q-band & W-band \\ \hline
Q $\times$ Q & 0 & 0.23 $\pm$ 0.09 & 0.06$\pm$0.19 \\ \hline
U $\times$ U & 0 & 0.22 $\pm$ 0.08 & 0.06$\pm$0.21 \\ \hline
Q $\times$ U & 0.5 & 0.54 $\pm$ 0.08 & 0.48$\pm$0.11 \\ \hline 
\end{tabular}
\tablecomments{The error for each measured value is the standard deviation of the correlation coefficients among modules.}
\label{tab:whitenoisecorr}
\end{table}
 
 \subsection{System Noise Temperature}
\label{sec:arraytesting-sn}
 
The system noise is given by:
\begin{eqnarray}
\lefteqn{T_{\rm system} = } & & \\ \nonumber
& & T'_{\textrm{CMB}} + T_{\mathrm{atm}} +\frac{T_{\mathrm{R}}}{G_{\mathrm{atm}}} + \frac{T_{\mathrm{W}}}{G_{\mathrm{atm}}G_{\mathrm{R}}} + \\ \nonumber
& & \frac{T_{\mathrm{IR}}}{G_{\mathrm{atm}}G_{\mathrm{R}}G_{\mathrm{W}}} + \frac{T_{\mathrm{H}}}{G_{\mathrm{atm}}G_{\mathrm{R}}G_{\mathrm{W}}G_{\mathrm{IR}}} + \\ \nonumber
& & \frac{T_{\mathrm{SP}}}{G_{\mathrm{atm}}G_{\mathrm{R}}G_{\mathrm{W}}G_{\mathrm{IR}}G_{\mathrm{H}}} + \\ \nonumber
& & \frac{T_{\mathrm{module}}}{G_{\mathrm{atm}}G_{\mathrm{R}}G_{\mathrm{W}}G_{\mathrm{IR}}G_{\mathrm{H}}G_{\mathrm{SP}}} ,
\end{eqnarray}
\noindent where $T_{\mathrm{atm}}$ is the effective atmospheric temperature, $G_{\mathrm{atm}}=e^{-\tau}$ is the transmission through the atmosphere where $\tau$ is atmospheric opacity, $T'_{\mathrm{CMB}}$ is the brightness temperature of the CMB, $T_{\mathrm{module}}$ is the noise temperature of a QUIET module, \{$T_{\mathrm{R}}$,$G_{\mathrm{R}}$\}, \{$T_{\mathrm{W}}$,$G_{\mathrm{W}}$\}, \{$T_{\mathrm{IR}}$,$G_{\mathrm{IR}}$\}, \{$T_{\mathrm{H}}$,$G_{\mathrm{H}}$\}, \{$T_{\mathrm{SP}}$,$G_{\mathrm{SP}}$\} are the effective noise temperatures and gains for both reflectors (including ohmic and spillover contributions), window, IR blocker, horns, and septum polarizers, respectively (Table~\ref{tab:noisetemps_breakdown}).  

\begin{figure}[]
\centering
\includegraphics[width=3in]{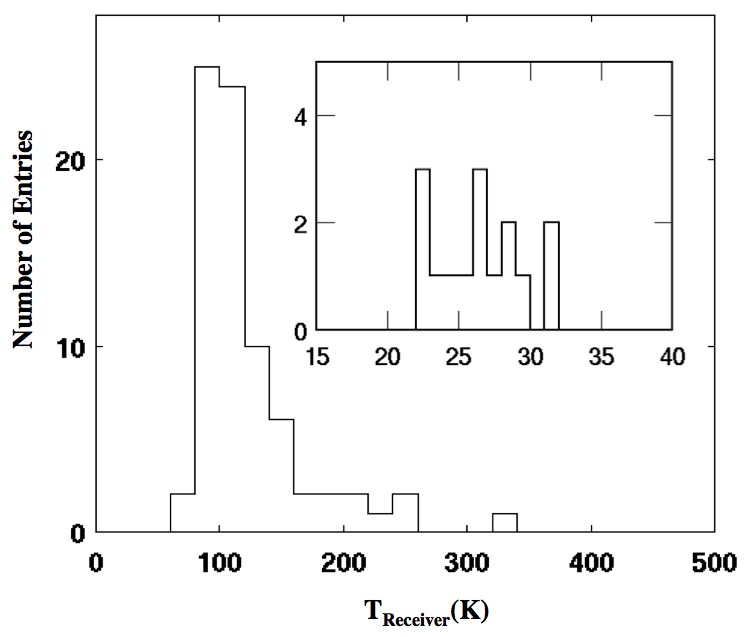}
\caption{System noise temperatures of W-band (main figure) and Q-band (inset) modules, after subtraction of the CMB temperature. For the Q-band values, the elevation-dependent atmospheric temperature was also subtracted. The Q-band (W-band) noise temperatures were obtained from normal (large) sky dips.} 
\label{fig:Trecs}
\end{figure}

The system noise can be found from the total power time streams taken during sky dips. During a sky dip, the sky temperature seen by the receiver changes with telescope elevation. Using an atmospheric model, the change in signal with this model-dependent change in sky temperature allows us to estimate the system noise. Histograms of receiver noise for both arrays are shown in Figure~\ref{fig:Trecs}. Receiver noise temperatures of 26\,K were determined for the Q-band array, and 106\,K for the W-band array, where the receiver noise was treated differently for the two (the atmosphere and CMB were removed for the Q-band histograms, while only the CMB was removed for the W-band array). 

The contribution to instrument noise due to the module alone can be estimated by subtracting assumed or measured values for all other known instrument noise sources (Table~\ref{tab:noisetemps_breakdown}). All components other than the modules are lossy; thus their noise temperatures are given by $(\frac{1}{G} - 1)\times T_{\mathrm{phys}}$, where $G$ is the gain of the component and $T_{\mathrm{phys}}$ is its physical temperature. The extrapolated module temperature is 15\,K for a Q-band module, and 77\,K for a W-band module. Measurements of the Q-band amplifiers give noise values of $\sim$ 18\,K; the most likely source of the discrepancy is that the loss in the septum polarizer was overestimated. Similar measurements for the W-band module give amplifier noise values of 50\,K. The discrepancy between the W-band module and amplifier temperatures stems from operating them uncompressed in the laboratory (this is explained in greater detail in Section~\ref{sec:arraytesting-responsivity}). 

\begin{table}[h]
\centering
\footnotesize
\caption{Estimated Contributions to the System Noise}
\begin{tabular}{lllll} \hline
Description & \multicolumn{2}{c}{Q-band} & \multicolumn{2}{c}{W-band} \\
& Gain & Noise(K) & Gain & Noise(K) \\ \hline
CMB+sky  & 0.96 & 11.1 & 0.98 & 5.9 \\
Reflectors & 0.99 & 2.7 & 0.99 & 2.7 \\  
Window & 0.99 & 2.8 & 0.983 & 4.8 \\
Horn & 0.99 & 0.2 & 0.99 & 0.2 \\
Septum Polarizer & 0.9 & 2.5 & 0.9 & 2.3 \\ 
TOTAL & 0.83  & 19 & 0.85 & 17 \\ \hline 
Measured $T_{\rm system}$ & - & 38 & - & 109 \\ 
Implied $T_{\rm module}$ & - & 15 & - & 77 \\ \hline
\end{tabular}
\tablecomments{The noise from each component has been divided by the gain of the previous elements in the optical chain. The values for the gain are not measured and are included for illustrative purposes; thus there are no associated error estimates. The atmospheric temperature and loss were computed for an elevation of 66$^{\circ}$ (the mid point of the CES elevation range), and a PWV of 1.2\,mm (Q-band) and 0.94\,mm (W-band).  All ambient temperatures are taken as 270\,K.}
\label{tab:noisetemps_breakdown}
\end{table}

\subsection{Instrument Sensitivity} 
\label{sec:sensitivity}

The sensitivity for the polarization response, $S_{\rm pol}$ ($\mu{\rm K}{\rm s}^{1/2}$), is calculated as the ratio of the white noise level to the responsivity. For the Q-band array, after data selection~\citep{quiet:2011}, the sensitivity is 69\,$\mu{\rm K}{\rm s}^{1/2}$ corresponding to an average module sensitivity of 275\,$\mu{\rm K}{\rm s}^{1/2}$. For the W-band array, the array sensitivity is 87\,$\mu {\rm K}{\rm s}^{1/2}$ \citep{quiet:2012}, corresponding to an average module sensitivity of 756\,$\mu {\rm K}{\rm s}^{1/2}$. Both values are given in thermodynamic units, so that the power detected by the receiver has been corrected from a Rayleigh-Jeans approximation to correspond to fluctuations in the blackbody temperature of the CMB. Functionally this is performed by dividing by $C_{\rm RJ}$, which is 0.95 (0.79) for the Q-band (W-band) central frequencies. These values can be compared to the expected sensitivity per module, $S_{\mathrm{pol}}$, from the radiometer equation \citep{Krauss:1986}:

\begin{equation}
S_{\mathrm{pol}} =  \frac{1}{C_{\mathrm{RJ}}} \times \frac{T_{\mathrm{instrument}}}{\sqrt{2 \Delta\nu} G_{\mathrm{total}}(1-f_{\mathrm{mask}})} .
\label{Trec_sensitivity}
\end{equation} 

\noindent Using the measured values for $T_{\rm system}$ and the atmospheric gain, $G_{\rm atm}$ (Table~\ref{tab:noisetemps_breakdown}), the bandwidths $\Delta\nu$ (Section~\ref{sec:bandpasses}), the Rayleigh-Jeans correction $C_{\rm RJ}$ for the CMB, and the fraction of the data masked during the phase switch transitions, $f_{\rm mask}$ (14\%, Section~\ref{sec:electronics}), sensitivity values of 310\,$\mu{\rm K}{\rm s}^{1/2}$ for the Q band, and 913\,$\mu{\rm K}{\rm s}^{1/2}$ for the W-band were found. Errors in bandpasses and the atmospheric temperature contribute directly to the difference between the two methods of computing the sensitivity $S_{\rm pol}$. A potential explanation for the greater discrepancy between these methods for the W-band array ($\sim$ 30\%) compared to the Q-band ($\sim$ 11\%) array is that $T_{\rm rec}$ is measured from the total power stream during sky dips, which could be compressed as much as 30\% (Appendix~\ref{sec:appendix:compression}) in the W-band data stream. This compression inflates the noise temperature by the same compression factor.

\section{Conclusions}
\label{sec:conclusions}
QUIET employs the largest HEMT-based receiver arrays to date.  
The 17-element Q-band array has a polarization sensitivity of 
69\,$\mu\mathrm{Ks}^{1/2}$, 
currently the most sensitive instrument in this band.
The 84-element W-band array has a 87\,$\mu\mathrm{Ks}^{1/2}$ sensitivity.  
Together the two arrays give the instrument sensitivity to angular scales
$\ell$ $\sim$ 25--975.
\par
The instrument design also achieves extremely low systematic errors.  
The optical design uses high-gain, low-crosspolar, and low-sidelobe 
corrugated feed horns and septum polarizers.  The receiver and mirrors 
are housed in an absorbing ground shield to reduce sidelobe pickup, and 
are mounted on a 3-axis telescope with boresight rotation.  The
polarimeter assemblies use electronic double demodulation to 
remove both $1/f$ noise and monopole leakage.  Finally, the differential-temperature assemblies
and calibration tools provide critical measurements and cross checks 
of the systematic errors.  The dominant systematic errors at $\ell \sim100$ 
are leakage for the Q-band instrument, and detector angle calibration for the W-band instrument.
QUIET's Q-band result has a 
systematic error of $r < 0.1$ at $\ell=100$ \citep{quiet:2011}, and $r < 0.01$ for the W-band result \citep{quiet:2012}, the lowest systematic uncertainty
on $r$ published to date.   

\acknowledgements
Support for the QUIET instrument and operation comes through the NSF
cooperative agreement AST-0506648. Support was also provided by NSF awards
PHY-0855887, PHY-0355328, AST-0448909, AST-1010016, and PHY-0551142; KAKENHI 20244041,
20740158, and 21111002; PRODEX C90284; a KIPAC Enterprise grant; and by the Strategic Alliance for
the Implementation of New Technologies (SAINT).  
This research used resources
of the National Energy Research Scientific Computing Center, which is supported by the
Office of Science of the U.S. Department of Energy under Contract No. DE-AC02-05CH11231.

Some work was performed on the Joint Fermilab-KICP Supercomputing
Cluster, supported by grants from Fermilab, the Kavli Institute for
Cosmological Physics, and the University of Chicago.  Some work was
performed on the Titan Cluster, owned and maintained by the University
of Oslo and NOTUR (the Norwegian High Performance Computing
Consortium), and on the Central Computing System, owned and operated
by the Computing Research Center at KEK.  Portions of this work were
performed at the Jet Propulsion Laboratory (JPL) and California
Institute of Technology, operating under a contract with the National
Aeronautics and Space Administration. The Q-band modules
were developed using funding from the JPL R\&TD program.  We acknowledge 
the Northrop Grumman Corporation for collaboration in the development and 
fabrication of HEMT-based cryogenic temperature-compatible MMICs.

C.D. acknowledges an STFC Advanced Fellowship and an ERC IRG grant under FP7.
R.B. acknowledges support from CONICYT project Basal PFB-06 and
ALMA-Conicyt 31070015.
A.D.M. acknowledges a Sloan foundation fellowship. H.K.E. acknowledges an
ERC Starting Grant under FP7.

PWV measurements were provided by the Atacama Pathfinder Experiment
(APEX). We thank CONICYT for granting permission to operate within the
Chajnantor Scientific Preserve in Chile, and ALMA for providing site
infrastructure support.
Field operations were based at the Don Esteban facility run by Astro-Norte.
We are particularly indebted to the engineers
and technician who maintained and operated the telescope: Jos\'e Cort\'es,
Cristobal Jara, Freddy Mu\~noz, and Carlos Verdugo.

In addition, we would like to acknowledge the following people for
their assistance in the instrument design, construction,
commissioning, operation, and in data analysis: Augusto Gutierrez
Aitken, Colin Baines, Phil Bannister, Hannah Barker, Matthew R. Becker, Alex
Blein, 
April Campbell, Anushya Chandra, Sea Moon Cho, Joelle Cooperrider,
Mike Crofts, Emma Curry, Maire Daly, Fritz Dejongh, Joy Didier, Greg
Dooley, Hans Eide, Will Grainger, Jonathon Goh, Peter Hamlington,
Takeo Higuchi, Seth Hillbrand, 
Ben Hooberman, Kathryn D. Huff, 
Norm Jarosik,
Eiichiro Komatsu,
Jostein Kristiansen, Donna Kubik,
Richard Lai, 
David Leibovitch, Kelly Lepo, Siqi Li, Martha Malin, Mark McCulloch, Oliver Montes, David Moore, 
Ian O'Dwyer, Gustavo Orellana, 
Stephen Osborne, Stephen Padin,
Felipe Pedreros,
Ashley Perko, Alan Robinson,
Jacklyn Sanders, Dale Sanford, Yunior Savon, 
Mary Soria, Alex Sugarbaker, 
David Sutton,
Matias Vidal, Liza Volkova, 
Stephanie Xenos, Octavio Zapata, and Mark Zaskowski.

\renewcommand{\theequation}{A\arabic{equation}}
\setcounter{equation}{0}  

\section{Appendix}
\label{sec:appendix}

\subsection{Compression}
\label{sec:appendix:compression}
This section explains some subtleties regarding nonlinearities, and how they affect
the polarization and total power measurements differently.  This complicates 
the use of periodic telescope sky dips to track the total power responsivity,
which is {\it assumed} to also track the polarization responsitivity.
During CMB operations, the receiver load temperature varies by $\sim$ 
2K due to changes in
the sky loading.  Nonlinearities also affect the use of large sky dip and Moon signals
to calibrate the total power responsivity.
For HEMT LNAs, compression (in which the amplifier gain depends on the 
input signal level) is the nonlinearity that is typically encountered in the 
QUIET operating regime. 
\par
The effect of compression on polarization responsivity is analyzed here.
Consider a horn looking at an unpolarized background at temperature $T_0$, where
$T_0 = T_{0x} = T_{0y}$, with axes $x$ and $y$ defined by the septum polarizer.  
Given below are the $Q_1$ diode measurements for the $0^\circ$ and 
$180^\circ$ leg B states, and the demod output (which is the polarization measurement):
\begin{eqnarray}
\nonumber
S_0(0^\circ) & = & g_0 \cdot \left(\frac{1}{2}(T_{0x} + T_{0y}) + \frac{1}{2}(T_{0x} - T_{0y})\right),\\
\nonumber
S_0(180^\circ) & = &  g_0\cdot \left(\frac{1}{2}(T_{0x} + T_{0y}) - \frac{1}{2}(T_{0x} - T_{0y})\right), \\
S_0(demod) & = & \frac{1}{2} (S_0(0^\circ) - S_0(180^\circ)) \nonumber \\ 
 & = & \frac{1}{2} \cdot g_0 \cdot (T_{0x} - T_{0y}) = 0, 
\label{eq:unpol}
\end{eqnarray}
where $g_0$ is the gain at temperature $T_0$.

Consider now the horn looking at a source on top of this background.
Without loss of generality, let the source be polarized in the $x$ direction at 
temperature $T_1$ such that $T_{1x} = T_{0x} + T_{Sx}$, $T_{1y} = T_{0y}$, 
$T_{avg} = \frac{1}{2} \cdot (T_{0x} + T_{Sx} + T_{0y})$.  Then:
\begin{eqnarray}
\nonumber
S_1(0^\circ) &  = & g_1 \cdot \left( T_{avg} + \frac{1}{2} (T_{0x} + T_{Sx} - T_{0y}) \right)\\
\nonumber
S_1(180^\circ) & = & g_1 \cdot \left( T_{avg} - \frac{1}{2} (T_{0x} + T_{Sx} - T_{0y}) \right) \\
\label{eq:compress-pol}
S_1(demod) & = & \frac{1}{2} \cdot g_1 \cdot T_{Sx} 
\end{eqnarray}
Note that the gain constant $g_1$ is relevant for the temperature $T_{avg}$, 
for the following reason. Since the incident E-fields at the horn input are 
linearly polarized, 
the septum polarizer splits the power equally between legs A and B.  Thus the legs see 
a constant input power given by $T_{avg}$, regardless of the instrumental position angle.  
Within the module, 
the LNAs are placed upstream of any phase-sensitive circuitry.  In this model, 
compression depends primarily on the input power at the first LNA.  Therefore, 
the first LNA sees power represented by $T_{avg}$, so a 
gain $g_1$ is associated to that input temperature.   
Thus, Equation~\ref{eq:compress-pol} shows that the polarization measurement is 
compressed by $(g_0 - g_1)/g_1$.  It is estimated that $(g_0 - g_1)/g_1$ changes 
by roughly $0.1\%$ per Kelvin for the W-band modules.
\par
Now consider the effect of compression on the total power responsivity.  
For an unpolarized
background source at temperature $T_0$, the $Q_1$ diode voltages 
for the leg B  $0^\circ$ and $180^\circ$  
states are as given in Equation~\ref{eq:unpol}, 
and the average (which gives the total power) is: 
\begin{eqnarray}
S_0(avg) & = & \frac{1}{2} \left( S_0(0^\circ) + S_0(180^\circ) \right) \nonumber \\ 
 & = & \frac{1}{2} \cdot g_0 \cdot (T_{0x} + T_{0y}) = g_0 \cdot T_{0}
\end{eqnarray}
Similarly, an unpolarized background source at temperature $T_1$ results in:
\begin{eqnarray}
S_1(avg) = g_1 \cdot T_1.
\end{eqnarray}
Here, $g_1$ and $g_0$ are the gains for temperatures $T_1$ and $T_0$ respectively.
It can be shown that:
\begin{eqnarray}
\label{eq:compress-tp}
S_1(avg) - S_0(avg) & = & g_1 (T_1 - T_0) c \\
\nonumber
c & = & \left( 1 - \frac{g_0 - g_1}{g_1}\frac{T_0}{T_1-T_0} \right) 
\end{eqnarray}
where $c$ is the ratio between the observed signal 
difference and the expected difference without 
compression.
\par
Comparing Equation~\ref{eq:compress-tp} with \ref{eq:compress-pol}, the total power
sensitivity compression is magnified by $T_0/(T_1 - T_0)$.
Assuming as an example, $T_1 - T_0$ = 2K (typical for a skydip), a system
temperature of $T_0$= 120K, and a typical gain
compression of $(g_0 - g_1)/g_1 = 0.002$ over that range, the 
resulting ratio is $c = 93\%$, or $7\%$ signal loss.
Therefore, in the data analysis, the absolute responsivities are derived from
polarized source measurements to avoid systematic biases of this type for the 
W-band diodes.

\subsection{Double demodulation}
\label{sec:appendix:dd}
\begin{figure}[t]
\centering
\includegraphics[width=3in]{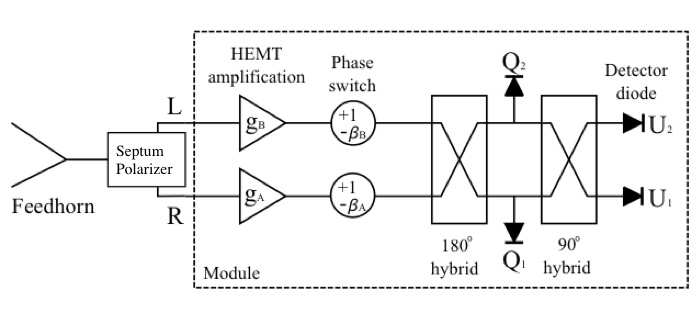}
\caption{Schematic of signal processing in a QUIET polarimeter assembly.}
\label{fig:mod_schem}
\end{figure}

This section discusses some imperfections in the module and their mitigation using double demodulation.
Table~\ref{ideal_diode_outputs.tbl} shows the detector diode outputs of an ideal module for the 
two leg B states, with the leg A state held fixed.   The idealization (see Figure~\ref{fig:mod-schem-ideal}) 
assumes equal transmission between the two leg B states, and between 
the two leg A states, and an ideal septum 
polarizer (see Section~\ref{sec:omtleakage}).    In practice, the transmissions are unequal, thus 
requiring extra parameters to describe the module.  Without loss of generality, let the transmission 
through the $0^\circ$($\uparrow$) state of legs A and B be equal to unity, and 
define $\beta_A$ and $\beta_B$ to be the transmissions through
these legs for the $180^\circ$($\downarrow$) state.  Using $g_A$ and $g_B$ as the effective voltage gains of the two legs (see Figure~ \ref{fig:mod_schem}), the detector diode voltages are given by:
\begin{widetext}
\begin{equation}
 V_{Q_{1}} (V_{Q_{2}})
  = \frac{1}{4}
  \left\{
   \begin{aligned}
    & \frac{1}{2}({g_A}^2 + {g_B}^2) I
     & & + \frac{1}{2}({g_A}^2 - {g_B}^2) V
     & & \pm g_A g_B Q
     \\
    & \frac{1}{2}({g_A}^2 + {g_B}^2 {\beta_B}^2) I
     & & + \frac{1}{2}({g_A}^2 - {g_B}^2 {\beta_B}^2) V
     & & \mp g_A g_B \beta_B Q
     \\
    & \frac{1}{2}({g_A}^2 {\beta_A}^2 + {g_B}^2) I
     & & + \frac{1}{2}({g_A}^2 {\beta_A}^2 - {g_B}^2) V
     & & \mp g_A g_B \beta_A Q
     \\
    & \frac{1}{2}({g_A}^2 {\beta_A}^2 + {g_B}^2 {\beta_B}^2) I
     & & + \frac{1}{2}({g_A}^2 {\beta_A}^2 - {g_B}^2 {\beta_B}^2) V
     & & \pm g_A g_B \beta_A \beta_B Q
   \end{aligned}
  \right\}
\end{equation}
\begin{equation}
 V_{U_{1}} (V_{U_{2}})
  = \frac{1}{4}
  \left\{
   \begin{aligned}
    & \frac{1}{2}({g_A}^2 + {g_B}^2) I
     & & + \frac{1}{2}({g_A}^2 - {g_B}^2) V
     & & \mp g_A g_B U
     \\
    & \frac{1}{2}({g_A}^2 + {g_B}^2 {\beta_B}^2) I
     & & + \frac{1}{2}({g_A}^2 - {g_B}^2 {\beta_B}^2) V
     & & \pm g_A g_B \beta_B U
     \\
    & \frac{1}{2}({g_A}^2 {\beta_A}^2 + {g_B}^2) I
     & & + \frac{1}{2}({g_A}^2 {\beta_A}^2 - {g_B}^2) V
     & & \pm g_A g_B \beta_A U
     \\
    & \frac{1}{2}({g_A}^2 {\beta_A}^2 + {g_B}^2 {\beta_B}^2) I
     & & + \frac{1}{2}({g_A}^2 {\beta_A}^2 - {g_B}^2 {\beta_B}^2) V
     & & \mp g_A g_B \beta_A \beta_B U
   \end{aligned}
  \right\}
\end{equation}
\end{widetext}
where the upper (lower) signs correspond to the signal of the diodes $Q_1$ and
$U_1$ ($Q_2$ and $U_2$).
The four rows for each $V_i$ correspond to the phase switch states of
$(\mathrm{A, B}) = (\uparrow, \uparrow)$, $(\uparrow, \downarrow)$,
$(\downarrow, \uparrow)$, and $(\downarrow, \downarrow)$, from the top
to the bottom.
Transmission imbalance between the phase switch states, 
signified by the deviation from unity of 
$\beta_{\rm A}$ and $\beta_{\rm B}$, causes
I $\rightarrow$ Q/U leakage. This can be seen in Table~\ref{demod.tbl}, 
showing the demodulated output dependences on $I \cdot (1-\beta_B^2)$. 
However, the difference between the $\uparrow$ and $\downarrow$ demodulated outputs 
is free from $I-$ dependence.  Calculating the time series of this difference is 
referred to as {\em double demodulation}.  
For the W-band, the rms of I $\rightarrow$ Q/U leakage distribution 
is reduced from roughly 0.8\% in the demodulated stream, to 0.4\% in the double-demodulated
stream.   A smaller reduction of $< 0.1\%$ is found for the Q-band, as the leakage is
dominated by other effects.

\begin{table}
\begin{center}       
\caption{Expressions for the Demodulated Output} 
\label{demod.tbl}
\begin{tabular}{|c|c|c|} 
\hline
\rule[-1ex]{0pt}{3.5ex} Leg A  &   &  Demodulated \\
\rule[-1ex]{0pt}{3.5ex} phase state & Diode  &  output  \\
\hline
\rule[-1ex]{0pt}{3.5ex} $\uparrow$ & $Q_1$ ($Q_2$) & $\frac{1-{\beta_{\rm B}}^2}{4}{g_{\rm B}}^2 I \pm \frac{1+\beta_{\rm B}}{2} g_{\rm A} g_{\rm B} Q$ \\
\hline
\rule[-1ex]{0pt}{3.5ex} $\uparrow$ & $U_1$ ($U_2$) & $\frac{1-{\beta_{\rm B}}^2}{4}{g_{\rm B}}^2 I \mp \frac{1+\beta_{\rm B}}{2} g_{\rm A} g_{\rm B} U$ \\
\hline 
\rule[-1ex]{0pt}{3.5ex} $\downarrow$ & $Q_1$ ($Q_2$) & $\frac{1-{\beta_{\rm B}}^2}{4} {g_{\rm B}}^2 I \mp \beta_{\rm A} \frac{1+\beta_{\rm B}}{2} g_{\rm A} g_{\rm B} Q$ \\
\hline
\rule[-1ex]{0pt}{3.5ex} $\downarrow$ & $U_1$ ($U_2$) & $\frac{1-{\beta_{\rm B}}^2}{4} {g_{\rm B}}^2 I \pm \beta_{\rm A} \frac{1+\beta_{\rm B}}{2} g_{\rm A} g_{\rm B} U$ \\
\hline
\end{tabular}
\tablecomments{Demodulated signal for each leg A phase state with 
the leg B phase state switching at 4 kHz. A factor of 1/4 has been omitted from each 
expression. The terms involving Stokes V are also omitted for simplicity. The 
upper (lower) signs correspond to detector diodes $Q_1$ and $U_1$ ($Q_2$ and $U_2$).}
\end{center}
\end{table} 

\subsection{Polarimeter Assembly Offset and $I\rightarrow Q/U$ Leakage}
\label{sec:omtleakage}
As shown in Section \ref{sec:appendix:dd}, it can be assumed that 
the module does not generate any instrumental polarization on its own 
since the double demodulation procedure nulls out this effect.
However, the interaction between the module and septum polarizer can cause 
irreducible instrumental polarization and offsets; this section derives these couplings.
Since the module does not generate instrumental polarization on its own, the module measures:
\begin{equation}
\label{eq:Qm}
Q_m = 2\Re(L_m^*R_m)
\end{equation}
and
\begin{equation}
\label{eq:Um}
U_m = -2\Im(L_m^*R_m),
\end{equation}
where $L_m$ and $R_m$ are the signals transmitted into the module inputs.  Without loss of generality,
all the constant factors are absorbed into the responsivity and set to unity.
The signals transmitted into the module inputs need not be the same as the $L$ and $R$ 
components at the septum polarizer input; this difference is a cause of 
instrumental polarization.

The effect of the septum polarizer is described by a 4x4 complex scattering matrix $S$:
\begin{equation}
\left( \begin{array}{c}
E_x' \\
L' \\
R' \\
E_y' \end{array} \right)
= S \cdot 
\left(\begin{array}{c}
E_x \\
L_r \\
R_r \\
E_y \end{array}\right), 
\end{equation}
\begin{equation}
S = 
\left( \begin{array}{cccc}
r_1 & \frac{e^{i\gamma}}{\sqrt{2}}\tau_{21} & \frac{e^{i\gamma}}{\sqrt{2}}\tau_{31} & r_{41} \\
\frac{e^{i\gamma}}{\sqrt{2}}\tau_{21} & r_2 & c &  i\frac{e^{i\gamma}}{\sqrt{2}}\tau_{24} \\
\frac{e^{i\gamma}}{\sqrt{2}}\tau_{31} & c & r_3 & -i\frac{e^{i\gamma}}{\sqrt{2}}\tau_{34} \\
r_{41} & i\frac{e^{i\gamma}}{\sqrt{2}}\tau_{24} & -i\frac{e^{i\gamma}}{\sqrt{2}}\tau_{34} & r_4
\end{array} \right),
\end{equation}
where $E_x$ and $E_y$ are electric field components at the septum polarizer input port;
$L'$ and $R'$ 
are the fields at the two septum polarizer output ports; $E_x'$ and $E_y'$ are 
the electric fields emitted from the septum polarizer back toward the feed horn;
$L_r$ and $R_r$ are signals reflected (or emitted) 
from the module inputs traveling back toward the septum polarizer output ports;
$e^{i\gamma}$ is the propagation phase shift; $\tau_{ij}$ and  $r$ are 
transmission and reflection coefficients respectively;
$c$ is a measure of the isolation between the output ports.
For an ideal septum polarizer, $\tau = 1$ and $r = c = 0$.
Symmetry across the septum implies $\tau_{21}=\tau_{31}$, $\tau_{24}=\tau_{34}$, 
$r_2=r_3$, and $r_{41}=0$, although manufacturing errors can cause these conditions to be violated.
As described in Sections \ref{sec:polarimeter-assemblies} and 
\ref{sec:characterization},
there are small departures from ideal operation.  In this section these departures are computed up to second order.
Note that the scattering matrix is frequency-dependent.  The analysis given here is strictly for a 
single frequency.  In practice, the result should be averaged with the effective bandpass.  The median value of the Q(W)-band, band-averaged return loss $(=-20\log|r|)$ for the septum polarizers is 19(30) dB, while the median value of the Q(W)-band, band-averaged isolation $(=-20\log|c|)$ is 22(28) dB.  Another quantity of interest is median value of the band averaged (linear) axial ratio of the septum polarizers.  This is measured to be 1.12(1.07) for Q(W)-band and implies a cross polar discrimination of 24.9(29.4) dB.
\par
A perturbative expansion is used to derive $L_m$ and $R_m$ which are the fields transmitted into the module inputs due to a 
sky source consisting of fields $E_x$ and $E_y$.  Here a noiseless module is assumed.  The case of a noise
signal from the module is described later.   To lowest order, the $S$ matrix applied to the 
column vector $(E_x,0,0,E_y)$ yields $(0,L',R',0)$, where 
\begin{equation}
\begin{aligned}
\label{eq:Lprime}
L' & = & \frac{e^{i\gamma}}{\sqrt{2}}\left( \tau_{21}\frac{L+R}{\sqrt{2}} + 
\tau_{24}\frac{L-R}{\sqrt{2}}\right) 
 \\
& = & \frac{e^{i\gamma}}{2}\left[ (\tau_{21}+\tau_{24})L + 
(\tau_{21}-\tau_{24})R\right].
\end{aligned}
\end{equation}
Similarly,
\begin{equation}
\label{eq:Rprime}
R' = \frac{e^{i\gamma}}{2}\left[ (\tau_{31} + \tau_{34})R + (\tau_{31} - \tau_{34})L\right].
\end{equation}
where $E_x = (L+R)/\sqrt{2}$ and 
$E_y = (L-R)/(i\sqrt{2})$.  However, $L_m$ and $R_m$ differ from $L'$ and $R'$ due to 
reflection at the module input.  
Let $r_L$ ($r_R$) be the reflection coefficient at the module's $L$ ($R$) input.  
Then the $S$ matrix applied to $(E_x,r_L L',r_R R',E_y)$ yields $(-,L_m,R_m,-)$ where
\begin{equation}
\label{eq:Lm}
L_m = (1 + r_2r_L)L' + cr_RR'.
\end{equation}
\begin{equation}
\label{eq:Rm}
R_m = (1 + r_3r_R)R' + cr_LL'.
\end{equation}
and for simplicity, the expressions for the first and fourth component are omitted.
The module output is 
\begin{equation}
\label{eq:LmRm}
L_m^{*}R_m = L'^{*}R'(1 + r_3r_R + r_2^{*}r_L^{*}) + L'^{*}L'cr_L + R'^{*}R'c^{*}r_R^{*}
\end{equation}
where $r_i$, $c$, $r_R$ and $r_L$ are assumed to be small, and terms above second order are dropped.
\par
In the following, the RHS of Equation \ref{eq:LmRm} is simplified into the underlying physics parameters $Q$ and $U$ in order
to identify the sources of instrumental polarization.  The terms $L'^{*}L'$ and $ R'^{*}R'$ need only be 
calculated to leading order since they appear in Equation \ref{eq:LmRm} multiplied by the 
second order terms $cr_L$ and $c^{*}r_R^{*}$.  To leading order, $ L'^{*}L'$  =  $L^{*}L$ and 
$R'^{*}R'$ = $R^{*}R$ since $\tau_{ij}\approx1$.
\par
The first term in Eq. \ref{eq:LmRm} is expanded by substituting Equation \ref{eq:Lprime} and 
\ref{eq:Rprime} and using $L^{*}R = (Q-iU)/2$, $LL^{*} = (I+V)/2$, and $RR^{*} = (I-V)/2$ to obtain
\begin{equation}
\begin{aligned}
L'^{*}R' = & \frac{1}{4}((\tau_{21}^{*}\tau_{31} + \tau_{24}^{*}\tau_{34})Q 
 -  i(\tau_{21}^{*}\tau_{34} + \tau_{24}^{*}\tau_{31})U \\
& + (\tau_{21}^{*}\tau_{31}-\tau_{24}^{*}\tau_{34})I 
 + (\tau_{24}^{*}\tau_{31} -\tau_{21}^{*}\tau_{34})V).
\end{aligned}
\end{equation}
The first two terms are the expected response to $Q$ and $U$.  The presence of $\tau_{ij}$ in these terms 
parameterizes the imperfections in the septum polarizer 
transmissions.  These terms reduce the gain to $Q$ and $U$, and in general 
cause mixing between $Q$ and $U$.  In practice, the gain is absorbed into the 
calibration of the total system 
responsivity,\footnote{The effect of the gain difference on the leakage terms 
is a neglected higher order effect.} and the $Q/U$ leakage is absorbed into the detector angle as defined in Equations~\ref{eqn:psi_q} and~\ref{eqn:psi_u}.  Therefore, these two terms do not cause instrumental polarization, and
 these imperfections can be neglected in the following discussion.  By the same argument, the 
terms $r_3r_R$ and $r_2^{*}r_L^{*}$ in Equation~\ref{eq:LmRm} can be ignored since 
their only effect is to change the gain and detector angle.
\par
The third and fourth terms represents $I \rightarrow Q/U$ and $V\rightarrow Q/U$ leakage respectively. 
Since $V\ll I$ for reasonable sources and the coefficients have the same order, these 
circular polarization leakages are neglected.
Combining these simplifications, the right-hand-side of 
Equation \ref{eq:LmRm} becomes:
\begin{equation}
\begin{aligned}
L_m^{*}R_m = & 
\frac{1}{4}\left[ 2\widetilde{Q} -2\widetilde{U} +  (\tau_{21}^{*}\tau_{31}-\tau_{24}^{*}\tau_{34})I \right] \\ 
  & + \frac{1}{2}(cr_L + c^{*}r_R^{*})I,
\end{aligned}
\end{equation}
where $2\widetilde{Q}=(\tau_{21}^{*}\tau_{31} + \tau_{24}^{*}\tau_{34})Q$ and $2\widetilde{U}=i(\tau_{21}^{*}\tau_{34} + \tau_{24}^{*}\tau_{31})U$.  Using Equation~\ref{eq:Qm} and ignoring $U \rightarrow Q$ leakage, the module output is:
\begin{equation}
Q_m = \Re(\widetilde{Q}) + \frac{1}{2}\Re(\tau_{21}^{*}\tau_{31}-\tau_{24}^{*}\tau_{34}) I + \Re(cr_L + c^{*}r_R^{*})I,
\end{equation}
where the first term is the expected response, the second term is $I\rightarrow Q$ leakage 
due to differential loss, and the third term is leakage caused by reflections at the module 
inputs coupling with the septum polarizer crosstalk.  Similarly, using Equation~\ref{eq:Um} and ignoring the $Q \rightarrow U$ leakage:
\begin{equation}
U_m = \Im(\widetilde{U}) -  \frac{1}{2}\Im(\tau_{21}^{*}\tau_{31}-\tau_{24}^{*}\tau_{34}) I - \Im(cr_L + c^{*}r_R^{*})I.
\end{equation}
In summary, the two equations above describe the measurements of a sky signal in the absence of noise
from the module. 
\par
Now consider the case of noise emitted from the module inputs, reflecting from the septum polarizer
and returning into the module.  Module noise stems primarily from the HEMT-based first stage LNAs.  
Since the sky signal and module noise are relatively incoherent, they decouple and the sky 
signal can be neglected in the following.
Let the module noise fields be given by the column vector $(0,L_r,R_r,0)$.   Applying the $S$ matrix, the vector $(-,L_m,R_m,-)$ is obtained where
\begin{equation}
L_m = L' = r_2L_r + cR_r 
\end{equation}
\begin{equation}
R_m = cL_r + r_3R_r.
\end{equation}
The output is 
\begin{equation}
L_m^{*}R_m = r_2^{*}L_r^{*}cL_r +  c^{*}R_r^{*} r_3R_r
\end{equation}
because the $L_rR_r$ terms average to zero due to the fact that the two amplifier noises are uncorrelated.
Thus each output acquires an offset
\begin{equation}
Q_m = 2L_r^{*}L_r\Re(r_2^{*}c) + 2 R_r^{*}R_r\Re(c^{*}r_3).
\end{equation}
\begin{equation}
U_m = -2L_r^{*}L_r\Im(r_2^{*}c) - 2 R_r^{*}R_r\Im(c^{*}r_3)
\end{equation}
The offset is independent of the input $I$; however, it is modulated by gain fluctuations so the offset also contributes to 1/f noise.

\bibliographystyle{apj}
\bibliography{instrument}

\end{document}